\documentclass[aps,prc,noshowpacs,superscriptaddress,floatfix,nofootinbib,10pt,twocolumn]{revtex4}

\newcommand{\ifproofpre}[2]{#1}

\makeatletter

\def\print@float#1#2{%
 \lengthcheck@sw{%
  \total@float{#1}%
 }{}%
 \@ifxundefined@cs{#1write}{}{%
  \begingroup
   \@booleanfalse\floats@sw
   #2%
   \raggedbottom
   \def\array@default{v}
   \let\@float\@float@LaTeX
   \let\@dblfloat\@dblfloat@LaTeX
   \let\trigger@float@par\triggerpar
   \let@environment{#1}{#1@float}%
   \let@environment{#1*}{#1@floats}%
   \expandafter\prepdef\csname#1\endcsname{\trigger@float@par}%
   \expandafter\prepdef\csname#1*\endcsname{\trigger@float@par}%
   \@namedef{fps@#1}{h!}%
   \expandafter\immediate
   \expandafter\closeout
               \csname#1write\endcsname
   \everypar{%
    \global\let\trigger@float@par\relax
    \global\everypar{}\setbox\z@\lastbox

   }%
   \input{\csname#1@stream\endcsname}%
  \endgroup
  \global\expandafter\let\csname#1write\endcsname\relax
 }%
}%

\def\oneapage{\def\begin@float@pagebreak{\newpage}\def\end@float@pagebreak{}}%

\makeatother

\usepackage{graphicx}
\usepackage{amsmath}
\usepackage{amssymb}
\usepackage{mathtools}
\usepackage{isotope}
\usepackage{dcolumn}  
\newcolumntype{x}[1]{D{.}{.}{#1}}
\usepackage{mathrsfs}

\usepackage{liealg}
\usepackage{wignert}
\usepackage{mcmath}

\newcommand{\reffigstrans}{\ref{fig-7be0-band1}--\ref{fig-12be0-band2}}
\newcommand{\reffigstransodd}{\ref{fig-7be0-band1}--\ref{fig-11be1-band1}}
\newcommand{\reffigstranseven}{\ref{fig-8be0-band1}--\ref{fig-12be0-band2}}



\newcommand{\Nmax}{{N_\text{max}}}
\newcommand{\scrD}{{\mathscr{D}}}
\newcommand{\scrA}{{\mathscr{A}}}
\newcommand{\scrR}{{\mathscr{R}}}
\newcommand{\MeV}{{\mathrm{MeV}}}

\newcommand{\fm}{{\mathrm{fm}}}
\newcommand{\hw}{{\hbar\Omega}}

\newcommand{\Jvec}{\vec{J}}
\newcommand{\Tvec}{\vec{T}}
\newcommand{\Qvec}{\vec{Q}}
\newcommand{\Mvec}{\vec{M}}
\newcommand{\lvec}{\vec{\ell}}
\newcommand{\svec}{\vec{s}}
\newcommand{\Dvec}{\vec{D}}

\newcommand{\glp}{{g_{\ell,p}}}
\newcommand{\gln}{{g_{\ell,n}}}
\newcommand{\gsp}{{g_{s,p}}}
\newcommand{\gsn}{{g_{s,n}}}
\newcommand{\gli}{{g_{\ell,i}}}
\newcommand{\gsi}{{g_{s,i}}}

\newcommand{\Dlp}{{\Dvec_{\ell,p}}}
\newcommand{\Dln}{{\Dvec_{\ell,n}}}
\newcommand{\Dsp}{{\Dvec_{s,p}}}
\newcommand{\Dsn}{{\Dvec_{s,n}}}

\begin{document}


\title{
Emergence of rotational bands in \textit{ab initio} no-core
configuration interaction calculations of the $\boldmath\isotope{Be}$
isotopes}

\author{P. Maris}
\affiliation{Department of Physics and Astronomy, Iowa State University, Ames, Iowa 50011-3160, USA}

\author{M. A. Caprio}
\affiliation{Department of Physics, University of Notre Dame, Notre Dame, Indiana 46556-5670, USA}

\author{J. P. Vary}
\affiliation{Department of Physics and Astronomy, Iowa State University, Ames, Iowa 50011-3160, USA}

\date{\today} 

\begin{abstract}
The emergence of rotational bands is observed in no-core configuration
interaction (NCCI) calculations for the $\isotope{Be}$ isotopes
($7\leq A \leq 12$), as evidenced by rotational patterns for
excitation energies, electromagnetic moments, and electromagnetic
transitions.  Yrast and low-lying excited bands are found.  The
results indicate well-developed rotational
structure in NCCI calculations, using the JISP16 realistic nucleon-nucleon
interaction within finite, computationally accessible
configuration spaces.
\end{abstract}


\maketitle

\section{Introduction}
\label{sec-intro}

Nuclei exhibit a wealth of collective phenomena, including clustering,
rotation, and
pairing~\cite{rowe2010:collective-motion,bohr1998:v1,bohr1998:v2}.
Collective dynamics have been extensively modeled in macroscopic
phenomenological
descriptions~\cite{rowe2010:collective-motion,bohr1998:v2,iachello1987:ibm,eisenberg1987:v1}.
Aspects of collectivity may also be obtained microscopically in the
conventional shell model, with an inert core and effective valence
interactions, \textit{e.g.}, Elliott $\grpsu{3}$
rotation~\cite{elliott1958:su3-part1,harvey1968:su3-shell} or rotation
in large-scale nuclear structure calculations of medium-mass
nuclei~\cite{horoi2006:56ni-shell}.  However, recent developments in
large-scale calculations have brought significant progress in the
$\textit{ab initio}$ description of light nuclei (\textit{e.g.},
Refs.~\cite{pieper2004:gfmc-a6-8,neff2004:cluster-fmd,hagen2007:coupled-cluster-benchmark,bacca2012:6he-hyperspherical,shimizu2012:mcsm,barrett2013:ncsm}).
We may now therefore hope to observe the emergence of collective
phenomena directly from first principles, that is, in a fully
\textit{ab initio} calculation of the nucleus, as a many-body system
in which all the constituent protons and neutrons participate, with
realistic interactions.

In \textit{ab initio} no-core configuration interaction (NCCI)
approaches~--- such as the no-core shell model
(NCSM)~\cite{navratil2000:12c-ab-initio,barrett2013:ncsm},
no-core full configuration (NCFC)~\cite{maris2009:ncfc},
importance-truncated NCSM
(IT-NCSM)~\cite{roth2007:it-ncsm-40ca,roth2009:it-ncsm-16o}, no-core
Monte Carlo shell model (MCSM)~\cite{abe2012:fci-mcsm-ncfc}, and
symmetry-adapted NCSM
(SA-NCSM)~\cite{dytrych2008:sp-ncsm,dytrych2013:su3ncsm} methods~---
the nuclear many-body bound-state eigenproblem is formulated as a
Hamiltonian matrix diagonalization problem.  The Hamiltonian is
represented with respect to a basis of antisymmetrized products of
single-particle states, generally harmonic oscillator states.  (For
the lightest nuclei, an antisymmetrized basis in Jacobi coordinates
has also been used.)  The problem is solved for the full system
of $A$ nucleons,
\textit{i.e.}, with no inert core.  In practice, such calculations
must be carried out in a finite space, typically obtained by truncating the
many-body basis according to a maximum allowed number $\Nmax$ of
oscillator excitations above the lowest oscillator configuration
(\textit{e.g.}, Ref.~\cite{navratil2009:ncsm}).  With increasing
$\Nmax$, the results converge towards those which would be achieved in
the full, infinite-dimensional space for the many-body system.

Computational restrictions limit the extent to which converged
calculations can be obtained for the observables needed to identify
collective phenomena.  In particular, the observables most indicative
of rotational collectivity~--- $E2$ moments and transition strengths~--- present
special challenges for convergence in an NCCI
approach~\cite{bogner2008:ncsm-converg-2N,cockrell2012:li-ncfc}, due
to their sensitivity to the large-radius asymptotic portions of the
nuclear wave function.  Nonetheless, signatures of collective
phenomena, \textit{e.g.}, deformation and clustering, have already
been obtained in \textit{ab initio} calculations of various
types~\cite{wiringa2000:gfmc-a8,dytrych2007:sp-ncsm-dominance,dytrych2008:sp-ncsm-deformation,neff2008:clustering-nuclei,maris2012:mfdn-ccp11,shimizu2012:mcsm}.

In this work, we observe the emergence of collective rotation in
\textit{ab initio} 
NCCI calculations for the $\isotope{Be}$ isotopes, with $7\leq A \leq
12$, using the realistic JISP16 nucleon-nucleon
interaction~\cite{shirokov2007:nn-jisp16}.  Evidence for rotational
band structure is found in the calculated excitation energies,
electric quadrupole moments, $E2$ transition matrix elements, magnetic
dipole moments, and $M1$ transition matrix elements.  In 
calculations of the even-mass $\isotope{Be}$ nuclei, yrast or
near-yrast sequences of angular momenta $0$, $2$, $4$, $\ldots$ arise
with calculated properties suggestive of $K=0$ rotational bands.
However, the most distinctive, well-developed, and systematic
rotational band structures are observed in calculations for odd-mass
nuclei. Given the same range of excitation energies and angular
momenta, the low-lying $\Delta J=1$ bands in the odd-mass nuclei
provide a richer set of energy and electromagnetic observables.
Bands are identified in both the natural (or valence-space) and unnatural parity spaces.

First, the properties expected in nuclear rotational structure are
reviewed, for the observables under consideration
(Sec.~\ref{sec-rot}).  Then, the results for rotational bands in NCCI
calculations of these $\isotope{Be}$ isotopes are presented
(Sec.~\ref{sec-results}): the calculations are outlined, results for
energies and electromagnetic matrix elements are presented,
and convergence is explored.  Finally, the calculated energies and
electromagnetic observables are examined in the context of rotational
band structure, and the calculated rotational bands are compared with experiment (Sec.~\ref{sec-disc}).
Preliminary results were reported in
Refs.~\cite{maris2012:mfdn-ccp11,caprio2013:berotor}.

\section{Background: Rotation}
\label{sec-rot}

\subsection{Rotational states}
\label{sec-rot-intro}

We first review the nature and expected signatures of
nuclear rotation~\cite{rowe2010:collective-motion,bohr1998:v2,rogers1965:nonspherical-nuclei}. 
Under the assumption of
adiabatic separation of the rotational degree of freedom,  a
rotational nuclear state may be described in terms of an 
\textit{intrinsic state}, as viewed in the non-inertial intrinsic frame, together with the rotational
motion of this intrinsic frame.    Here we consider, in particular, 
axially symmetric structure, for which the
intrinsic state $\tket{\phi_K}$ is characterized by definite angular momentum projection $K$ along
the intrinsic symmetry axis.  The full nuclear state $\tket{\psi_{JKM}}$,
with total angular momentum $J$ and projection $M$, then has the form
\begin{multline}
\label{eqn-psi}
\tket{\psi_{JKM}}=\Bigl[\frac{2J+1}{16\pi^2(1+\delta_{K0})}\Bigr]^{1/2}
\int d\vartheta\,\bigl[\scrD^J_{MK}(\vartheta)\tket{\phi_K;\vartheta}
\ifproofpre{\\}{}
+(-)^{J+K}\scrD^J_{M-K}(\vartheta)\tket{\phi_{\bar{K}};\vartheta}\bigr],
\end{multline}
where $\vartheta$ represents the Euler angles for rotation of the
intrinsic state, and $\tket{\phi_{\bar{K}}}$ is the
$\scrR_2$-conjugate intrinsic state (the $\scrR_2$ operator
induces a rotation by $\pi$ around the intrinsic-frame $y$ axis), which has angular momentum
projection $-K$ along the symmetry axis.  

The most recognizable features in the spectroscopy of rotational
states reside not in the states taken individually but in the
relationships~--- relative energies and electromagnetic multipole
operator matrix elements~--- among different rotational states
$\tket{\psi_{JKM}}$ sharing the same intrinsic state $\tket{\phi_K}$.
These states constitute members of a rotational band, with angular
momenta $J=K$, $K+1$, $\ldots$, except that, for $K=0$ bands, only
even $J$ are present, in the case of positive $\scrR_2$ symmetry (or
only odd $J$, in the case of negative $\scrR_2$ symmetry).

\subsection{Energies}
\label{sec-rot-energy}

Within a rotational band, energies follow the pattern
\begin{equation}
\label{eqn-EJ}
E(J)=E_0+AJ(J+1),
\end{equation}
where, in terms of the moment of inertia $\cal{J}$ about an axis
perpendicular to the symmetry axis, the rotational energy constant is
$A\equiv\hbar^2/(2\cal{J})$.  For $K=1/2$ bands, deviation from the
adiabatic rotational energy formula~(\ref{eqn-EJ}) is generally
substantial, due to the influence of the Coriolis contribution to the
kinetic energy. If one assumes that the intrinsic state may be
well described as a single nucleon coupled to an axially symmetric
core, the result is an energy
staggering given by
\begin{equation}
\label{eqn-EJ-stagger}
E(J)=E_0+A\bigl[J(J+1)+a(-)^{J+1/2}(J+\tfrac12)\bigr],
\end{equation}
where $a$ is the Coriolis decoupling parameter.  

\subsection{Quadrupole matrix elements}
\label{sec-rot-e2}

Reduced matrix elements $\trme{\psi_{J_fK}}{\Qvec_2}{\psi_{J_iK}}$ of
the electric quadrupole ($E2$) operator $\Qvec_2$ between states within
a band\footnote{We follow the reduced matrix element normalization and
phase conventions of
Edmonds~\cite{edmonds1960:am,varshalovich1988:am}.  Therefore,
quadrupole moments are related to reduced matrix elements by
$eQ(J)\equiv(16\pi/5)^{1/2}\tme{JJ}{Q_{2,0}}{JJ}=(16\pi/5)^{1/2}(2J+1)^{-1/2}\tcg{J}{J}{2}{0}{J}{J}\trme{J}{\Qvec_2}{J}$,
and reduced transition probabilities are related to reduced matrix
elements by $B(E2;J_i\rightarrow
J_f)=(2J_i+1)^{-1}\abs{\trme{J_f}{\Qvec_2}{J_i}}^2$.\label{fn-e2}} are
entirely determined by the intrinsic matrix elements and the
rotational structure.  These reduced matrix elements are given
by~\cite{rowe2010:collective-motion}
\begin{multline}
\label{eqn-e2-rme}
\trme{\psi_{J_fK}}{\Qvec_2}{\psi_{J_iK}}
\ifproofpre{\\}{}
=
\frac{({2J_i+1})^{1/2}}{1+\delta_{K0}}
\Bigl[
\tcg{J_i}{K}{2}{0}{J_f}{K} \tme{\phi_K}{Q_{2,0}}{\phi_K}
 \\
 +(-)^{J_i+K}
 \tcg{J_i,}{-K,}{2,}{2K}{J_f}{K} \tme{\phi_K}{Q_{2,2K}}{\phi_{\bar{K}}}\Bigr],
\end{multline}
where $J_i$ and $J_f$ are the initial and final angular momenta, respectively.  The
second (or cross) term in~(\ref{eqn-e2-rme}), involving
$\tme{\phi_K}{Q_{2,2K}}{\phi_{\bar{K}}}$, contributes only for $0\leq
K\leq1$, in which case the quadrupole operator can connect the
intrinsic state $\tket{\phi_K}$ with its conjugate state
$\tket{\phi_{\bar{K}}}$.  For
$K=0$, the contribution of this term is identical to that of the first
term, and it therefore simply enters into the normalization of the
expression, its effect canceling that of the $1+\delta_{K0}$ factor. Let us
set
aside, for the moment, the possible contribution of this cross term
for bands with $K=1/2$ or $1$.
\begin{figure*}
\centerline{\includegraphics[width=\ifproofpre{0.90}{1}\hsize]{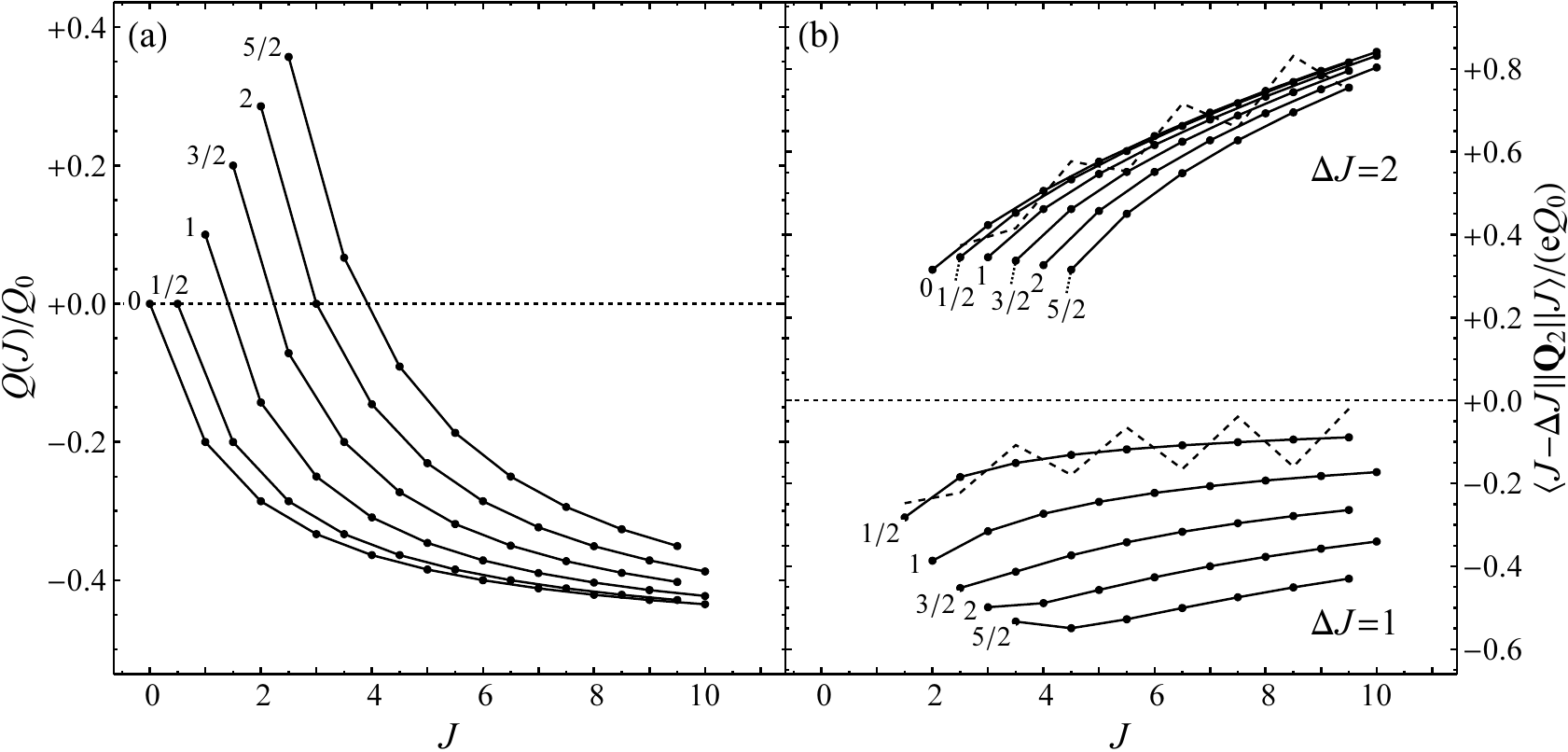}}
\caption{Rotational predictions for~(a) electric quadrupole moments and
(b)~electric quadrupole transition reduced matrix elements, within a rotational band,
normalized to the intrinsic quadrupole moment $Q_0$, following from~(\ref{eqn-e2-rme}).  Predictions are
shown for bands with $0\leq K \leq 5/2$, as indicated.  
The staggering induced in transitions within a $K=1/2$ band, 
taking $\tme{\phi_K}{Q_{2,2K}}{\phi_{\bar{K}}}/
\tme{\phi_K}{Q_{2,0}}{\phi_K}=+0.1$ for illustration, is indicated by
the dotted lines.
}
\label{fig-rotor-e2}      
\end{figure*}

Then, all reduced matrix
elements within a band are proportional to the intrinsic
quadrupole moment
$eQ_0\equiv(16\pi/5)^{1/2}\tme{\phi_K}{Q_{2,0}}{\phi_K}$,
\textit{i.e.}, the quadrupole moment of the intrinsic state, as calculated with respect to the
intrinsic symmetry axis.
The spectroscopic quadrupole moments of band members
are obtained in terms of $Q_0$ as
\begin{equation}
\label{eqn-Q}
Q(J)=\frac{3K^2-J(J+1)}{(J+1)(2J+3)}Q_0,
\end{equation}
and reduced transition probabilities within a band are obtained as
\begin{equation}
\label{eqn-BE2}
B(E2;J_i\rightarrow J_f)=\frac{5}{16\pi} \tcg{J_i}{K}{2}{0}{J_f}{K}^2 (eQ_0)^2.
\end{equation}
However, in the present work, rather than considering these reduced
transition probabilities, we find it more informative to
consider the
\textit{signed} reduced matrix elements,
\begin{multline}
\label{eqn-MEE2}
\ifproofpre{}{\hfill}  
\trme{\psi_{J_fK}}{\Qvec_2}{\psi_{J_iK}}
\ifproofpre{\\}{}
=\sqrt{\frac{5}{16\pi}} (2J_i+1)^{1/2}\tcg{J_i}{K}{2}{0}{J_f}{K} (eQ_0),
\ifproofpre{}{\hfill}  
\end{multline}
to retain
further meaningful phase information as we examine the 
rotational structure.  
The values of $Q(J)$ and $\trme{\psi_{J_fK}}{\Qvec_2}{\psi_{J_iK}}$ within a
rotational band, normalized to $Q_0$, are shown for reference in
Fig.~\ref{fig-rotor-e2}.

In obtaining these results for rotational matrix elements, $\Qvec_2$ may
be taken to be any operator of the form $Q_{2\mu}=\sum_{i=1}^Ae_i
r_{i}^2Y_{2\mu}(\uvec{r}_{i})$, where $e_i$ is a ``charge'' for the
$i$th nucleon, and $\vec{r}_{i}$ its
position~\cite{suhonen2007:nucleons-nucleus}.  Depending on the choice
of the coefficients $e_i$ for protons and neutrons, $\Qvec_2$ may
therefore variously represent the proton quadrupole~--- or physical
electric
quadrupole~--- tensor $\Qvec_p=e\sum_{i=1}^Zr_{p,i}^2Y_{2}(\uvec{r}_{p,i})$ (for $e_p=e$ and $e_n=0$), the neutron quadrupole
tensor $\Qvec_n=e\sum_{i=1}^Nr_{n,i}^2Y_{2}(\uvec{r}_{n,i})$ (for $e_p=0$ and $e_n=e$), or the
mass quadrupole tensor $\Qvec_m$ (\textit{i.e.}, their sum).\footnote{For consistency with common notation in the context of
electromagnetic transitions~\cite{rowe2010:collective-motion}, we
retain the electron charge $e$ in the normalization for both
$\Qvec_p$ and $\Qvec_n$. To the extent that isospin symmetry is maintained,
calculations of neutron quadrupole matrix elements in the
$\isotope{Be}$ isotopes are equivalent to calculations of electric
quadrupole matrix elements for isospin partner states in the $N=4$
isotones.}  While matrix elements of the electric quadrupole operator
$\Qvec_p$ are most immediately accessible in experiment, through
electromagnetic observables, matrix elements of $\Qvec_n$ may be viewed on
an equal footing in the rotational analysis of calculated wave
functions.  Therefore, these neutron matrix elements are considered
alongside the proton matrix elements throughout
Sec.~\ref{sec-results}.  They provide a valuable complementary set of
observables for the purpose of investigating whether or not the
nuclear wave functions satisfy the conditions of adiabatic
rotational separation.  

Let us now return to the cross term in the rotational
expression~(\ref{eqn-e2-rme}) for the reduced matrix element.  In
the analysis of $K=1/2$ or $1$ bands in well-deformed rotor nuclei, in
heavier mass regions, this cross term is commonly neglected, since
$Q_0\sim\tme{\phi_K}{Q_{2,0}}{\phi_K}$ is strongly enhanced, while
$\tme{\phi_K}{Q_{2,2K}}{\phi_{\bar{K}}}$ is presumed to be of typical
single-particle strength~\cite{rowe2010:collective-motion}.  However,
it may have greater significance for the light nuclei considered here,
where the number of nucleons participating in collective motion is
more limited, and rotational quadrupole strengths for candidate rotational
states can therefore be expected to be less dramatically enhanced over
single-particle strengths.  

The alternating sign of the cross term,
for successive $J_i$ values, may therefore be expected to introduce a
nonnegligible staggering in matrix elements for bands with $K=1/2$ or $1$.  That said, for the
particular case of quadrupole moments (diagonal matrix elements) in
$K=1/2$ bands, of principal interest in the present work, it should be
noted that the cross term does not actually contribute to the
rotational predictions, since the Clebsch-Gordan coefficients
$\tcg{J,}{-\tfrac12,}{2,}{1}{J}{\tfrac12}$ vanish identically, and thus
no staggering is obtained.  The cross-term can still
induce staggering in the transition matrix elements
within these $K=1/2$ bands.  For illustration, the staggering induced
by intrinsic matrix elements in the ratio  $\tme{\phi_K}{Q_{2,2K}}{\phi_{\bar{K}}}/
\tme{\phi_K}{Q_{2,0}}{\phi_K}=+0.1$ is indicated in
Fig.~\ref{fig-rotor-e2}(b) (see dotted curves).  Staggering of
approximately this scale may be suggested by (or is at least not
inconsistent with) some of the calculated results for matrix elements
in Sec.~\ref{sec-results-trans}.

\subsection{Dipole matrix elements}
\label{sec-rot-m1}
\begin{figure*}
\centerline{\includegraphics[width=\ifproofpre{0.90}{1}\hsize]{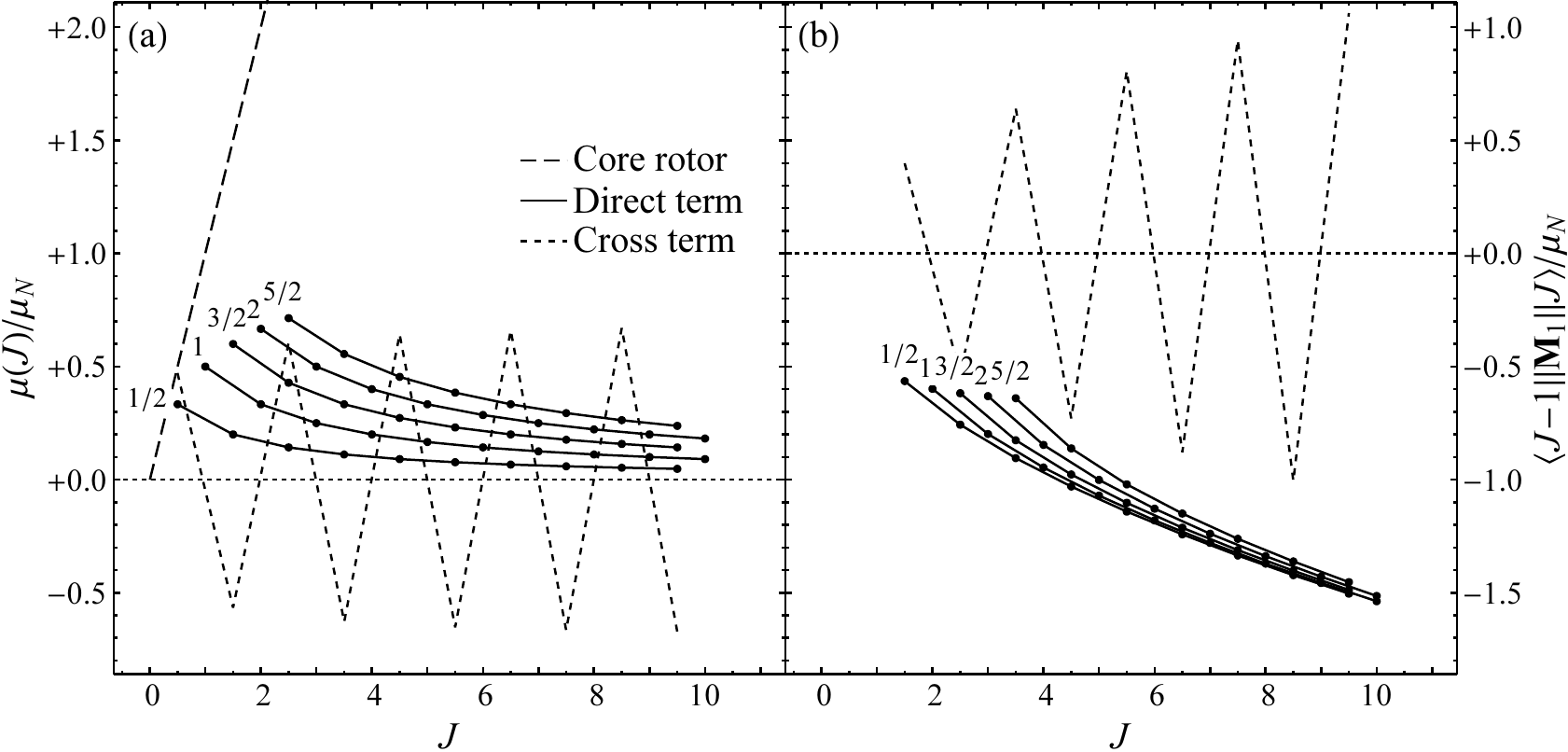}}
\caption{Rotational predictions for each of the terms 
contributing, in~(\ref{eqn-mu}) and~(\ref{eqn-MEM1}), to~(a) magnetic
dipole moments and (b)~magnetic dipole transition reduced matrix
elements, within a rotational band: the core rotor term for dipole
moments (dashed line),
direct term (solid lines, $K\geq1/2$ only, as indicated), and cross term (dotted
lines, $K=1/2$ only).  For purposes of comparison, these terms are shown with equal coefficients, \textit{i.e.},
with $a_1=a_2=a_3=\mu_N$.  Predictions are shown for bands with $0\leq K
\leq 5/2$.}
\label{fig-rotor-m1}      
\end{figure*}

The standard rotational analysis for magnetic dipole ($M1$) matrix elements~\cite{rowe2010:collective-motion,bohr1998:v2}
relies upon the assumption that the nucleus may be divided into a
``core'' rotor and extra-core particles.  First, we recall that the
magnetic dipole operator may be written as 
\begin{equation}
\label{eqn-m1-micro}
\Mvec_1=\sqrt{\frac{3}{4\pi}} \mu_N
\sum_{i=1}^A\bigl(\gli \lvec_i +\gsi \svec_i\bigr),
\end{equation}
where $\mu_N$ is the
nuclear magneton, $\gli$ and $\gsi$ are the orbital and
spin $g$ factors for the $i$th
nucleon, $\lvec_i$ and $\svec_i$ are the dimensionless orbital and spin
angular momentum operators, and we have adopted Gaussian
units~\cite{suhonen2007:nucleons-nucleus}.  Then a decomposition of the dipole operator
into core and particle terms is accomplished as $\Mvec_1=[3/(4\pi)]^{1/2} g_R \mu_N \Jvec + \Mvec_1'$, where
$g_R$ is an effective gyromagnetic ratio for the core rotor, $\Jvec$ is
the dimensionless total angular momentum operator, and
$\Mvec_1'=[3/(4\pi)]^{1/2} \mu_N
\sum_{i=1}^A\bigl[(\gli-g_R)\lvec_i +(\gsi-g_R)
\svec_i\bigr]$.  The matrix elements of the first term include the entire
contribution of the core rotor.  Matrix elements of the second,
residual term $\Mvec_1'$ receive contributions only from the extra-core
particles.  The rotational predictions for these latter matrix
elements may be expressed in terms of intrinsic matrix elements, much as for the quadrupole operator
in~(\ref{eqn-e2-rme}).  

Reduced matrix elements\footnote{Under the present reduced matrix
element convention (see footnote~\ref{fn-e2}), dipole moments are
related to reduced matrix elements by
$\mu(J)\equiv(4\pi/3)^{1/2}\tme{JJ}{M_{1,0}}{JJ}=(4\pi/3)^{1/2}(2J+1)^{-1/2}\tcg{J}{J}{1}{0}{J}{J}\trme{J}{\Mvec_1}{J}$,
and reduced transition probabilities by $B(M1;J_i\rightarrow
J_f)=(2J_i+1)^{-1}\abs{\trme{J_f}{\Mvec_1}{J_i}}^2$.} of the full
magnetic dipole operator between states within a band are given
by~\cite{rowe2010:collective-motion}
\begin{multline}
\label{eqn-m1-rme}
\trme{\psi_{J_fK}}{\Mvec_1}{\psi_{J_iK}}
=
\sqrt{\frac{3}{4\pi}}g_R \mu_N \trme{J_f}{\Jvec}{J_i}
\delta_{J_iJ_f}
\ifproofpre{\\}{}
+
({2J_i+1})^{1/2}\Bigl[
\tcg{J_i}{K}{1}{0}{J_f}{K} \tme{\phi_K}{M_{1,0}'}{\phi_K}
 \\
  +\delta_{K,\tfrac12}(-)^{J_i+\tfrac12}
  \tcg{J_i,}{-\tfrac12,}{1,}{1}{J_f}{\tfrac12} \tme{\phi_{1/2}}{M_{1,1}'}{\phi_{\overline{1/2}}}\Bigr].
\end{multline}
The first (rotor) term contributes only to the diagonal matrix
elements, \textit{i.e.}, with $J_f=J_i$, for which case we may use
the standard identity for the reduced matrix element of the angular
momentum operator within a state of angular momentum $J$, $\trme{J}{\Jvec}{J}=[J(J+1)(2J+1)]^{1/2}$. 
The first (or direct)
term within the brackets, involving the matrix element
$\tme{\phi_K}{M_{1,0}'}{\phi_K}$, contributes in general (except for
$K=0$ dipole moments), while the second (or cross) term within
the brackets, involving the matrix element
$\tme{\phi_{1/2}}{M_{1,1}'}{\phi_{\overline{1/2}}}$ between conjugate
intrinsic states, contributes only for $K=1/2$.  In this case, the cross term may
contribute with a strength comparable to that of the direct term~\cite{bohr1998:v2}.

For later reference, in Sec.~\ref{sec-results-trans}, we note the explicit forms
of the rotational
predictions which are obtained from~(\ref{eqn-m1-rme}), after
evaluating the Clebsch-Gordan coefficients (\textit{e.g.}, Sec.~8.5 of
Ref.~\cite{varshalovich1988:am}).  For dipole
moments,
\begin{equation}
\label{eqn-mu}
\mu(J)=a_0 J + a_1\frac{K}{J+1}+a_2\delta_{K,\tfrac12}\frac{(-)^{J-1/2}}{2\sqrt{2}}\frac{2J+1}{J+1},
\end{equation}
and, for the reduced matrix elements for $\Delta J=1$ transitions,
\begin{multline}
\ifproofpre{}{\hfill}  
\label{eqn-MEM1}
\trme{\psi_{J-1,K}}{\Mvec_1}{\psi_{JK}}
\ifproofpre{\\}{}
=
-\sqrt{\frac{3}{4\pi}}\sqrt{\frac{J^2-K^2}{J}}
\Biggl[a_1+a_2\delta_{K,1/2}\frac{(-)^{J-1/2}}{\sqrt{2}}\Biggr],
\ifproofpre{}{\hfill}  
\end{multline}
where $a_0=g_R\mu_N$,
$a_1=(4\pi/3)^{1/2}\tme{\phi_K}{M_{1,0}'}{\phi_K}$, and $a_2=(4\pi/3)^{1/2}\tme{\phi_{1/2}}{M_{1,1}'}{\phi_{\overline{1/2}}}$.
The terms contributing to $\mu(J)$ in~(\ref{eqn-mu}) are shown in
Fig.~\ref{fig-rotor-m1}(a), while those contributing to
$\trme{\psi_{J-1,K}}{\Mvec_1}{\psi_{JK}}$ in~(\ref{eqn-MEM1}) are shown in
Fig.~\ref{fig-rotor-m1}(b).
 
In obtaining these results for rotational matrix elements, $\Mvec_1$
may be any operator of the form~(\ref{eqn-m1-micro}).  The structure
of this operator is more
apparent if it is
decomposed into proton/neutron and orbital/spin contributions, as
\begin{equation}
\label{eqn-m1-D}
\Mvec_1=\glp \Dlp + \gln \Dln + \gsp \Dsp + \gsn \Dsn,
\end{equation}
where we define the \textit{dipole terms}\footnote{In, \textit{e.g.},
Ref.~\cite{suhonen2007:nucleons-nucleus}, the expression
\textit{dipole term} is used to refer to each of the contributing
proton/neutron and orbital/spin
\textit{matrix elements}, when the $M1$ operator is applied at the one-body
level, \textit{i.e.}, between single-particle orbitals.  Here we
generalize the use of \textit{dipole term} to refer to each of the contributing
proton/neutron and orbital/spin \textit{operators}, and to the
application of these operators at the many-body level.}
\begin{equation}
\begin{aligned}
\label{eqn-D}
\Dlp&=\sqrt{\frac{3}{4\pi}} \mu_N \lvec_p
&\quad
\Dln&=\sqrt{\frac{3}{4\pi}} \mu_N \lvec_n
\ifproofpre{\\}{&}\quad
\Dsp&=\sqrt{\frac{3}{4\pi}} \mu_N \svec_p
&\quad
\Dsn&=\sqrt{\frac{3}{4\pi}} \mu_N \svec_n,
\end{aligned}
\end{equation}
where $\lvec_p$ is the total proton orbital angular momentum operator
[$\lvec_p\equiv\sum_{i=1}^Z\lvec_{p,i}$], $\lvec_n$ is the total neutron
orbital angular momentum operator
[$\lvec_n\equiv\sum_{i=1}^N\lvec_{n,i}$], $\svec_p$ is the total proton spin
angular momentum operator [$\svec_p\equiv\sum_{i=1}^Z\svec_{p,i}$], and $\svec_n$
is the total neutron spin angular momentum operator
[$\svec_n\equiv\sum_{i=1}^N\svec_{n,i}$].  The electromagnetic $M1$ operator
is obtained for the particular choice $\glp=1$,
$\gln=0$, $\gsp\approx 5.586$, and
$\gsn \approx-3.826$.  However, the rotational
results~(\ref{eqn-mu})--(\ref{eqn-MEM1}) apply more generally to
matrix elements of any linear combination of these terms, including,
as we consider in Sec.~\ref{sec-results}, each of the four
dipole terms individually.

These different dipole operators provide complementary probes of the
proton/neutron and orbital/spin contributions to the rotational
structure.\footnote{In particular, the diagonal matrix elements, or
dipole moments, calculated using each of the individual dipole terms
are proportional to the $\lvec_p$, $\svec_p$, $\lvec_n$, and $\svec_n$
contributions, respectively, to the total angular momentum of the
nuclear state~\cite{maris2013:ncsm-pshell} (see also Sec.~\ref{sec-disc-intrinsic}).}  Thus, while matrix
elements of the traditional electromagnetic $M1$ operator are most
immediately accessible in experiment, they do not exhaust the
available structural information for calculated wave functions.  For
instance, the electromagnetic operator is blind to the neutron orbital
motion, which may be expected to be a dominant contributor to total
angular momentum in a neutron-rich rotational nucleus.

\section{Results}
\label{sec-results}

\subsection{Calculations}
\label{sec-results-calc}

We consider here the results of NCCI calculations for the
$\isotope{Be}$ isotopes with $7\leq A \leq 12$.  In practice, nuclear
many-body calculations must be carried out in a truncated space.  A
conventional harmonic oscillator basis is used for the present
calculations, with truncation according to the $\Nmax$ scheme, that
is, by the number of oscillator excitations relative to the lowest
Pauli-allowed configuration.  The eigenvalues
and wave functions, and thus calculated observables, obtained in such
calculations are in general dependent both upon the basis truncation
$\Nmax$ and on the oscillator length for the basis, which is specified
by the oscillator energy $\hw$. Detailed illustrations may be
found in, \textit{e.g.},
Refs.~\cite{bogner2008:ncsm-converg-2N,maris2009:ncfc,cockrell2012:li-ncfc,maris2013:ncsm-pshell,shirokov2014:jisp16-binding},
including calculations for some of the isotopes and observables under
consideration here.  In the present calculations, basis truncations
$\Nmax=10$ or $11$ have been used, depending on the parity under
consideration, as discussed further below.  The basis $\hw$
parameters have been chosen near the variational minimum for the
ground state energy (specifically, $\hw=20\,\MeV$ for
$\isotope[7,8]{Be}$ and $\hw=22.5\,\MeV$ for
$\isotope[9\text{--}12]{Be}$).  The calculations are carried out in
the proton-neutron $M$ scheme~\cite{whitehead1977:shell-methods},
using the code
MFDn~\cite{sternberg2008:ncsm-mfdn-sc08,maris2010:ncsm-mfdn-iccs10,aktulga2013:mfdn-scalability}.

These calculations are based on the JISP16
interaction~\cite{shirokov2007:nn-jisp16}, which is a charge-independent two-body
interaction derived from nucleon-nucleon scattering data and adjusted
via a phase-shift equivalent transformation to describe light nuclei
without explicit three-body interactions.  The bare JISP16 interaction
is used, without renormalization to the truncated space (see
Ref.~\cite{maris2009:ncfc}).  The Coulomb interaction has been omitted
from the Hamiltonian, to ensure exact conservation of isospin, thereby
simplifying the spectrum (we consider states of minimal isospin
$T=T_z$).  However, the primary effect of the Coulomb interaction, if
included, is simply to induce a shift in the overall binding energies,
which is irrelevant to the analysis of rotational band observables.

Due to the parity conserving nature of the nuclear interaction in
these calculations, the eigenproblem separates into positive and
negative parity sectors.  The parity of the lowest allowed oscillator
configuration may be termed the
\textit{natural parity},  and that obtained by
promoting one nucleon by one shell the \textit{unnatural parity}.
Thus, the natural parity is negative for the
odd-mass isotopes $\isotope[7,9,11]{Be}$, and positive for the
even-mass isotopes $\isotope[8,10,12]{Be}$.
While the lowest
unnatural parity states normally lie at significantly higher energy
than those of natural parity, they are calculated to lie within a few
$\MeV$ of the lowest natural parity states in the isotopes
$\isotope[9,11]{Be}$~\cite{maris2012:mfdn-ccp11}, and are thus
included in the present discussion of these nuclei.  Note that parity
inversion arises for $\isotope[11]{Be}$, \textit{i.e.}, the ground
state is experimentally~\cite{npa1990:011-012} in the unnatural parity
space, and both spaces are near-degenerate in calculations at finite
$\Nmax$ (see Ref.~\cite{navratil2009:ncsm}).
The NCCI basis states with an even number of excitations above the
lowest oscillator configuration span the natural parity space, while the
basis states with an odd number of excitations span the unnatural
parity space~--- hence the application of even and odd $\Nmax$
truncations to these spaces, respectively.

Diagonalization of the large Hamiltonian matrices
encountered in these NCCI calculations relies upon the Lanczos
algorithm~\cite{lanczos1950:algorithm}.  For a given number of Lanczos
iterations, only a limited set of energy eigenvalues (and
corresponding eigenvectors) are converged, starting from the bottom of
the spectrum, giving the lowest energies within the given many-body
space.  Within the framework of an $M$-scheme calculation, where
many-body basis states are restricted to a given value of the total
angular momentum projection $M$, all states with $J\geq M$ are
included in the space (\textit{e.g.},
Ref.~\cite{whitehead1977:shell-methods}).  Thus, in practice, to
address the entire yrast region, results of calculations in spaces
with different $M$ values must be aggregated.  For example, the yrast
state of a given $J$ might be too high in the energy spectrum to be
practically obtained from calculations in the $M=0$ space (or $M=1/2$
for odd-mass nuclei).  However, in the $M=J-2$ space, this lowest
state of angular momentum $J$ might be one of the highest states
actually converged (the $20$th, say).  And, in the $M=J$ space, it
may be expected to be one of the lowest states obtained (albeit not
necessarily \textit{the} lowest, since the yrast line is not
necessarily monotonic, particularly in the presence of Coriolis
staggering).

\subsection{Energy levels}
\label{sec-results-energy}
\begin{figure*}[p]
\centerline{\includegraphics[width=\ifproofpre{0.95}{0.95}\hsize]{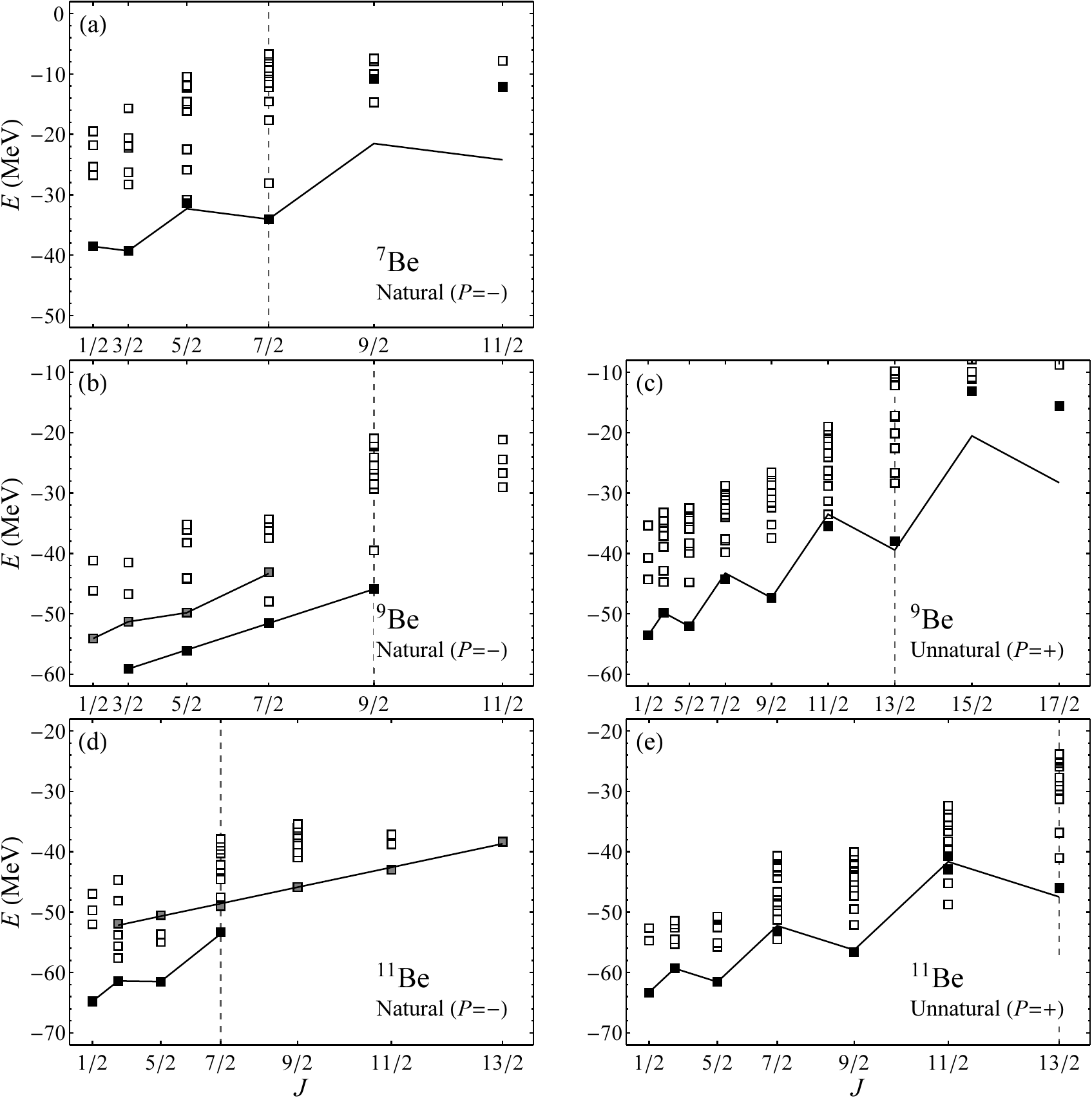}}
\caption{Energy eigenvalues obtained for states in  the
natural~(left) and unnatural~(right) parity spaces of the odd-mass
$\isotope{Be}$ isotopes ($7\leq A \leq11)$.  Energies are plotted with
respect to an angular momentum axis which is scaled to be linear in
$J(J+1)$, to facilitate identification of rotational energy patterns.
Solid symbols indicate candidate 
yrast band members, and shaded symbols indicate candidate excited band members.  The lines
indicate the corresponding best fits for rotational energies (see
text).  Where quadrupole
transition strengths indicate significant fragmentation,
more than one state of a given $J$ may be indicated as a band member.
The vertical dashed lines indicate the maximal angular
momentum accessible within the lowest harmonic oscillator
configuration (or valence space).}
\label{fig-energy-odd}      
\end{figure*}
\begin{figure}
\centerline{\includegraphics[width=\ifproofpre{0.95}{0.55}\hsize]{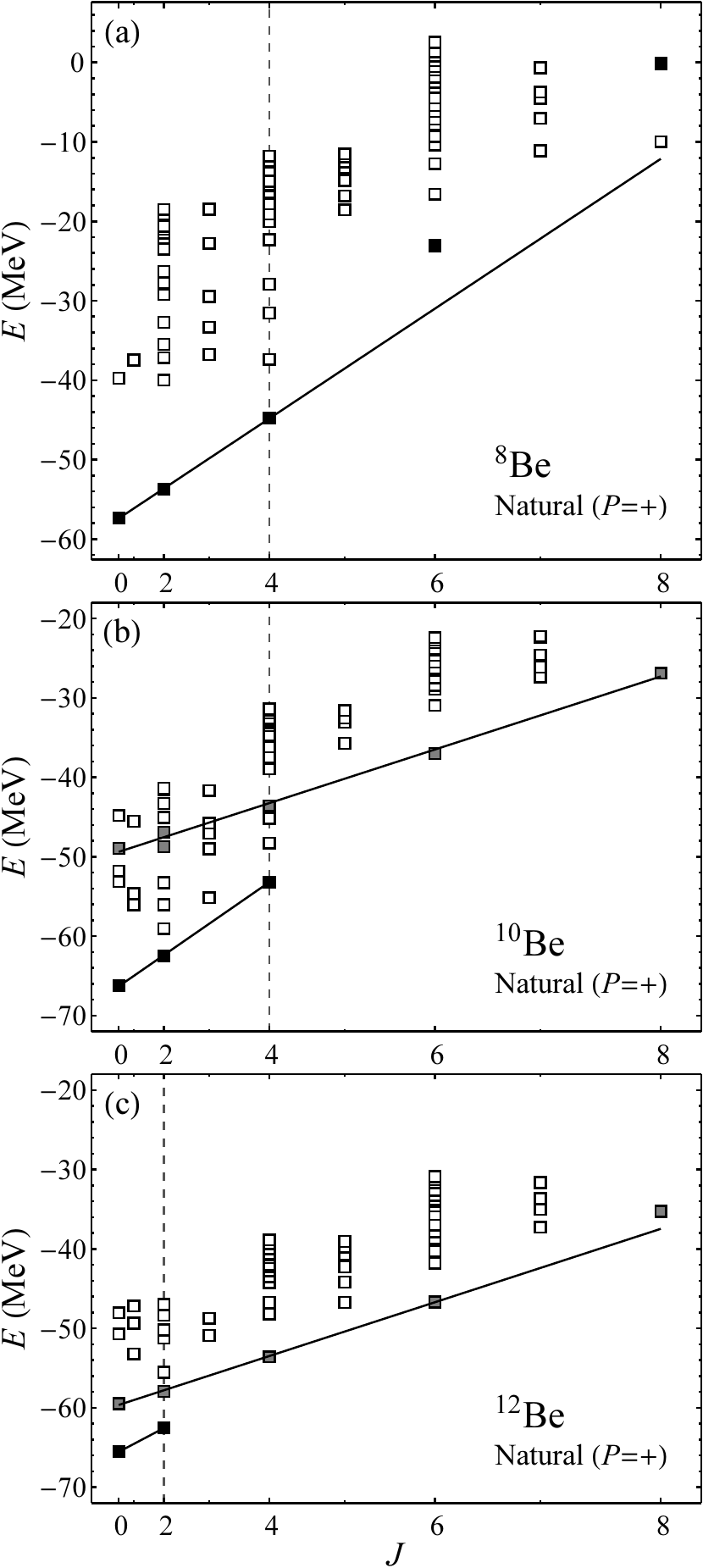}}
\caption{Energy eigenvalues obtained for states in  the
\textit{natural} parity spaces of the even-mass $\isotope{Be}$ isotopes ($8\leq A \leq12)$.  
See Fig.~\ref{fig-energy-odd} caption for discussion of the plot contents and labeling.
}
\label{fig-energy-even}      
\end{figure}

Identification of candidate rotational band members relies not only on
recognizing rotational energy patterns, but also on identifying
collective enhancement of quadrupole transition strengths and
verifying rotational patterns of electromagnetic moments and
transition matrix elements.  Nonetheless, it is natural to begin the
discussion of band structure by considering energies.

The calculated energy eigenvalues for low-lying states of the odd-mass
$\isotope{Be}$ isotopes (with $7\leq A \leq11$) are shown in
Fig.~\ref{fig-energy-odd}.  Where both natural and
unnatural parity spaces have been calculated (see Sec.~\ref{sec-results-calc}), these are shown separately.  Low-lying states of
even-mass $\isotope{Be}$ isotopes (with $8\leq A \leq12$) are shown in
Fig.~\ref{fig-energy-even}.  To facilitate identification of
rotational bands, it is helpful to plot the calculated excitation
energies against an angular momentum axis which is scaled as $J(J+1)$,
so that energies within an ideal rotational band lie on a straight
line or, for $K=1/2$ bands, staggered about a straight line.  For the
candidate bands in Figs.~\ref{fig-energy-odd}
and~\ref{fig-energy-even}, a straight line fit~(\ref{eqn-EJ}) to all
band members is shown, except that, for $K=1/2$ bands, an energy fit
is obtained by adjusting the parameters of~(\ref{eqn-EJ-stagger}) to
the lowest three band members (the remainder of the line is thus an
extrapolation).

The yrast states are noticeably isolated in energy from the remaining
states at low angular momentum, though not necessarily at the higher
angular momenta shown in Figs.~\ref{fig-energy-odd}
and~\ref{fig-energy-even}.  This separation is by
$\sim20\,\MeV$ for the ground state of $\isotope[8]{Be}$
[Fig.~\ref{fig-energy-even}(a)].  The yrast and near-yrast states
therefore yield the most immediately recognizable sets of candidate
band members.  Candidate yrast band members, in particular, are
indicated by the solid black squares in Figs.~\ref{fig-energy-odd}
and~\ref{fig-energy-even}.  The yrast band members can, for the most
part, be identified simply from the rotational pattern of their
energies, although in the present analysis the energy is taken only as
a basis for identification of
\textit{candidate} band members, pending further analysis of
electromagnetic observables.  ``Yrast'' bands may be identified separately
in the natural [Fig.~\ref{fig-energy-odd}(a,b,d)] and unnatural [Fig.~\ref{fig-energy-odd}(c,e)] parity spaces.
In the odd-mass $\isotope{Be}$ isotopes
considered (Fig.~\ref{fig-energy-odd}), yrast rotational bands are
found with $K=1/2$ or $3/2$.
In the even-mass $\isotope{Be}$ isotopes, the yrast states
(Fig.~\ref{fig-energy-even}) constitute prospective $K=0$ ground state
rotational bands, albeit in some instances severely truncated.

The density of states rapidly increases off the yrast line, leading to
the possibility of fragmentation of rotational states by mixing with
nearby states and, in general, hindering the identification of band
members.  For the yrast $K=1/2$ bands, alternate bandmembers are
raised in energy into this region of higher density of states, as a
result of the Coriolis staggering in energies.  Any excited bands must
also be sought in this region of higher density of states.
Nonetheless, several excited candidate bands can be
identified,\footnote{We focus, in the present discussion, on the most
clearly identifiable yrast or near-yrast bands.  However, this should
not be taken to exclude the possibility of additional excited bands,
suggestions of which may be found in the calculated spectra and matrix
elements.  See, \textit{e.g.}, Fig.~2(b) of
Ref.~\cite{caprio2013:berotor}, for a low-lying example.} once
electromagnetic moments and transition matrix elements have been taken
into account, as indicated by shaded squares in
Figs.~\ref{fig-energy-odd} and~\ref{fig-energy-even}.  Excited
candidate bands are identified in the natural parity spaces of
$\isotope[9]{Be}$ [Fig.~\ref{fig-energy-odd}(b)], $\isotope[11]{Be}$
[Fig.~\ref{fig-energy-odd}(d)], $\isotope[10]{Be}$
[Fig.~\ref{fig-energy-even}(b)], and $\isotope[12]{Be}$
[Fig.~\ref{fig-energy-even}(c)], again with $K=1/2$ or $3/2$ for the
odd-mass isotopes, or $K=0$ for the even-mass isotopes.  Examples of
fragmentation of rotational candidate band members, as indicated by
quadrupole transition strengths, may be seen in the present
calculations at $J=2$ in the excited band of $\isotope[10]{Be}$
[Fig.~\ref{fig-energy-even}(b)] and at $J=11/2$ in the unnatural
parity yrast band of $\isotope[11]{Be}$
[Fig.~\ref{fig-energy-odd}(e)].

A basic question to be addressed (taken up in view of the full set of
calculated data in Sec.~\ref{sec-disc-patterns}) is whether or not the
rotational bands exhibit termination.  If so, we then wish to understand the relation between the
terminating angular momentum and the angular momenta accessible in a
traditional valence shell description for these nuclei.  In general,
the maximal angular momentum available in the $\Nmax=0$ space,
\textit{i.e.}, the lowest oscillator configuration, of the NCCI scheme
is, equivalently, the maximal angular momentum possible in a
traditional shell model description using the last
partially-occupied oscillator shell (here the $p$ shell) as the valence space.
This maximal valence angular momentum is indicated by the dashed vertical lines in
Figs.~\ref{fig-energy-odd} and~\ref{fig-energy-even}.  For the
unnatural parity spaces 
[Fig.~\ref{fig-energy-odd}(c,e)], the maximal angular momentum
accessible in the $\Nmax=1$ space, \textit{i.e.}, the lowest oscillator configuration of
\textit{unnatural} parity, is indicated instead.
This angular momentum is, in general, higher than for the
corresponding natural parity space, since promotion of a nucleon to a
higher shell relaxes Pauli constraints on the allowed angular momentum
couplings, while also making higher-$j$ orbitals accessible.  The
maximal angular momenta are summarized in
Table~\ref{tab-shell-maximal}.  

\begin{table}
  \caption{Maximal angular momenta accessible for the $\isotope{Be}$
  isotopes in the valence, or $\Nmax=0$ space, \textit{i.e.}, lowest oscillator configurations of natural parity.  For $\isotope[9,11]{Be}$, the maximal
  angular momenta accessible in the $\Nmax=1$ space, or lowest oscillator configurations of unnatural parity, are also shown.  }
\label{tab-shell-maximal}
\begin{center}
\begin{ruledtabular}
\begin{tabular}{lrrrrrr}
Parity & $\isotope[7]{Be}$&$\isotope[9]{Be}$&$\isotope[11]{Be}$&$\isotope[8]{Be}$&$\isotope[10]{Be}$&$\isotope[12]{Be}$\\
\hline
Natural & $\sfrac72$&$\sfrac92$&$\sfrac72$ &4 & 4 & 2 \\
Unnatural &  & $\sfrac{13}2$ &  $\sfrac{13}2$
\end{tabular}
\end{ruledtabular}
\raggedright
\end{center}
\end{table}


Candidate yrast bands in
Figs.~\ref{fig-energy-odd} and~\ref{fig-energy-even} exhibit some form
of discontinuity in the evolution of the energies with $J$, or else
clear termination, at the maximal valence angular momentum.  In
contrast, several of the excited candidate bands extend to higher
angular momentum without apparent disruption to the rotational energy
pattern.  These observations are revisited below
(Sec.~\ref{sec-disc-patterns}) in light of electromagnetic observables.

\subsection{Electromagnetic matrix elements}
\label{sec-results-trans}

For each candidate rotational band highlighted in
Figs.~\ref{fig-energy-odd} and~\ref{fig-energy-even}, we now compare
the calculated electric quadrupole and magnetic dipole matrix
elements~--- that is, moments and transition matrix elements~--- with
the rotational predictions.  The calculated results are shown in
Figs.~\reffigstransodd{} for the odd-mass
$\isotope{Be}$ isotopes and in
Figs.~\reffigstranseven{} for the even-mass
$\isotope{Be}$ isotopes.\footnote{The radically truncated $K=0$ yrast
band of $\isotope[12]{Be}$, which terminates at $J=2$, as indicated in
Fig.~\ref{fig-energy-even}(c), is not included among these figures but
does appear in Sec.~\ref{sec-disc}.}  Let us for now focus on
defining the contents of these figures, while deferring analysis to
Sec.~\ref{sec-disc}.

In each of Figs.~\reffigstrans{}, the quadrupole moments for all
states within the considered band are shown in panel~(a).  The values
are normalized to $Q_0$, to faciliate uniform comparison with the
rotational predictions for $Q(J)/Q_0$ from~(\ref{eqn-Q}), which are
shown as a curve in each plot, appropriate to the $K$ quantum number
of the given band.  The value of $Q_0$ used for normalization has in
each case been obtained simply from the quadrupole moment of the
lowest-energy bandmember with nonvanishing quadrupole moment.  Thus,
for $K=1/2$ bands, since the quadrupole moment of the $J=1/2$ band
head vanishes identically, either the $J=3/2$ or $5/2$ band member is
used for normalization, according to the energy staggering. Similarly,
for $K=0$ bands, the quadrupole moment of the $J=2$ band member is
used for normalization.  The exceptions are the excited $K=0$ bands in
$\isotope[10]{Be}$ [Fig.~\ref{fig-10be0-band2}(a)] and
$\isotope[12]{Be}$ [Fig.~\ref{fig-12be0-band2}(a)], for which the
$J=4$ band member is found to be less fragmented and therefore used
for normalization.  Quadrupole moments in Figs.~\reffigstrans{} are
calculated using both the proton (solid symbols) and neutron (open
symbols) quadrupole tensors (as discussed in Sec.~\ref{sec-rot-e2}).
(In some cases, data points for the neutron results may not be separately visible
in these figures, when they are indistinguishable from the
corresponding proton results.)  The proton and neutron quadrupole
moments are normalized separately, since no
\textit{a priori} relation exists between the intrinsic matrix
elements of the $\Qvec_p$ and $\Qvec_n$ operators (except in $\isotope[8]{Be}$,
where isospin symmetry makes the proton and neutron quadrupole moments
trivially identical).  

\newcommand{\bandwidth}{0.80}


\begin{figure*}
\centerline{\includegraphics[width=\ifproofpre{\bandwidth}{1.0}\hsize]{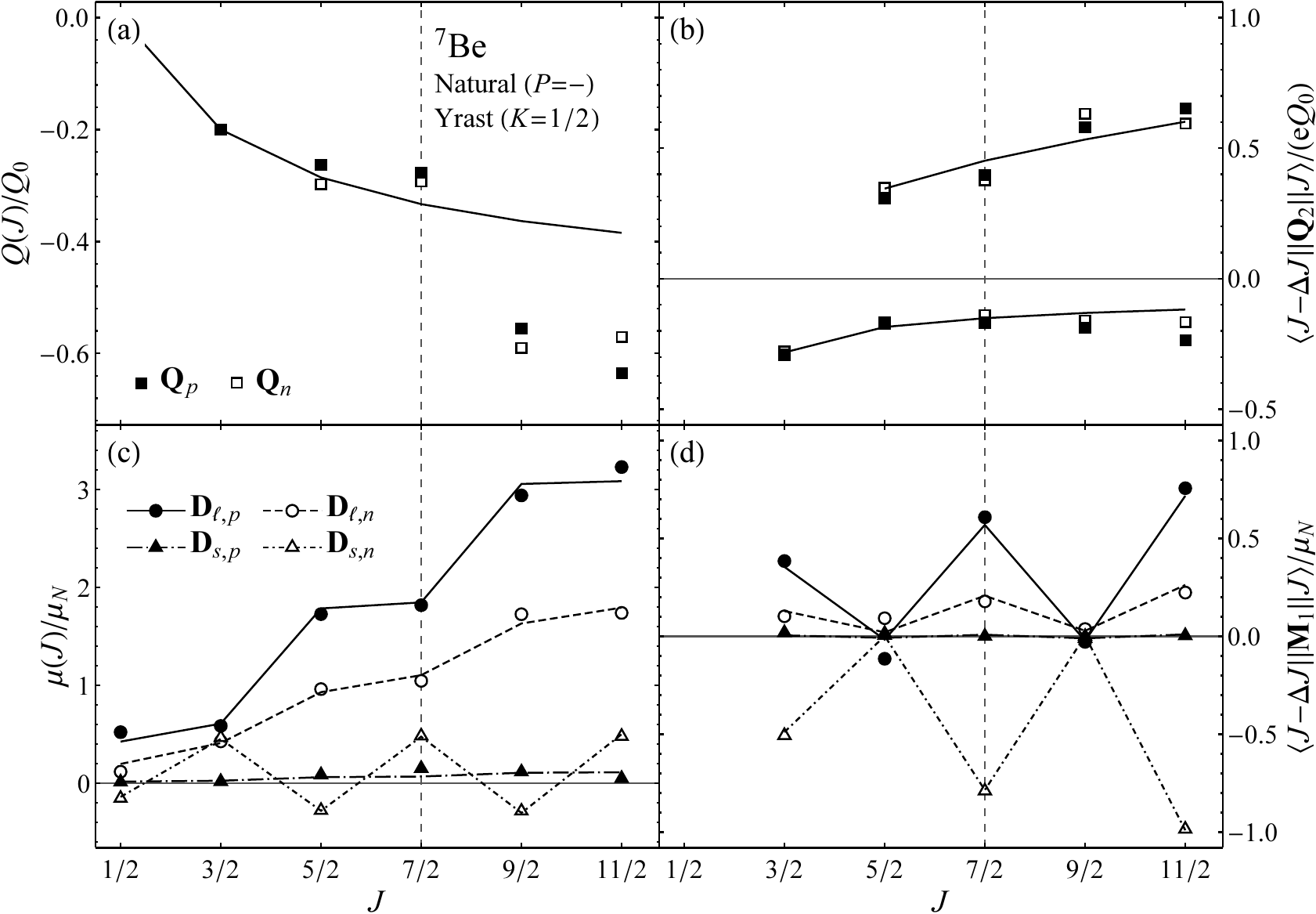}}
\caption{Dipole and quadrupole matrix element observables for the
  $\isotope[7]{Be}$ natural parity yrast band: (a)~quadrupole ($E2$)
  moments, (b)~quadrupole transition reduced matrix elements,
  (c)~dipole ($M1$) moments, and (d)~dipole transition reduced matrix
  elements.  For quadrupole moments and matrix elements, values
  calculated using both proton and neutron operators are shown.
  Similarly, for dipole moments and matrix elements, values calculated
  using all four dipole terms (\textit{i.e.}, proton/neutron and
  orbital/spin contributions) are shown.  Quadrupole moments and
  matrix elements are normalized to $Q_0$, obtained from the lowest
  band member's quadrupole moment.  The curves indicate the rotational
  values for the quadrupole moments and matrix elements (normalized to
  $Q_0$) and for the dipole moments and matrix elements (using band
  best-fit parameters obtained from the dipole moments).  The vertical
  dashed lines indicate the maximal angular momentum accessible within
  the lowest harmonic oscillator configuration (or valence space).  }
\label{fig-7be0-band1}      
\end{figure*}


\begin{figure*}[p]
\centerline{\includegraphics[width=\ifproofpre{\bandwidth}{1.0}\hsize]{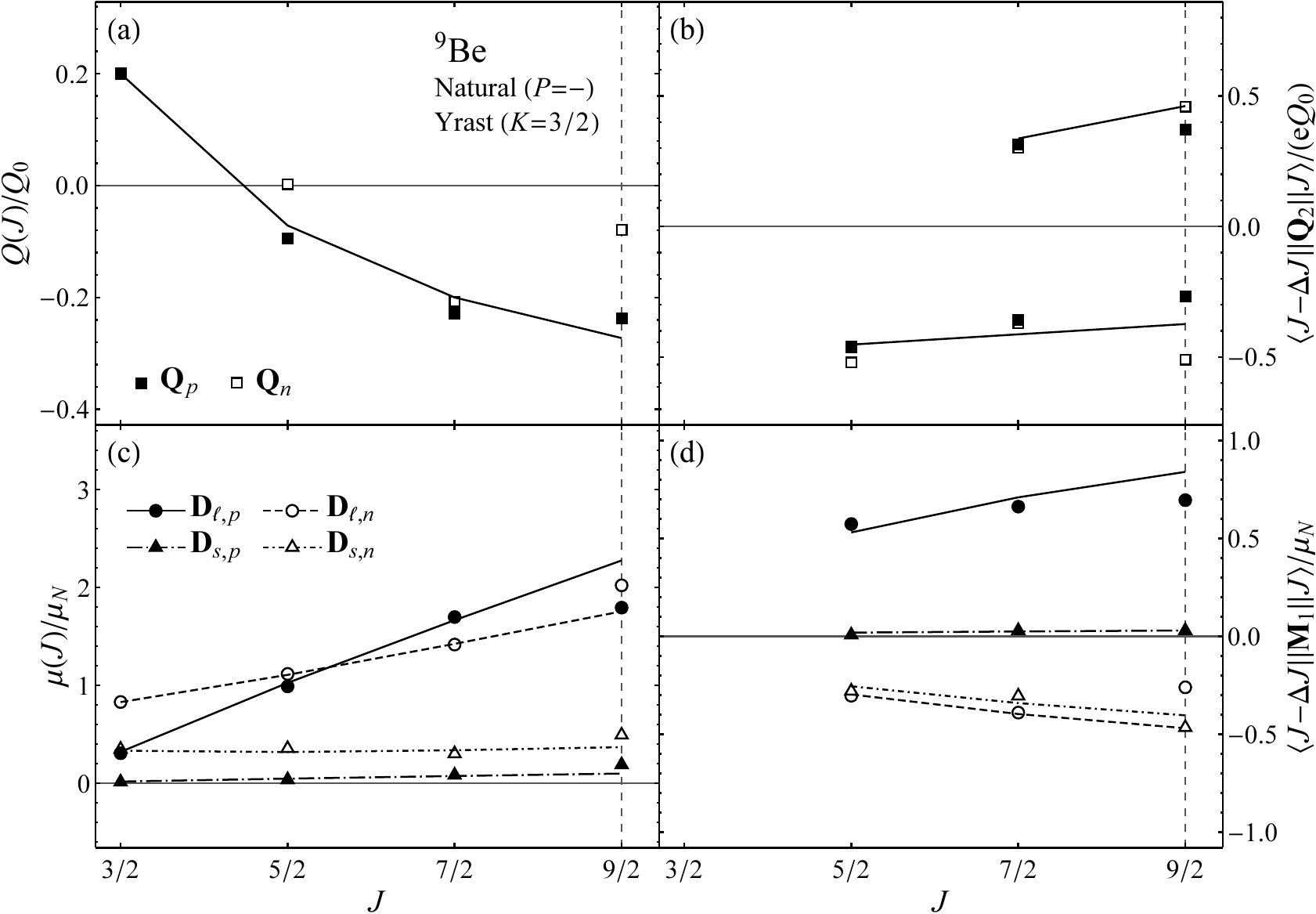}}
\caption{Dipole and quadrupole matrix element observables for the 
$\isotope[9]{Be}$ natural parity yrast
band. See Fig.~\ref{fig-7be0-band1} caption for discussion of the plot contents and labeling.
}
\label{fig-9be0-band1}      
\end{figure*}
\begin{figure*}
\centerline{\includegraphics[width=\ifproofpre{\bandwidth}{1.0}\hsize]{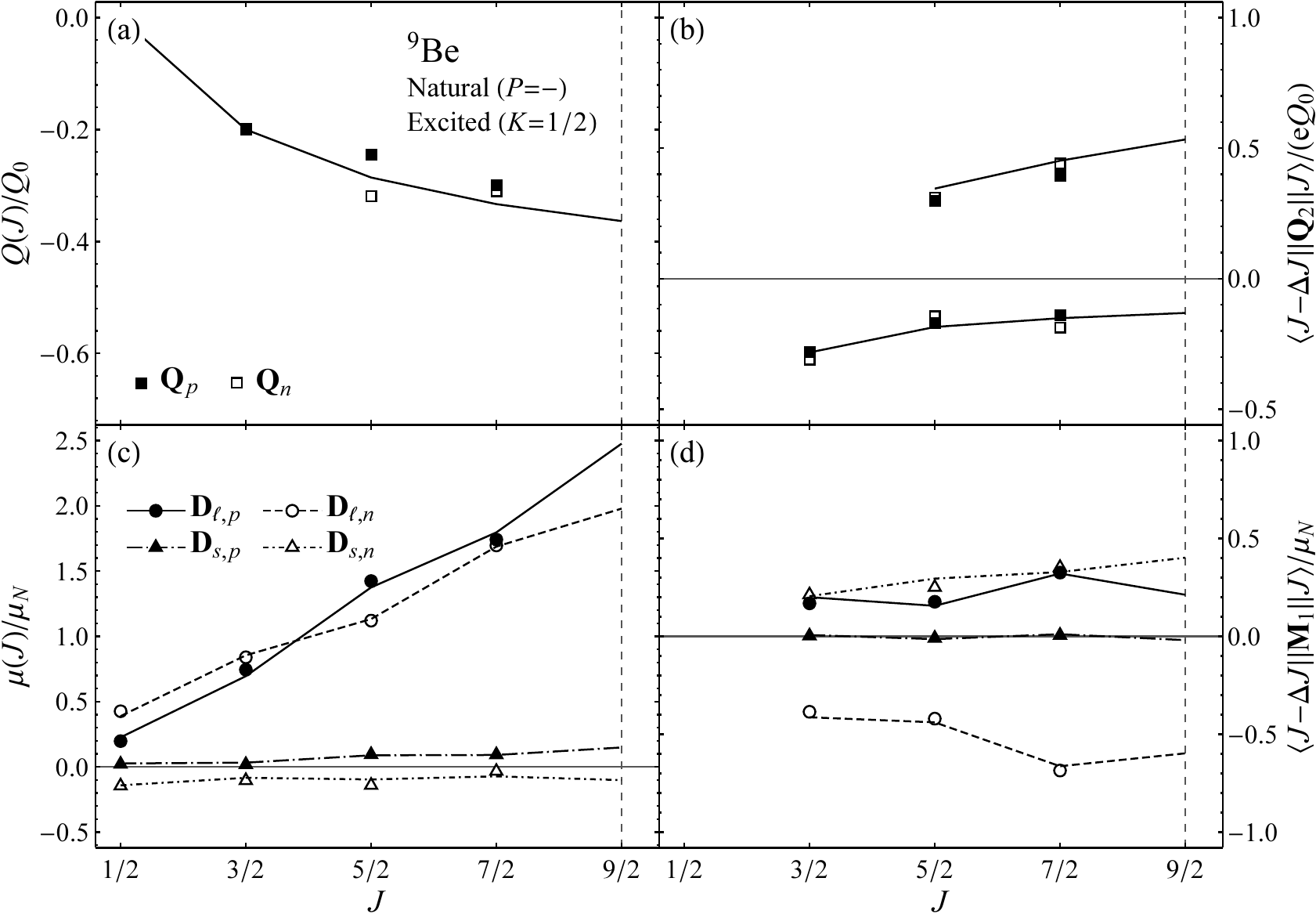}}
\caption{Dipole and quadrupole matrix element observables for the 
$\isotope[9]{Be}$ natural parity excited
band. See Fig.~\ref{fig-7be0-band1} caption for discussion of the plot contents and labeling.
}
\label{fig-9be0-band2}      
\end{figure*}
\begin{figure*}[p]
\centerline{\includegraphics[width=\ifproofpre{\bandwidth}{1.0}\hsize]{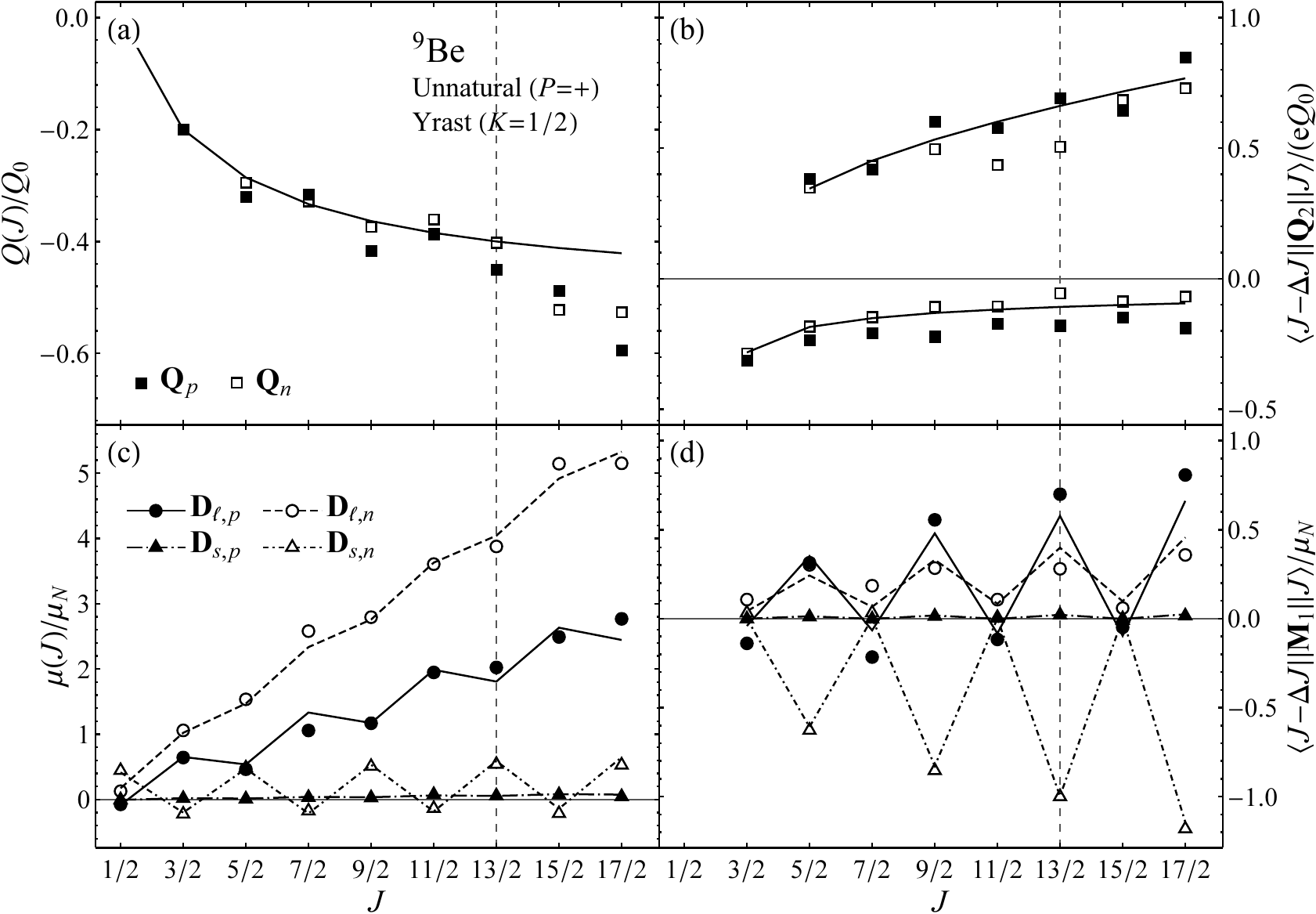}}
\caption{Dipole and quadrupole matrix element observables for the 
$\isotope[9]{Be}$ unnatural parity yrast
band. See Fig.~\ref{fig-7be0-band1} caption for discussion of the plot contents and labeling.
}
\label{fig-9be1-band1}      
\end{figure*}


\begin{figure*}[p]
\centerline{\includegraphics[width=\ifproofpre{\bandwidth}{1.0}\hsize]{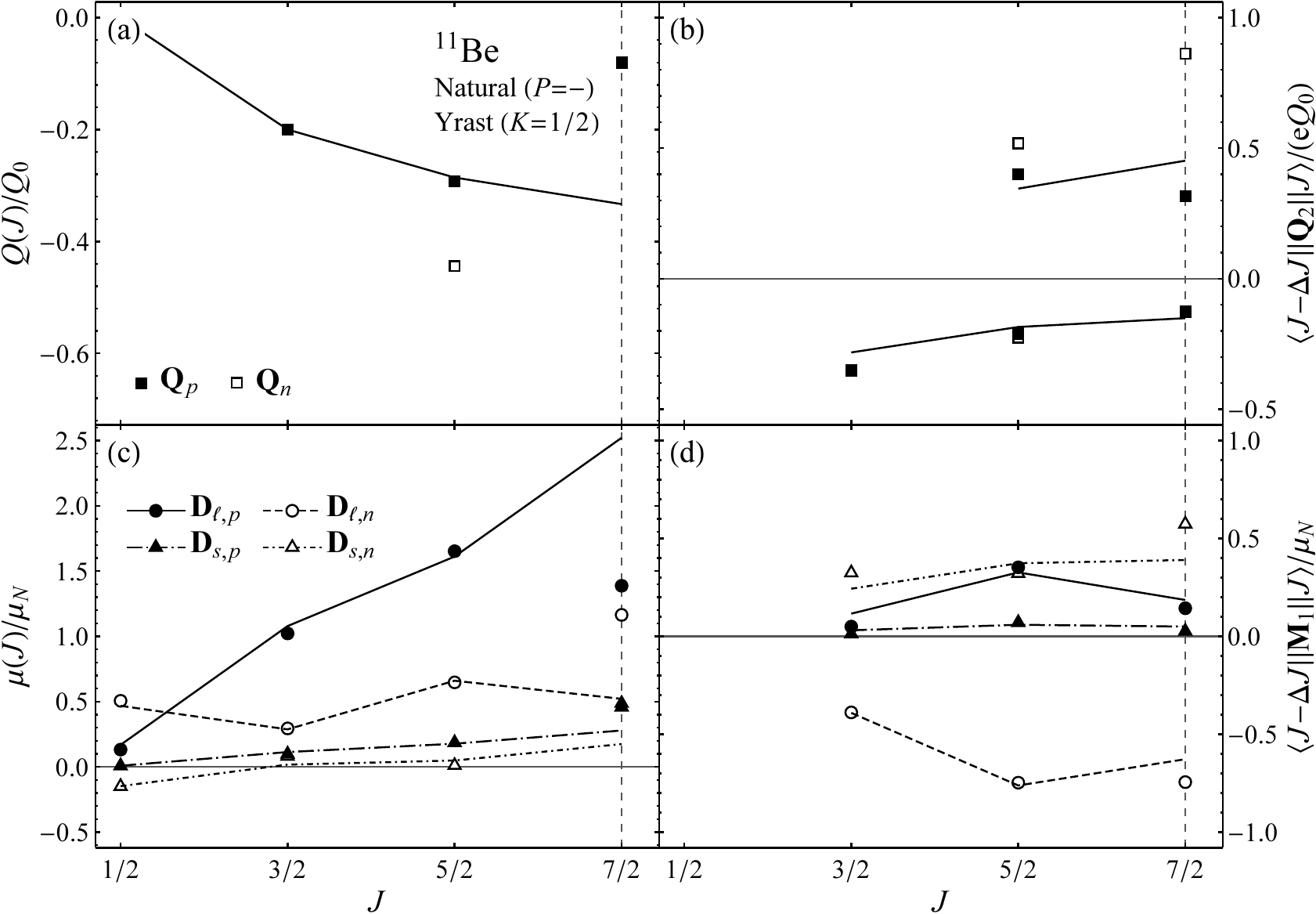}}
\caption{Dipole and quadrupole matrix element observables for the 
$\isotope[11]{Be}$ natural parity yrast
band. See Fig.~\ref{fig-7be0-band1} caption for discussion of the plot contents and labeling.
}
\label{fig-11be0-band1}      
\end{figure*}
\begin{figure*}[p]
\centerline{\includegraphics[width=\ifproofpre{\bandwidth}{1.0}\hsize]{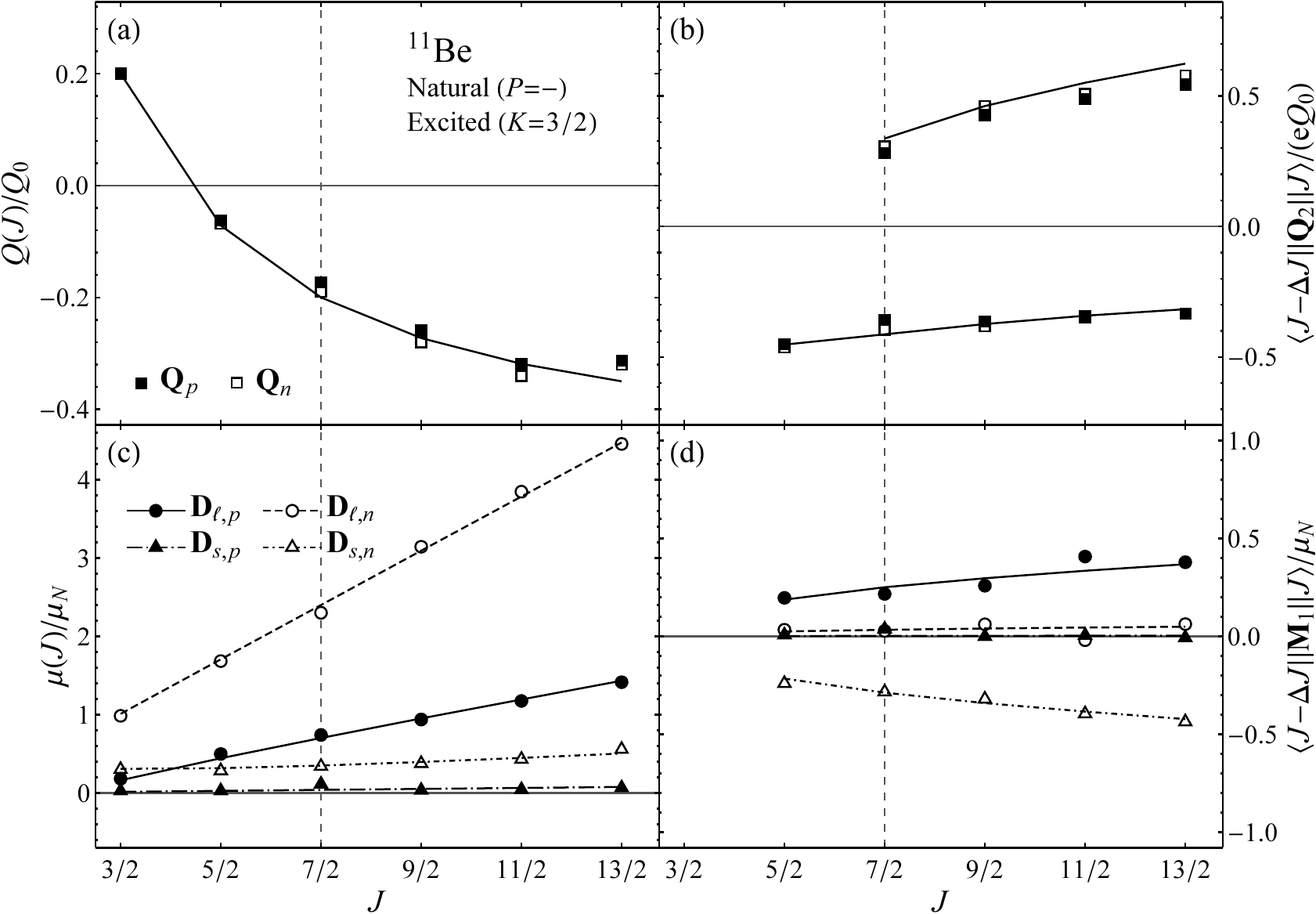}}
\caption{Dipole and quadrupole matrix element observables for the 
$\isotope[11]{Be}$ natural parity excited
band. See Fig.~\ref{fig-7be0-band1} caption for discussion of the plot contents and labeling.
}
\label{fig-11be0-band2}      
\end{figure*}
\begin{figure*}[p]
\centerline{\includegraphics[width=\ifproofpre{\bandwidth}{1.0}\hsize]{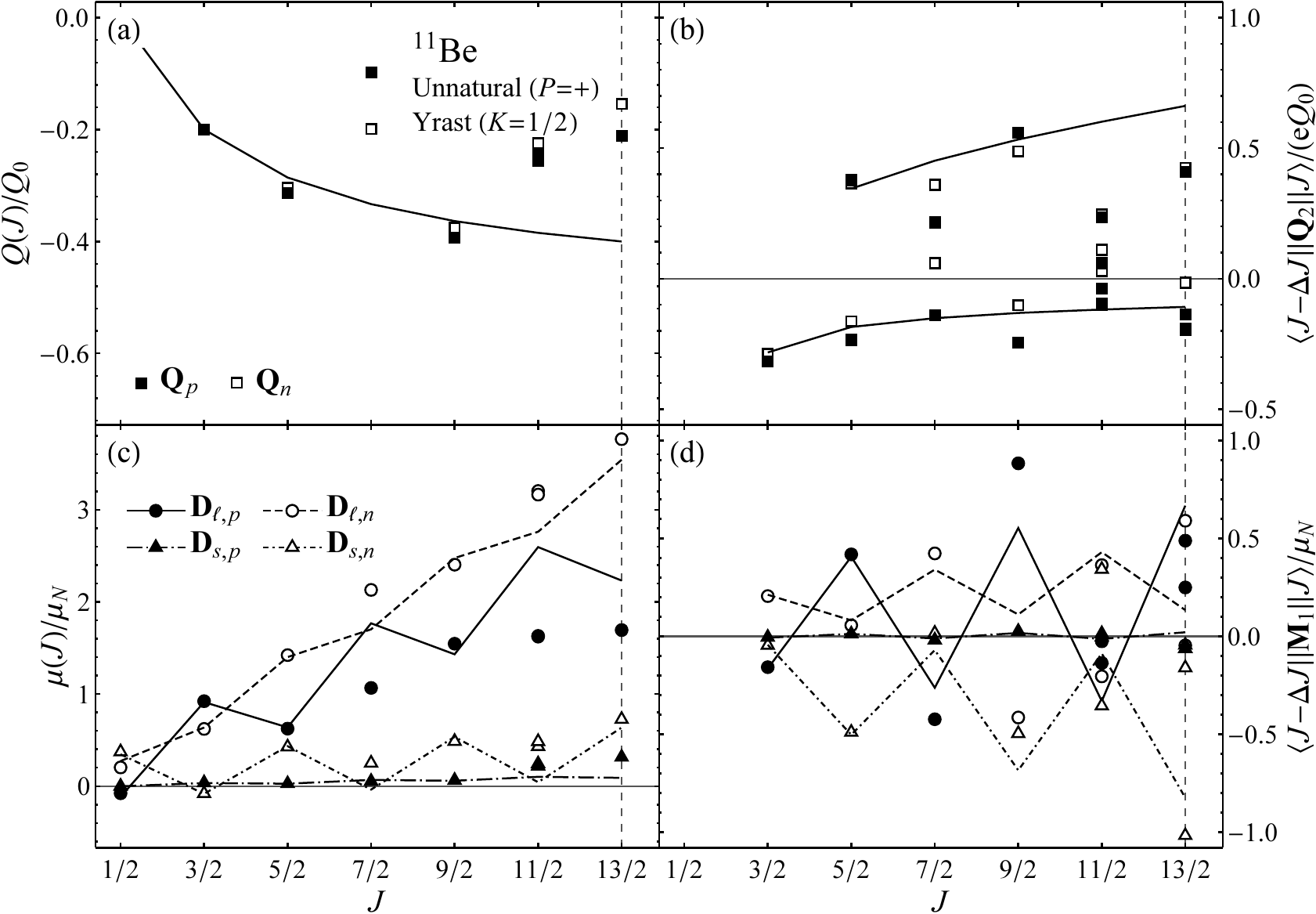}}
\caption{Dipole and quadrupole matrix element observables for the 
$\isotope[11]{Be}$ unnatural parity yrast
band. See Fig.~\ref{fig-7be0-band1} caption for discussion of the plot contents and labeling.
}
\label{fig-11be1-band1}      
\end{figure*}


\begin{figure*}[p]
\centerline{\includegraphics[width=\ifproofpre{\bandwidth}{1.0}\hsize]{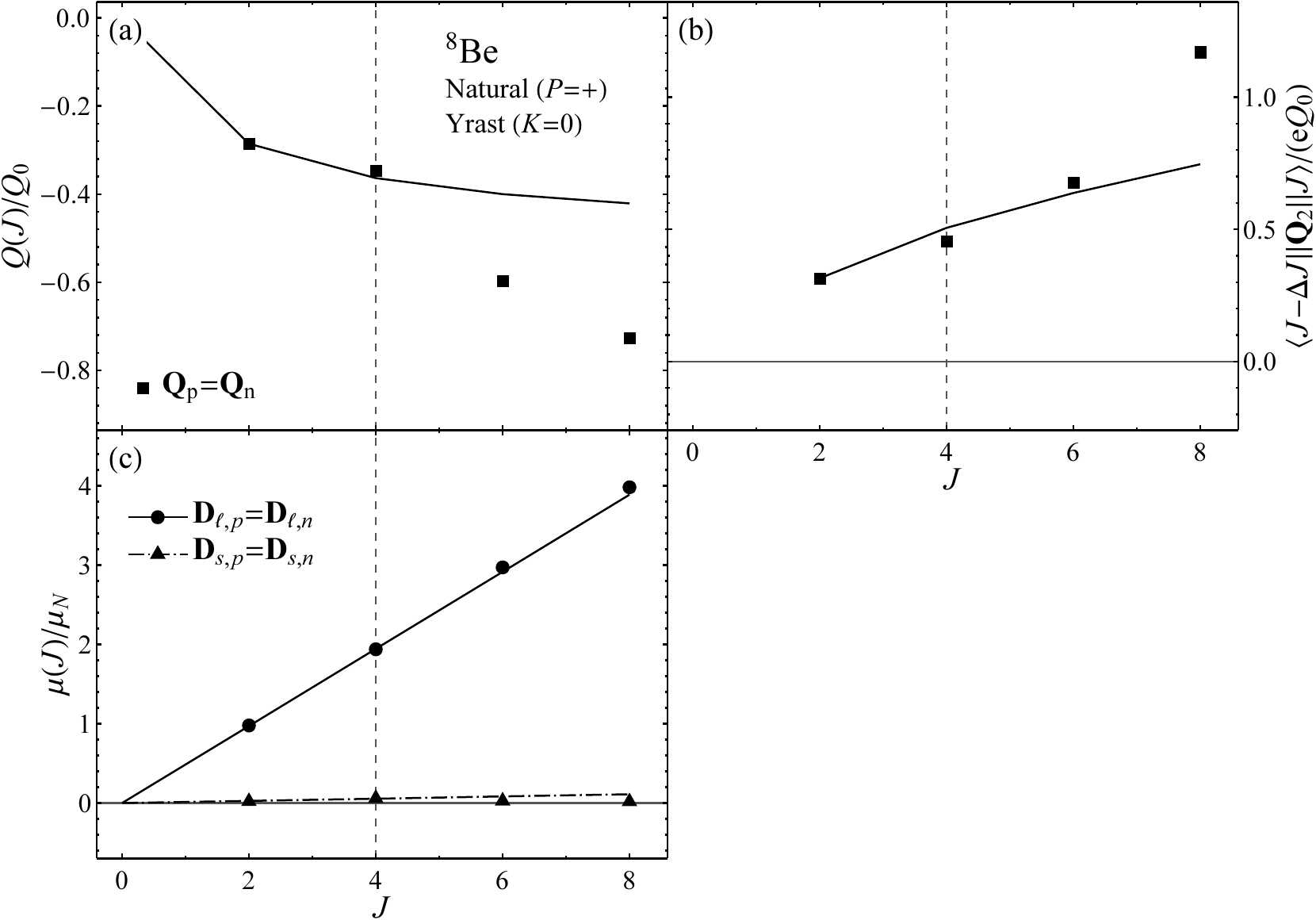}}
\caption{Dipole and quadrupole matrix element observables for the 
$\isotope[8]{Be}$ natural parity yrast
band. See Fig.~\ref{fig-7be0-band1} caption for discussion of the plot
contents and labeling.  Proton and neutron observables are identical
due to isospin symmetry for $\isotope[8]{Be}$.
}
\label{fig-8be0-band1}      
\end{figure*}


\begin{figure*}[p]
\centerline{\includegraphics[width=\ifproofpre{\bandwidth}{1.0}\hsize]{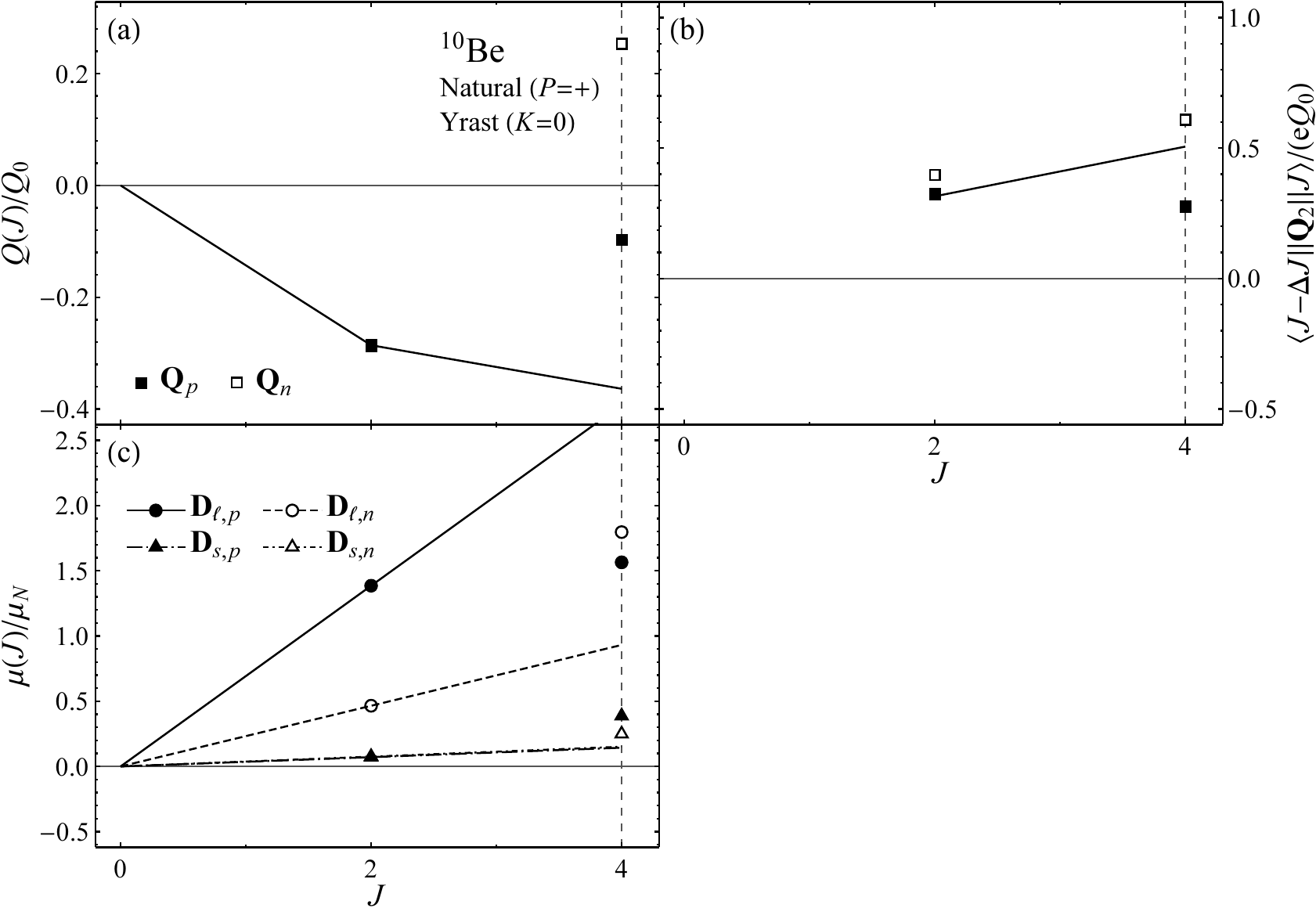}}
\caption{Dipole and quadrupole matrix element observables for the 
$\isotope[10]{Be}$ natural parity yrast
band. See Fig.~\ref{fig-7be0-band1} caption for discussion of the plot contents and labeling.
}
\label{fig-10be0-band1}      
\end{figure*}
\begin{figure*}[p]
\centerline{\includegraphics[width=\ifproofpre{\bandwidth}{1.0}\hsize]{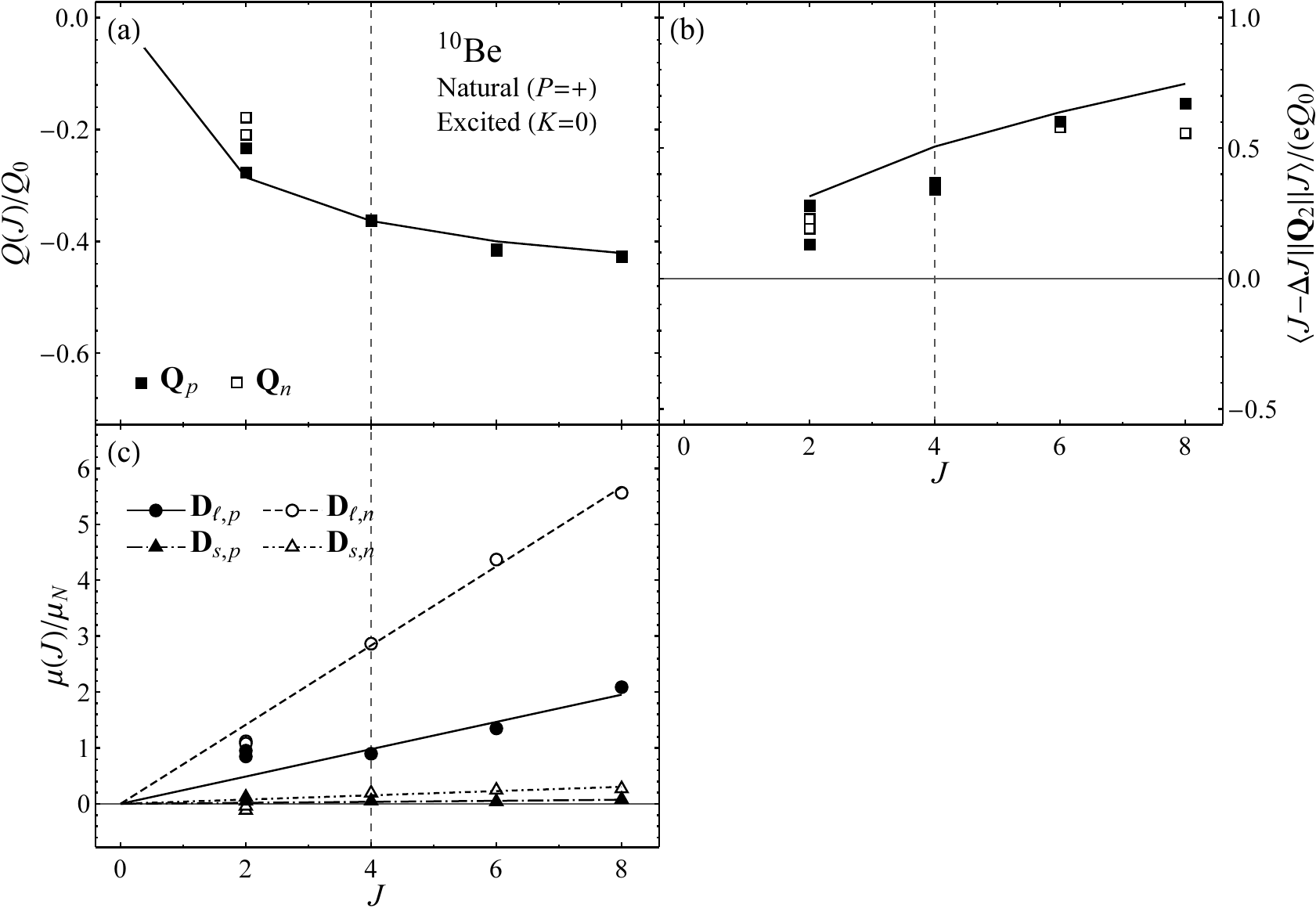}}
\caption{Dipole and quadrupole matrix element observables for the 
$\isotope[10]{Be}$ natural parity excited
band. See Fig.~\ref{fig-7be0-band1} caption for discussion of the plot contents and labeling.
}
\label{fig-10be0-band2}      
\end{figure*}


\begin{figure*}[p]
\centerline{\includegraphics[width=\ifproofpre{\bandwidth}{1.0}\hsize]{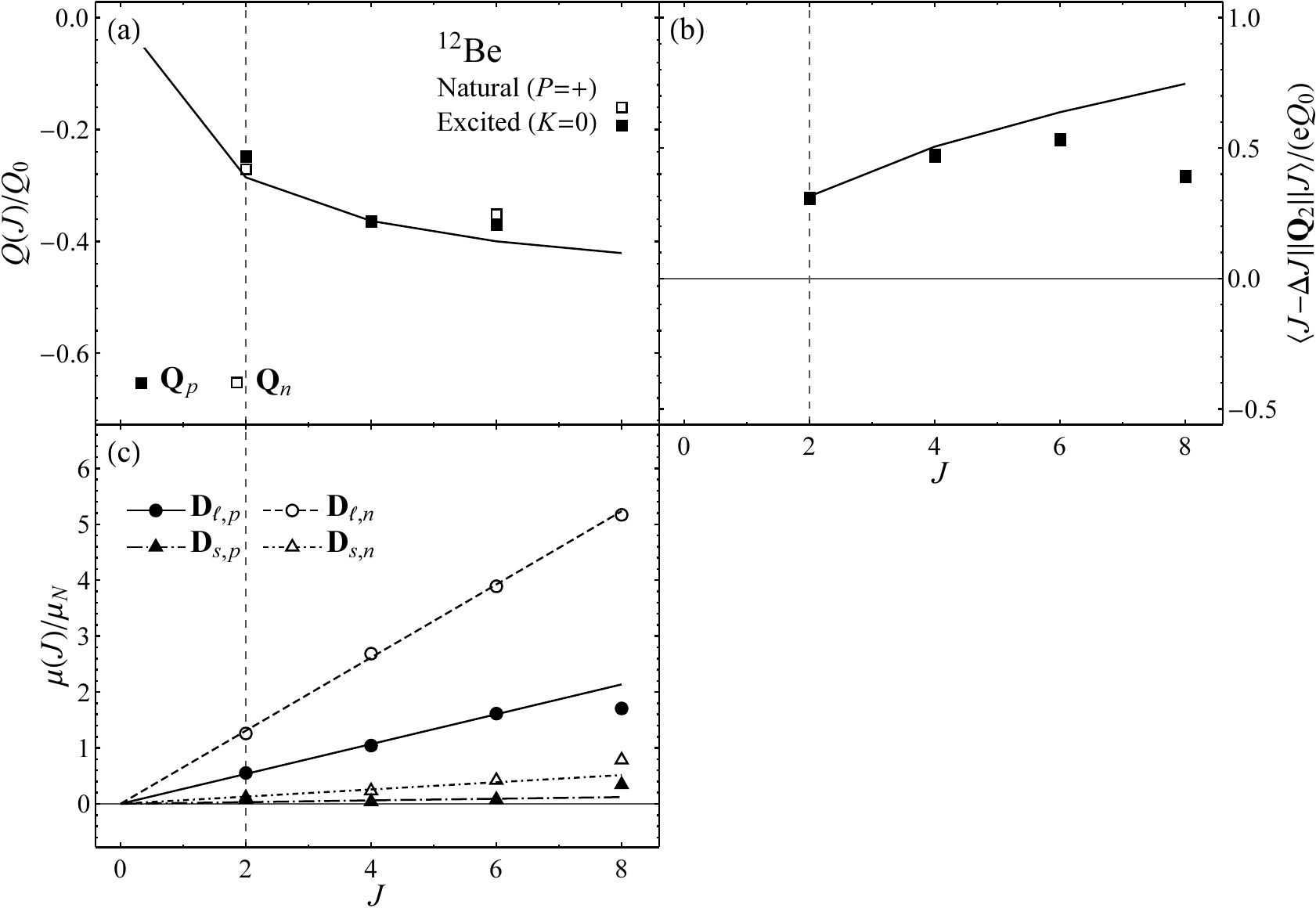}}
\caption{Dipole and quadrupole matrix element observables for the 
$\isotope[12]{Be}$ natural parity excited
band. See Fig.~\ref{fig-7be0-band1} caption for discussion of the plot
contents and labeling.
}
\label{fig-12be0-band2}      
\end{figure*}

In-band quadrupole transition reduced matrix elements are shown in
panel~(b) of each of Figs.~\reffigstrans{}, again as obtained for both
proton (solid symbols) and neutron (open symbols) quadrupole
operators, and for $\Delta J=2$ transitions (upper curves) and $\Delta
J=1$ transitions (lower curves), with the latter not being applicable
to $K=0$ bands [Figs.~\reffigstranseven{}].  The matrix elements are
shown normalized as $\trme{J-\Delta J}{\Qvec_2}{J}/(eQ_0)$, for comparison
with rotational values from~(\ref{eqn-MEE2}).  The same $Q_0$ values
are used as in panel~(a) of each of the figures, \textit{i.e.},
deduced from the quadrupole moment of a suitable low-lying band
member.  Therefore, no free normalization parameter remains for the
transition matrix elements in panel~(b).  For $K=1/2$ bands, recall
that the quadrupole moments are insensitive to the cross term
connecting $\scrR_2$-conjugate intrinsic states in~(\ref{eqn-e2-rme})
and therefore cannot be used to extract the corresponding intrinsic
matrix element $\tme{\phi_K}{Q_{2,2K}}{\phi_{\bar{K}}}$.  No attempt
is made to ``fit'' any possible staggering in the calculated
quadrupole transition matrix elements to determine this matrix
element.  Rather, the rotational values of matrix elements
from~(\ref{eqn-MEE2}) are simply shown, without the cross term, as a
baseline for any possible staggering.

The four distinct ``dipole moments'' for each band member, calculated
using each of the separate dipole term operators $\Dlp$,
$\Dln$,  $\Dsp$, and $\Dsn$ (defined in Sec.~\ref{sec-rot-m1}) are shown in
panel~(c) of each of
Figs.~\reffigstrans{}.  Similarly, for the
odd-mass $\isotope{Be}$ isotopes, the four distinct $\Delta J=1$
dipole transition matrix elements, calculated using each of the dipole
term operators, are shown in panel~(d) of each of
Figs.~\reffigstransodd{}.  (No dipole
transitions are possible in the $K=0$ bands of the even-mass
$\isotope{Be}$ isotopes, due to the $\Delta J=2$ angular momentum
difference between levels.)
The magnetic dipole moment or magnetic dipole transition matrix
element pertinent to physical electromagnetic transitions can, of
course, be recovered as a particular linear
combination~(\ref{eqn-m1-D}) of these values, as determined by the
physical $g$ factors (Sec.~\ref{sec-rot-m1}).  However, for purposes
of rotational analysis, we retain the more complete information provided by the matrix elements
of the four physically distinct dipole terms.

The rotational predictions for the dipole moments from~(\ref{eqn-mu})
involve up to three parameters ($a_0$, $a_1$, and $a_2$), representing
the different core rotor and intrinsic matrix elements, in contrast
to the situation for the quadrupole moments, for which the rotational
predictions have as their only parameter a simple overall normalization. As noted in
Sec.~\ref{sec-rot-m1}, the cross term may contribute with a strength
comparable to that of the direct term, so no simplification can be
obtained through its omission.  Similarly, the rotational predictions for the dipole
transitions involve up to two parameters ($a_1$ and $a_2$).
Therefore, the calculated values are shown directly in
Figs.~\reffigstrans{}, without
normalization to any fitted intrinsic matrix element.

The curves representing the rotational predictions are thus not
uniquely determined by the $K$ quantum number but rather are
determined by a simultaneous global fit of the parameters $a_0$,
$a_1$, and $a_2$ (as applicable, depending upon the $K$ quantum
number) to the dipole moments and transition matrix elements (again,
as applicable) within the band.  For $K=0$ bands, the expected angular
momentum dependence for the dipole moments is indeed simply linear,
and the fit consists of determining the core rotor normalization
parameter $a_0$.  For $K=1/2$ bands, all three terms (involving $a_0$,
$a_1$, and $a_2$) are involved, introducing staggering, while, of
these, only the two involving $a_1$ and $a_2$ enter into the expected
transition matrix elements.  Finally, for bands with $K>1/2$, only two
terms (involving $a_0$ and $a_1$) are in play, while only the term
involving $a_1$ enters into the expected transition matrix elements.
These parameters for the rotational predictions are fitted
independently for each of the dipole term operators, within a band, as
there is no \textit{a priori} relation among the intrinsic matrix
elements for the different dipole terms.  These fits give equal
weights to all moments and transitions, where all levels within the
band are considered up to the point at which substantial anomalies
(band termination) or fragmentation effects are found to occur.

Recall that our purpose is to determine the extent to which the
calculated eigenstates in the \textit{ab initio} calculation conform
to the expectation of an adiabatic rotational scheme, \textit{i.e.},
one yielding wave functions of the form~(\ref{eqn-psi}), and that, to
do so, we are relying upon the indirect evidence provided by energies
and matrix elements of electromagnetic operators~--- these latter, the
matrix elements, being concretely related to the hypothesized
form~(\ref{eqn-psi}) for the wave functions.  Since \textit{signed}
reduced matrix elements $\trme{J_f}{\Tvec_\lambda}{J_i}$, rather than
\textit{unsigned} reduced transition probabilities
$B(T\lambda;J_i\rightarrow J_f)$, of a transition operator
$\Tvec_\lambda$, are considered~--- for both quadrupole and dipole
transitions~--- in Figs.~\reffigstrans{}, it is necessary for us to
elaborate on the extraction of the signs on these quantities and their
interpretation in the rotational context.  The rotational predictions
for the matrix elements~--- as given in~(\ref{eqn-e2-rme})
and~(\ref{eqn-m1-rme})~--- entail definite relations among the signs
of the transition matrix elements (which are also related to the signs
of the moments).  These signs are lost (through taking the square) in
going from matrix elements to transition probabilities.  When
comparing with experiment, such loss of sign information is
inconsquential, since the signs of transition matrix elements are
impossible to extract experimentally (a possible exception, in
principle, being through interference between different excitation
pathways in multi-step Coulomb excitation~\cite{alder1975:book}).
However, the relevant matrix elements are fully accessible for the
present \textit{ab initio} calculated wave functions.

In considering the signs of calculated matrix elements, it must be
borne in mind that the eigenfunctions of the Hamiltonian are
determined only to within an overall phase (\textit{i.e.}, sign, for
real wave functions).  Thus, if the state $\tket{\psi_{JKM}}$
of~(\ref{eqn-psi}) is characterized by a rotational wave function,
then $-\tket{\psi_{JKM}}$ is an equally valid rotational state, the
choice of one over the other being simply a matter of convention,
implicitly embodying also the arbitrary choices of convention entailed
in the definitions of the $\scrD$ functions and Clebsch-Gordan
coefficients~\cite{edmonds1960:am}.  Numerical diagonalization of the
many-body Hamiltonian will arbitrarily select either sign for the
eigenvector (in general, irreproducibly, depending upon the choice of
initial Lanczos trial states or other variations in algorithmic
details).  Diagonal matrix elements (or moments) are insensitive to
the sign of the wave function.  The sign choices on two states of
angular momenta $J$ and $J'$~--- which we may denote by $\sigma_J$ and
$\sigma_{J'}$, respectively~--- enter into the transition reduced
matrix elements $\trme{J'}{\Tvec_\lambda}{J}$ between these states as
the product $\sigma_{J'} \sigma_J$.

In comparing the results for matrix elements from the calculated
eigenvectors with the values expected from the rotational model
formulas, one must therefore attempt to choose the phases on the
calculated eigenvectors to best correlate with the phase convention
embodied in~(\ref{eqn-psi}).  The present analysis considers matrix
elements of a multitude of operators among the states within a band,
which suffice to fully determine (or, rather, overdetermine) the
comparatively few arbitrary phase choices arising from the signs on
the states.  For instance, if we consider specifically the transitions
among three adjacent band members of angular momenta $J-2$, $J-1$, and
$J$, there are $14$ independent transition reduced matrix elements
considered in this work, namely, four from the $\Delta J=1$
transitions for the two independent quadrupole operators ($\Qvec_p$
and $\Qvec_n$), two more from the $\Delta J=2$ transitions for these
quadrupole operators, and eight from the $\Delta J=1$ transitions for
the four independent dipole operators ($\Dlp$, $\Dln$, $\Dsp$, and
$\Dsn$): 
\begin{displaymath}
\renewcommand{\arraystretch}{0}
\begin{array}{ccccc}
 &\xleftarrow{2\,E2\:+\:4\,M1}& & \xleftarrow{2\,E2\:+\:4\,M1} &
\\
\overset{\smash{J-2}}\bigcirc && \overset{\smash{J-1}}\bigcirc & & \overset{\smash{J}}\bigcirc 
\\
&\multicolumn{3}{c}{\xleftarrow[~~\quad\quad\quad\quad
2\,E2\quad\quad\quad\quad~~]{\quad}} & 
\end{array}.
\end{displaymath}
The transition matrix elements are insensitive to a global reversal of
sign, \textit{i.e.}, of $\sigma_{J-2}$, $\sigma_{J-1}$, and
$\sigma_{J}$ simultaneously.  Thus, effectively, only two arbitrary
sign degrees of freedom are present~--- these may be taken as the
choices of the relative phases $\sigma_{J-2}/\sigma_{J}$ and
$\sigma_{J-1}/\sigma_{J}$ of the band members~--- amongst the signs of
the $14$ transition matrix elements.  The two arbitrary signs may be
determined by matching the signs of two transition matrix elements to
the signs expected from the rotational formulas: \textit{e.g.}, once
the sign of $Q_{0,p}$ has been determined from the quadrupole moments,
we may choose the two relative phases $\sigma_{J-2}/\sigma_{J}$ and
$\sigma_{J-1}/\sigma_{J}$ so that the $\Delta J=2$ and $\Delta J=1$
proton quadrupole transition matrix elements from the state $J$ to the
lower band members match those of the rotational predictions.  Then,
the signs of the remaining $12$ matrix elements serve as unambiguous
predictions from the many-body calculation, to be tested against the
rotational expectations.  To allow consistent presentation of signs
over the different sets of calculations in Figs.~\reffigstrans{}, the
free signs in the transition analysis are chosen to enforce
consistency with the positive sign of $Q_{0,p}$ from the quadrupole
moment analysis, \textit{i.e.}, to make
$\trme{J-1}{\Qvec_{2,p}}{J}/(eQ_{0,p})$ negative and/or
$\trme{J-2}{\Qvec_{2,p}}{J}/(eQ_{0,p})$ positive [see
Fig.~\ref{fig-rotor-e2}(b)], as applicable.

\subsection{Convergence}
\label{sec-results-conv}
\begin{figure*}[p]
\centerline{\includegraphics[width=\ifproofpre{0.85}{0.9}\hsize]{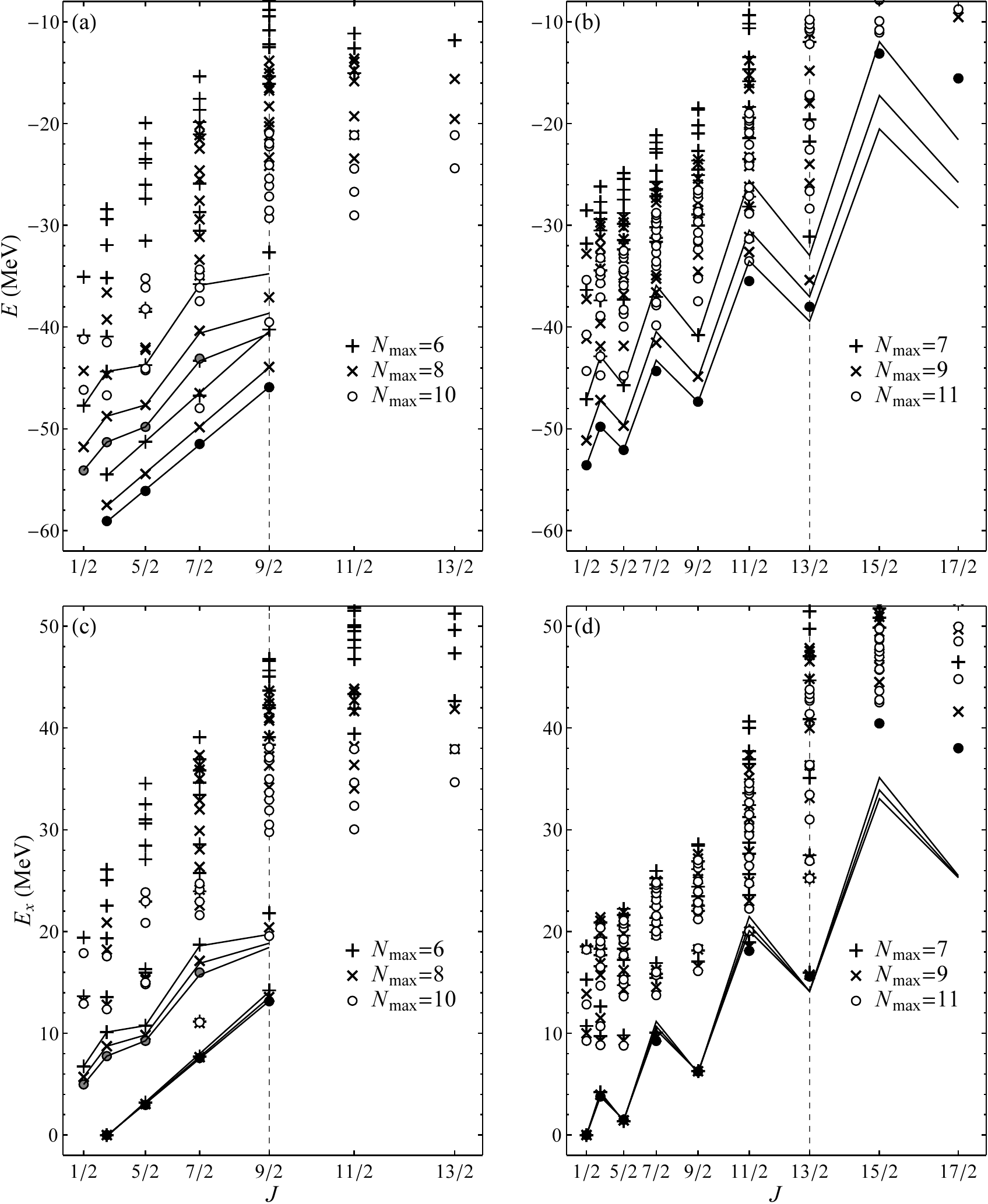}}
\caption{Dependence of calculated energies for $\isotope[9]{Be}$
on the basis truncation $\Nmax$: absolute energies $E$ (top) and
excitation energies $E_x$ (bottom).  Low-lying states in the
natural~(left) and unnatural~(right) parity spaces are shown.
Calculations are for $6\leq \Nmax \leq 10$ for natural parity, or
$7\leq \Nmax \leq 11$ for unnatural parity (calculations of increasing
$\Nmax$ are indicated by plusses, crosses, and diamonds,
respectively). Excitation energies are taken separately for each
parity.  See Fig.~\ref{fig-energy-odd} caption for discussion of other
aspects of the plot contents and labeling.  }
\label{fig-energy-9be-conv}      
\end{figure*}

The calculations presented in the preceding sections are obtained at a
particular level of truncation ($\Nmax=10$ or $11$) of the many-body
configuration space and therefore represent a snapshot along the path
to convergence, \textit{i.e.}, to the actual predictions of the
many-body Schr\"odinger eigenproblem with the JISP16 interaction.
Therefore, in interpreting the results, it is important to have
insight into the nature of the dependence on $\Nmax$ exhibited by
observables for the rotational states, and therefore into the impact
of incomplete convergence on the rotational features observed in the
calculations.

We focus on $\isotope[9]{Be}$ for illustration, beginning with the
energies within rotational bands.  Let us first consider the energy
eigenvalues (these eigenvalues represent the total binding of the
nuclear system).  The calculated eigenvalues for several successive
values of $\Nmax$ are shown at top in Fig.~\ref{fig-energy-9be-conv},
both for the natural parity space ($\Nmax=6$, $8$, and $10$)
[Fig.~\ref{fig-energy-9be-conv}(a)] and for the unnatural parity space
($\Nmax=7$, $ 9$, and $11$) [Fig.~\ref{fig-energy-9be-conv}(b)].  For
each of these $\Nmax$ values, the curves indicate rotational formula
fits to the energies, obtained as described in
Sec.~\ref{sec-results-energy}.  For each step in $\Nmax$, it may be
observed that the energies shift lower by several $\MeV$, an amount
which is large compared to the rotational energy scale.  Thus,
naively, the energies would seem inadequately converged to permit an
analysis of rotational properties.

Nevertheless, the energies of different members of the same band may
be observed to converge at similar rates, and thus the
\textit{relative energies} of levels within a band are far less
dependent upon the truncation than are the energy eigenvalues
themselves.  Excitation energies for $\isotope[9]{Be}$ are shown at
bottom in Fig.~\ref{fig-energy-9be-conv}.  (These are calculated
separately for each parity, \textit{i.e.}, relative to the ``ground
state'' of that parity.)  In examining the unnatural parity band
[Fig.~\ref{fig-energy-9be-conv}(d)], it may be noted that there are
candidate band members at $J=15/2$ and $17/2$, based on
electromagnetic observables, beyond the maximal angular momentum
permitted by the valence configuration
(Sec.~\ref{sec-results-energy}), with energies which lie substantially
above the rotational energy predictions (by $\sim10\,\MeV$, for
$\Nmax=11$).  However, if we consider the $\Nmax$ dependence, we see
that these band members are also falling rapidly in energy relative to
lower band members.  Thus, as convergence continues, the discontinuity
in the behavior of the energies at the maximal valence angular
momentum may be substantially altered, perhaps even disappearing,
representing a qualitative change in the band properties from those
observed at lower $\Nmax$ values.

%
\begin{figure*}[tb]
\centerline{\includegraphics[width=\ifproofpre{0.85}{1.0}\hsize]{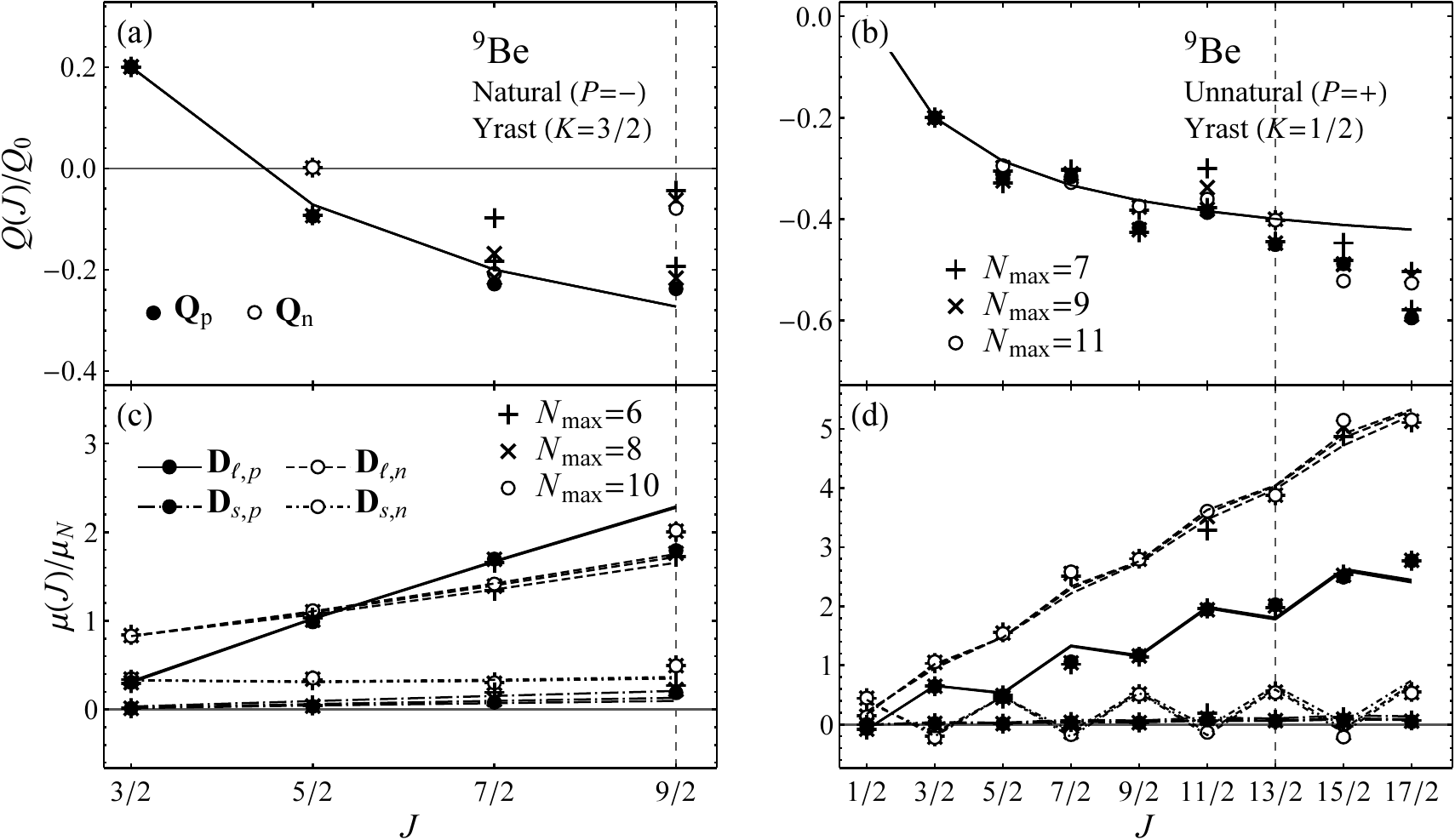}}
\caption{Dependence of calculated electromagnetic moments for $\isotope[9]{Be}$
on the basis truncation $\Nmax$: quadrupole moments~(top) and dipole
moments~(bottom), of the natural~(left) and unnatural~(right) parity
yrast bands.  Calculations are for $6\leq \Nmax \leq 10$ for natural
parity, or $7\leq \Nmax \leq 11$ for unnatural parity (calculations of
increasing $\Nmax$ are indicated by plusses, crosses, and circles,
respectively).  See Fig.~\ref{fig-7be0-band1} caption for discussion
of other aspects of the plot contents and labeling.}
\label{fig-trans-9be-conv}      
\end{figure*}

While the \textit{excitation energy} of the excited natural parity
band of $\isotope[9]{Be}$ [Fig.~\ref{fig-energy-9be-conv}(c)], as
measured relative to the yrast band, decreases with $\Nmax$, this
relative motion is less rapid than that of the eigenvalues themselves.
The change in excitation energy for successive steps in $\Nmax$ is
decreasing, suggesting that the excitation energy of the band might be
approaching a converged value.

The calculated quadrupole matrix elements are highly dependent on both
$\Nmax$ and $\hw$,\footnote{For representative examples of the $\Nmax$
and $\hw$ dependence of calculated quadrupole moments and transition
strengths in a neighboring nuclide, see Fig.~9 of
Ref.~\cite{maris2013:ncsm-pshell}.  Results are shown for states in
$\isotope[7]{Li}$, the mirror nucleus to $\isotope[7]{Be}$.}  and at
present no definitive procedure is available for extracting estimates
for the true, converged values.  Nonetheless, \textit{ratios} of
calculated quadrupole matrix elements within a band are sufficiently
stable with respect to $\Nmax$ to permit identification of the
rotational patterns.  Again, $\isotope[9]{Be}$ is taken for
illustration, at top in Fig.~\ref{fig-trans-9be-conv}, where the
quadrupole moments of the natural parity
[Fig.~\ref{fig-trans-9be-conv}(a)] and unnatural parity
[Fig.~\ref{fig-trans-9be-conv}(b)] yrast bands are shown, for
successive values of $\Nmax$.  As in Figs.~\reffigstrans{}, these
values are taken in ratio to the intrinsic quadrupole moment, which is
calculated from the quadrupole moment of the lowest possible band
member in each case (see Sec.~\ref{sec-results-trans}).  The
calculated ratios are extremely stable with respect to $\Nmax$ for the
lowest band members.  Note especially the proton and neutron
quadrupole moments of the natural parity $J=5/2$ band members in
Fig.~\ref{fig-trans-9be-conv}(a).  These values remain stable with
$\Nmax$ even though they deviate noticeably from the rotational
predictions, so stability is not to be taken to be uniquely associated
with strict adiabatic rotation.  Greater $\Nmax$ dependence is found
for some of the band members at higher angular momentum.

Convergence of magnetic dipole matrix elements is much more rapid than
for electric quadrupole matrix elements.  The magnetic dipole moments
and transition matrix elements are much less sensitive to the basis
$\hw$.\footnote{For the $\Nmax$ and $\hw$ dependence of the calculated
dipole moments for several states in $\isotope[9]{Be}$, see Fig.~6 of
Ref.~\cite{maris2013:ncsm-pshell}.  The quantities labeled in that
figure as spin contributions are equivalent to the dipole terms
considered here.} The calculated magnetic dipole matrix elements for
successive values of $\Nmax$ are shown in
Fig.~\ref{fig-trans-9be-conv}~(bottom).  Note that these dipole
moments are virtually insensitive to $\Nmax$, at the $\Nmax$ values shown.  In the unnatural parity
band [Fig.~\ref{fig-trans-9be-conv}(d)], recall that the energy
convergence properties change markedly for the band members above the
maximal valence angular momentum, but no such discontinuity is
observed for the magnetic dipole moments, for which good
convergence~--- and agreement with the rotational fit~--- is
maintained for all candidate band members.
\begin{figure*}
\centerline{\includegraphics[width=\ifproofpre{0.8}{1.0}\hsize]{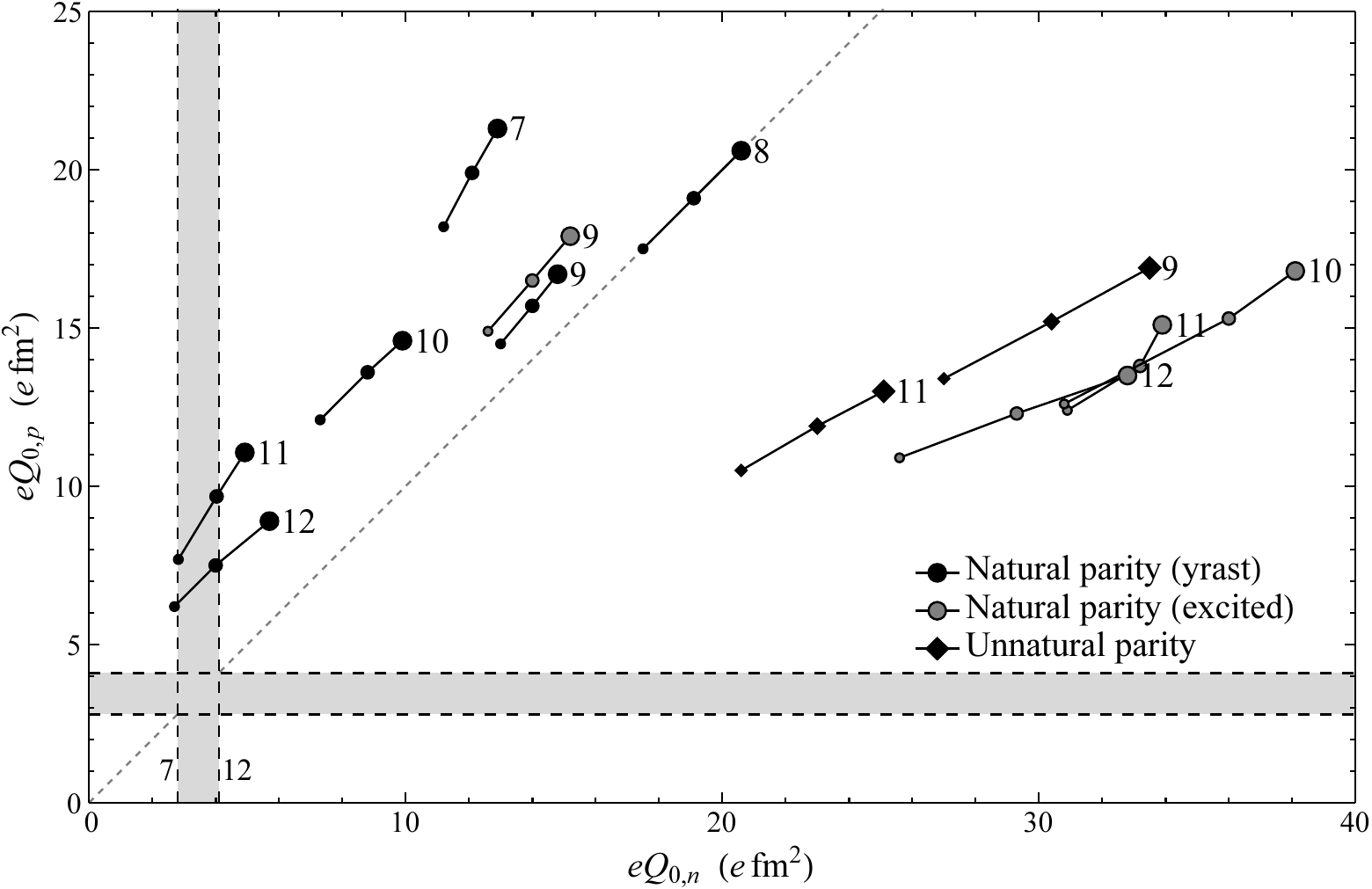}}
\caption{Intrinsic quadrupole moments plotted as $eQ_{0,p}$
  \textit{vs.} $eQ_{0,n}$, for all bands considered in the present
  work.  Bands are distinguished as natural parity yrast (circles,
  solid), natural parity excited (circles, shaded), and unnatural
  parity yrast (diamonds).  Data points are labeled by the mass number
  of the $\isotope{Be}$ isotope.  Values are shown for successive
  $\Nmax$ values ($6\leq \Nmax \leq 10$ for natural parity or $7\leq
  \Nmax \leq 11$ for unnatural parity) to provide an indication of
  convergence (the larger symbols indicate higher $\Nmax$ values) and
  of the stability of the ratio $Q_{0,p}/Q_{0,n}$.  The line
  $Q_{0,p}/Q_{0,n}=1$ is marked (dashed diagonal line), as is the
  range of Weisskopf estimates as $A$ ranges from $7$ to $12$ (shaded
  bands).}
\label{fig-Q0}      
\end{figure*}

Finally, let us return to the quadrupole moments, but consider them
now on an absolute scale (\textit{i.e.}, not as ratios, or normalized
to an intrinsic quadrupole moment), despite their lack of convergence.
Since the ratios within the bands have already been seen to be largely
independent of $\Nmax$, the absolute values may be summarized by the
overall normalization, as given by the intrinsic quadrupole moment.
The intrinsic quadrupole moments, as extracted for each of the bands
we are considering in the $\isotope{Be}$ isotopes, from calculations
with successive values of $\Nmax$ (again, $\Nmax=6$, $8$, and $10$ for
natural parity bands, or $\Nmax=7$, $9$, and $11$ for unnatural parity
bands) are indicated in Fig.~\ref{fig-Q0}.  Successive $\Nmax$
calculations for a given band are joined by a line, and the value
calculated at the highest $\Nmax$ is indicated by the largest symbol.

While the absolute magnitudes of $Q_{0,p}$ and $Q_{0,n}$ are far from
converged in these calculations, they are increasing with $\Nmax$ (as
is typically the case for NCCI calculations carried out near the
variational minimum in $\hw$), and they may therefore may be
interpreted, at least heuristically, as providing a lower bound on the
true converged values.  The calculated values are already $\sim2$ to
$10$ times a single-particle estimate for the intrinsic quadrupole
moment obtained following Weisskopf's
approach~\cite{weisskopf1951:estimate} (see
Appendix~\ref{app-weisskopf} for the derivation).  This single-particle estimate varies from
$\sim2.8\,e\fm^2$ for $A=7$ to $\sim4.1\,e\fm^2$ for $A=12$ , as indicated by the
shaded bands in Fig.~\ref{fig-Q0}.\footnote{In considering the present
  enhancements, one should note that these comparisons are of
  \textit{matrix elements}.  Comparisons of nuclear transition data
  with Weisskopf estimates are often carried out for $B(E2)$ values,
  \textit{i.e.}, proportional to \textit{squared} matrix elements.
  Taking the present matrix elements in square would, of course,
  significantly amplify the quoted ratios of the calculated value to
  the single particle value.}  

By comparison, although the available experimental electromagnetic
moment and transition data for the $\isotope{Be}$
isotopes~\cite{npa2002:005-007,npa2004:008-010,npa1990:011-012} are
limited, we may estimate values for the electric (\textit{i.e.},
proton) intrinsic quadrupole moment $eQ_{0,p}$ for the yrast bands of
$\isotope[8\text{--}10]{Be}$ from the data, using the rotational
relations~(\ref{eqn-Q}) and~(\ref{eqn-BE2}).  The low-lying states in
$\isotope[8]{Be}$ decay primarily to $\alpha$ particles, and the
$2^+\rightarrow0^+$ $\gamma$-ray branch in the $\isotope[8]{Be}$ yrast
band is not observed~\cite{npa2004:008-010}.  However, the measured
$4^+\rightarrow2^+$ $\gamma$ width~\cite{datar2013:8be-radiative}
implies a reduced matrix element which corresponds to
$eQ_{0,p}=23.0(15)\,e\fm^2$ for this band.\footnote{The $\gamma$ width
  must first be transformed to a $B(E2)$ value, which is then simply
  related to the intrinsic quadrupole moment by~(\ref{eqn-BE2}).
  However, the conversion involves a factor or $E_\gamma^5$, where
  $E_\gamma$ is the transition $\gamma$-ray energy.  Thus, the deduced
  $B(E2)$ and, hence, $eQ_{0,p}$ is highly sensitive to the energy
  taken for the $4^+$ resonance.  While we have used the evaluated
  energy~\cite{npa2004:008-010} to deduce $E_\gamma$, Datar~\textit{et
    al.}~\cite{datar2013:8be-radiative} use the resonance parameters
  from $\alpha$ scattering, yielding a larger $B(E2)$ value, from
  which one obtains $eQ_{0,p}=27.2(13)\,e\fm^2$.} The measured $3/2^-$
ground-state quadrupole moment in the $\isotope[9]{Be}$ yrast
band~\cite{npa2004:008-010} similarly corresponds to
$eQ_{0,p}=26.44(19)\,e\fm^2$ for this band.  [The
  $5/2^-\rightarrow3/2^-$ and $7/2^-\rightarrow3/2^-$ $\gamma$
  widths~\cite{npa2004:008-010} are then approximately consistent with
  the expected rotational values.  Specifically, taking the evaluated
  widths and $E2/M1$ division for the transitions in
  $\isotope[9]{Be}$~\cite{npa2004:008-010} yields
  $B(E2;5/2^-\rightarrow3/2^-)/(eQ_{0,p})^2=4.0(3)\times10^{-2}$,
  compared to the rotational value $\approx3.41\times10^{-2}$, and
  $B(E2;7/2^-\rightarrow3/2^-)/(eQ_{0,p})^2=1.4(6)\times10^{-2}$,
  compared to the rotational value $\approx1.42\times10^{-2}$.] The
measured $2^+\rightarrow0^+$ $\gamma$ width in the $\isotope[10]{Be}$
yrast band~\cite{npa2004:008-010} implies a reduced matrix element
which correponds to $eQ_{0,p}=23.0(11)\,e\fm^2$ for this band.

Note that, as $\Nmax$ increases, the \textit{ratio} of the proton and
neutron intrinsic quadrupole moments remains approximately constant, as
may be seen by observing that successive calculated points move
approximately radially outward from the origin of the plot, along a
line of fixed $Q_{0,p}/Q_{0,n}$.  The ratio $Q_{0,p}/Q_{0,n}$ may
therefore be examined as a stable observable of the band (considered further
in Sec.~\ref{sec-disc-intrinsic}).

\section{Band discussion}
\label{sec-disc}

\subsection{Rotational patterns}
\label{sec-disc-patterns}

We are now in a position to examine the energies and electromagnetic
observables, as laid out Sec.~\ref{sec-results}, in the context of
rotational structure and, in particular, to survey the patterns
emerging in the rotational bands along the $\isotope{Be}$ isotopic
chain.  We are interested not only in agreement of the calculated
observables with a rotational picture but also in the nature of the
deviations from rotational behavior.  While some of these
discrepancies may simply be numerical artifacts of incomplete
convergence, as explored in Sec.~\ref{sec-results-conv}, others
indicate physical deviations from the simplest picture of adiabatic
rotation.  Such deviations are to be expected from the various
reasonable physical scenarios,
\textit{e.g.}, that the eigenfunctions reflect
valence shell physics, or that they involve $\alpha$ clusters
surrounded by weakly coupled neutrons.  It is perhaps more remarkable,
not that these deviations arise, but that they are not so large as to
overwhelm the underlying rotational signatures.

Although rotational band structure is robustly apparent, persisting
across calculations of varying basis sizes, the rotational patterns
are being observed in incompletely converged values of observables.
Many of the details~--- certainly many of the quantitative measures
and also likely, to some extent, qualitative features of the bands~---
are therefore still in flux,
\textit{i.e.}, dependent upon the basis truncation parameter $\Nmax$ 
and the oscillator length scale.  Therefore, our interest in the
following discussion principally resides in examining the qualitative
properties of the rotational patterns which emerge from these
\textit{ab initio} calculations and recognizing the different
varieties of band structure which are found, rather than in extracting
definitive numerical predictions for comparison with experiment.
Detailed quantitative comparisons with experimentally identified
rotational bands in the $\isotope{Be}$ isotopes
will likely require the application of basis extrapolation
methods~\cite{maris2009:ncfc,coon2012:nscm-ho-regulator,furnstahl2012:ho-extrapolation,shirokov2014:jisp16-binding}
to deduce the converged values for energies and electromagnetic
observables.

Recall that candidate band members are identified based on both
energies and electric quadrupole strengths
(Sec.~\ref{sec-results-energy}).  The general pattern of agreement
between the energies of these band members and the expected rotational
values for the different bands (Figs.~\ref{fig-energy-odd}
and~\ref{fig-energy-even}) broadly falls into one of three
qualitatively different categories:

(1)~The band terminates at (or below) the maximal angular momentum
permitted within the valence space (see
Sec.~\ref{sec-results-energy}), \textit{i.e.}, no further candidate
band members are identified beyond this angular momentum, and good
agreement in energies is obtained for all band members.  This category
includes the $\isotope[9]{Be}$ natural parity yrast and excited bands
[Fig.~\ref{fig-energy-odd}(b)], $\isotope[11]{Be}$ natural parity
yrast band [Fig.~\ref{fig-energy-odd}(d)], $\isotope[11]{Be}$
unnatural parity yrast band [Fig.~\ref{fig-energy-odd}(e)], and
$\isotope[10]{Be}$ yrast band [Fig.~\ref{fig-energy-even}(b)].  The
severely truncated yrast ``band'' of $\isotope[12]{Be}$
[Fig.~\ref{fig-energy-even}(c)] may also be associated with this
category, to the extent that it can meaningfully be interpreted as a
rotational band.

(2)~Further candidate band members are identified above the maximal
valence angular momentum, but, while good agreement in energies is
obtained up to this angular momentum, the band members above this
angular momentum deviate from rotational energies.  In particular, the
band members above the valence cutoff are found to lie high in energy
relative to the rotational formula.  This category includes the
$\isotope[7]{Be}$ natural parity yrast band
[Fig.~\ref{fig-energy-odd}(a)], $\isotope[9]{Be}$ unnatural parity
yrast band [Fig.~\ref{fig-energy-odd}(c)], and $\isotope[8]{Be}$ yrast
band [Fig.~\ref{fig-energy-even}(a)].

(3)~Further candidate band members are identified above the maximal
valence angular momentum, and good agreement in energies is obtained
for all band members, persisting above the maximal valence angular
momentum with no noticeable discontinuity.  This category includes the
$\isotope[11]{Be}$ natural parity excited band
[Fig.~\ref{fig-energy-odd}(d)], $\isotope[10]{Be}$ excited band
[Fig.~\ref{fig-energy-even}(b)], and $\isotope[12]{Be}$ excited band
[Fig.~\ref{fig-energy-even}(c)].

For bands in the second category, the discontinuity at the maximal
valence angular momentum might not necessarily reflect a true
difference in the \textit{ab initio} description of the band members
below this angular momentum and above it~--- as would be obtained in a
fully converged calculation~--- but rather may, at least in part,
reflect the difference in convergence rates of these levels, as
already noted in Sec.~\ref{sec-results-conv}.  The discontinuity will
clearly be modified as convergence continues with increasing $\Nmax$, and it could conceivably
disappear for sufficiently large $\Nmax$.  Therefore, the distinction
between bands in the second and third categories is not necessarily
rigid.  Similarly, it is conceivable that some bands which, in the
present analysis, are classified into the first category,
\textit{i.e.}, terminating at the maximal valence angular momentum,
could develop further identifiable band members in more completely
converged calculations, and therefore be reclassified into the second
or third category.

For the $K=1/2$ bands, as a result of the Coriolis contribution to the
kinetic energy, alternate band members are raised into a region of
higher density of states (Sec.~\ref{sec-results-energy}).  This may
result in fragmentation, as in the $\isotope[11]{Be}$
unnatural parity yrast band at $J=11/2$
[Fig.~\ref{fig-energy-odd}(e)], or in a disturbance in energy for a
raised band member relative to the rotational expectation but without
manifest mixing, as in the $\isotope[9]{Be}$ unnatural parity yrast
band at $J=11/2$ [Fig.~\ref{fig-energy-odd}(c)].  However,
comparatively close spacing in energies between levels ($<1\,\MeV$) is
also possible, without obvious fragmentation or perturbation arising,
as at $J=5/2$ for the $\isotope[7]{Be}$ natural parity yrast band
[Fig.~\ref{fig-energy-odd}(a)].  It must be realized that these
specifics are of interest only as examples of what \textit{can} happen
in the calculation, not as necessarily robust predictions for
fragmentation in these particular rotational bands.  Due to the
differences in convergence rates of different levels, chance
proximities of the band members with background states may arise at
specific $\Nmax$ values.  The disturbances to individual levels are
therefore highly ephemeral features of the calculations, appearing and
disappearing as $\Nmax$ increases.

Further considering the energies within these $K=1/2$ bands, recall
that the expected rotational curves shown in Fig.~\ref{fig-energy-odd}
are based on a simple estimate of the rotational band energy
parameters in~(\ref{eqn-EJ-stagger}), extracted from just the lowest
three band members.  This simple fit suffices to reproduce the
energies of higher bands members to a remarkable degree, when
compared, \textit{e.g.}, to the energy spacings between successive
levels within the band (tens of $\MeV$) or to the changes in the
calculated energies with each step in $\Nmax$ due to incomplete
convergence (several $\MeV$).  For instance, for the $\isotope[9]{Be}$
natural parity excited $K=1/2$ band [Fig.~\ref{fig-energy-odd}(b)],
the NCCI calculation for the energy of the $J=7/2$ band member matches
the rotational estimate, based on the $J=1/2$ through $5/2$ band
member energies, to within $\sim0.2\,\MeV$.

The $\isotope[9]{Be}$ natural parity bands
[Fig.~\ref{fig-energy-odd}(b)] merit special comment, among the
candidate bands considered here, in that both natural parity bands are
confined to the angular momenta permitted within the valence space.
The excited band terminates at $J=7/2$, while the yrast band extends
to the maximal valence angular momentum, that is, $J=9/2$.  Although
the terminating $J=9/2$ band member is most closely identified with
the yrast band, on the basis of quadrupole transitions, there is also
significant $B(E2)$ strength from this state to the excited band
members: mainly to the $J=5/2$ member of the excited band (at
$\sim0.4$ times the in-band strength), but also somewhat to the
$J=7/2$ member (at $\sim0.1$ times the in-band strength).  In
contrast, the second $J=9/2$ state does not appear to be significantly
connected to the excited band, despite being near the energy which
would be expected from the rotational formula for a $J=9/2$ member of
the excited band, instead primarily decaying to the remaining (second)
low-lying $J=7/2$ state.

The agreement between the calculated electric quadrupole matrix
elements and the expected rotational values, as laid out in
Figs.~\reffigstrans{}(a,b), while certainly not exact, nonetheless
suggests a remarkably clean separation of rotational and intrinsic
degrees of freedom in the \textit{ab initio} NCCI calculations.  This
agreement is typically found to be best at lower angular momentum,
deteriorating for higher angular momentum band members, but to an
extent which varies greatly among the bands.  Several patterns may be
noted:

(1)~For the bands in the first category above, \textit{i.e.}, which
terminate at (or below) the maximal valence angular momentum, there is
a tendency for the quadrupole observables involving the terminating
band member to exhibit larger deviations from the rotational
expectations: consider the $J=9/2$ band member of the
$\isotope[9]{Be}$ natural parity yrast band
[Fig.~\ref{fig-9be0-band1}~(top)], $J=7/2$ band member of the
$\isotope[11]{Be}$ natural parity yrast band
[Fig.~\ref{fig-11be0-band1}~(top)], and $J=4$ band member of the
$\isotope[10]{Be}$ natural parity yrast band
[Fig.~\ref{fig-10be0-band1}~(top)].  However, the quadrupole
observables do not exhibit noticeable termination effects at $J=7/2$
in the $\isotope[9]{Be}$ natural parity excited band
[Fig.~\ref{fig-9be0-band2}~(top)].  It is worth noting that
termination effects are to be expected in the Elliott $\grpsu{3}$
shell model description of nuclear rotation, in which quadrupole
strengths within a band decline as band termination is approached (see
discussion in Ref.~\cite{harvey1968:su3-shell}).  A similar fall-off
can be obtained in $\grpsptr$ symplectic
calculations~\cite{draayer1984:spsm-20ne} of rotational bands (see
Fig.~6 of Ref.~\cite{rowe1985:micro-collective-sp6r}).  However, it is
difficult to isolate signs of such systematic mechanisms for band
termination phenomena from the possible confounding effects of mixing,
since the yrast states also become less well isolated in energy at
these higher angular momenta.

(2)~For the bands in the second and third categories above,
\textit{i.e.}, which extend beyond their corresponding maximal valence angular
momenta, there may (or may not) be a discontinuity in the quadrupole
observables at the maximal valence angular momentum.  This behavior is
somewhat, but not entirely, correlated with that of the energies.
Recall that there is a discontinuity in energies in the
$\isotope[7]{Be}$ natural parity yrast band
[Fig.~\ref{fig-energy-odd}(a)], $\isotope[9]{Be}$ unnatural parity
yrast band [Fig.~\ref{fig-energy-odd}(c)], and $\isotope[8]{Be}$
natural parity yrast band [Fig.~\ref{fig-energy-even}(a)] at the
maximal valence angular momentum.  There is a corresponding
discontinuity in the quadrupole \textit{moments} for these bands in
$\isotope[7]{Be}$ [Fig.~\ref{fig-7be0-band1}(a)], possibly
$\isotope[9]{Be}$ [Fig.~\ref{fig-9be1-band1}(a)], and
$\isotope[8]{Be}$ [Fig.~\ref{fig-8be0-band1}(a)].  However, there is
no apparent discontinuity in the quadrupole \textit{transition matrix
elements} in $\isotope[7]{Be}$ [Fig.~\ref{fig-7be0-band1}(b)] and
$\isotope[9]{Be}$ [Fig.~\ref{fig-9be1-band1}(b)].  For
$\isotope[8]{Be}$ [Fig.~\ref{fig-8be0-band1}(b)], a significant
deviation arises only at $J=8$, by which point the band has moved off
the yrast line, although a change in curvature might putatively
already be evident at $J=6$.  The deviations of the calculated
quadrupole moments and transition matrix elements, for the band
members above the maximal valence angular momentum, are uniformly
toward larger magnitudes relative to the rotational values.  For the
bands in the third category above,
\textit{i.e.}, for which the energies continue to follow the
rotational expectation past the maximal valence angular momentum~---
namely, the $\isotope[11]{Be}$ natural parity excited band
[Fig.~\ref{fig-energy-odd}(d)], $\isotope[10]{Be}$ natural parity
excited band [Fig.~\ref{fig-energy-even}(b)], and $\isotope[12]{Be}$
natural parity excited band [Fig.~\ref{fig-energy-even}(c)]~--- the
quadruple moments and transition matrix elements are similarly
consistent with the expected rotational values
(Figs.~\ref{fig-11be0-band2},
\ref{fig-10be0-band2}, and
\ref{fig-12be0-band2}, respectively), with the
exception that reduced strengths are found for the highest identified
band member, at $J=8$, in $\isotope[12]{Be}$.

(3)~Not unexpectedly, states in regions of higher level density are
susceptible to fragmentation (or mixing) effects, which are reflected
in the quadrupole moments of those states and in transitions involving
those states, as either the initial or final state.  For instance, for
the unnatural parity yrast band of $\isotope[11]{Be}$
[Fig.~\ref{fig-11be1-band1}~(top)], note that it is the band members
which have been raised in energy by the Coriolis staggering~--- with
$J=7/2$ and $11/2$ [Fig.~\ref{fig-energy-odd}(e)]~--- for which the
values for the quadrupole moments and transition matrix elements
deviate from the rotational values (or, rather, for the $J=11/2$ band
member, significant fragmentation over two states precludes
unambiguous comparison).

For the $K=0$ bands, in the even-mass isotopes, it is natural to
consider the calculated $E(4^+)/E(2^+)$ energy ratios (\textit{e.g.},
Ref.~\cite{casten2000:ns}).  Specifically, we must restrict our
attention to the yrast bands, in $\isotope[8]{Be}$ and
$\isotope[10]{Be}$, due to fragmentation of the $J=2$ state in the
excited band.  The expected rotational ratio is
$E(4^+)/E(2^+)=10/3\approx3.33$.  For both $\isotope[8]{Be}$ and
$\isotope[10]{Be}$, the energy ratios are somewhat higher, at $3.41$
and $3.50$, respectively.  The experimental
values~\cite{npa2004:008-010} similarly lie above the rotational
ratio, at $3.75(5)$ and $3.49$, respectively.\footnote{The
$E(4^+)/E(2^+)$ ratio indicated for $\isotope[10]{Be}$ is based on a
tentative spin-parity assignment of $4^+$ for the $11.7\,\MeV$
level~\cite{npa2004:008-010}.}

Similarly, for these bands, it is natural to compare the quadrupole
moment of the $J=2$ state and the $2\rightarrow0$ transition matrix
element, or, customarily, the $B(E2;2\rightarrow0)$ reduced transition
probability, against a rotational description.  This is essentially
the comparison already being made by comparison of the data point at
$J=2$ with the rotational curve in the plots of the $\isotope[8]{Be}$
yrast band [Fig.~\ref{fig-8be0-band1}(b)] and $\isotope[10]{Be}$ yrast
band [Fig.~\ref{fig-10be0-band1}(b)] transition matrix elements.  The
normalization of the curve indicating rotational values is, for these
bands, determined from $Q(2)$, so the proximity of the calculated data
points to the rotational curve indicates agreement between $Q(2)$ and
$B(E2;2\rightarrow0)$.  Specifically, $Q_0$ is obtained from $Q(2)$
via the rotational relation $Q(2)/Q_0=-2/7$, and the rotational
expectation for the transition strength is, in turn, given in terms of
this $Q_0$ by $B(E2;2\rightarrow0)/(eQ_0)^2=
1/(16\pi)\approx1.99\times10^{-2}$.  In the present calculations, the
yrast band of $\isotope[8]{Be}$ [Fig.~\ref{fig-8be0-band1}(b)] has
$B(E2;2\rightarrow0)/(eQ_0)^2=1.96\times10^{-2}$, in agreement with
the rotational value to within $\sim1.3\%$.\footnote{Note that
considering the square of the transition matrix element, to obtain the
$B(E2)$ strength, approximately doubles the relative deviation between
the calculated and rotational values~---
\textit{e.g.}, for $\isotope[8]{Be}$ the agreement of the transition matrix
element itself with the rotational value is at the $\sim0.6\%$ level.}
The yrast band of $\isotope[10]{Be}$ [Fig.~\ref{fig-10be0-band1}(b)]
has $B(E2;2\rightarrow0)/(eQ_0)^2=2.10\times10^{-2}$ for the proton
quadrupole operator, in agreement with the rotational value to within
$\sim6\%$, but $3.13\times10^{-2}$ for the neutron quadrupole
operator, deviating more substantially (as already apparent from
Fig.~\ref{fig-10be0-band1}).  However, even the severely truncated
yrast band in $\isotope[12]{Be}$ yields ratios
$B(E2;2\rightarrow0)/(eQ_0)^2=3.7\times10^{-2}$ for the proton
quadrupole operator and $3.2\times10^{-2}$ for the neutron quadrupole
operator, still within a factor of two of the rotational values.  For
the excited $K=0$ bands in $\isotope[10]{Be}$ and $\isotope[12]{Be}$,
recall that $Q(4)$ is used instead of  $Q(2)$ for normalization, due to
fragmentation of the $J=2$ band member, and the same comparison cannot
directly be made.

The magnetic dipole moments and transition matrix elements bring a
physically complementary set of observables to the problem of
identifying rotational structure in these bands, with more robust
convergence properties than the quadrupole observables (Sec.~\ref{sec-results-conv}).  For the
odd-mass isotopes, the analysis of the dipole observables is more
complicated than for the quadrupole observables, due to the greater
number of relevant parameters (as discussed in
Sec.~\ref{sec-results-trans}).  However, the dipole observables are
the most cleanly rotational, even at high angular momenta.  Recall
that the curves indicating the expected rotational values for these
observables, shown at bottom in Figs.~\reffigstrans{}, are based on a
simultaneous fit to moments and transition matrix elements, generally
restricted to the lower angular momentum band members.

For the odd-mass isotopes, let us consider the behavior of the
magnetic dipole observables under the three different band termination
scenarios.  In the bands in the first category, with termination at
(or below) the maximal valence angular momenta, devations from the
expected rotational values arise in roughly the same terminating
states as for the electric quadrupole matrix elements.  There are
modest deviations at the terminating $J=9/2$ state in the
$\isotope[9]{Be}$ natural parity yrast band
[Fig.~\ref{fig-9be0-band1}(c,d)], but not for the terminating $J=7/2$
state of the excited band [Fig.~\ref{fig-9be0-band2}(c,d)], consistent
with the pattern for the quadrupole observables.  Similarly,
deviations occur at the $J=7/2$ termination for the $\isotope[11]{Be}$
natural parity yrast band [Fig.~\ref{fig-11be0-band1}(c,d)], and
fragmentation again obscures the situation for the $\isotope[11]{Be}$
unnatural parity yrast band [Fig.~\ref{fig-11be1-band1}(c,d)].  For
the remaining bands, in the second and third categories,
\textit{i.e.}, which continue through the maximal valence
angular momentum, the calculated dipole term moments and transition
matrix elements are consistent with rotational values to the highest
angular momentum considered.  This agreement holds regardless of the
deviations found earlier for quadrupole moments or transition matrix
elements (Figs.~\ref{fig-7be0-band1},
\ref{fig-9be1-band1}, and~\ref{fig-11be0-band1}).
 
\subsection{Observables reflecting intrinsic structure}
\label{sec-disc-intrinsic}

So far we have focused on the \textit{existence} of rotational
structure rather than the
\textit{intrinsic structure} underlying this rotation.  That is, we
have been concerned with the extent to which the observable patterns
are consistent with an adiabatic separation of the wave function as
in~(\ref{eqn-psi}), into rotational and intrinsic factors, rather than
exploring the the nature of the intrinsic state itself.  Nonetheless,
some of the observables we are considering do have the potential to
shed light on the intrinsic state.

Although the intrinsic quadrupole moment itself, taken on an absolute
scale, is an obvious choice for an intrinsic structural indicator, as
a measure of deformation, it is rendered largely uninformative by its
incomplete convergence (Sec.~\ref{sec-results-conv}).  In contrast, as
noted in Sec.~\ref{sec-results-conv}, the ratio $Q_{0,p}/Q_{0,n}$
(Fig.~\ref{fig-Q0}) appears to be relatively converged and thus
provides a stable measure comparing the proton and neutron structure
within the intrinsic state.  For instance, in Fig.~\ref{fig-Q0},
consider the behavior of $Q_{0,p}$ and $Q_{0,n}$ as functions of
$\Nmax$ for the natural parity yrast and excited bands of
$\isotope[9]{Be}$.  As noted earlier, these two bands both terminate
at angular momenta consistent with valence space structure.  From
Fig.~\ref{fig-Q0}, it is apparent that these bands have $Q_{0,p}$
values which closely track each other as the basis $\Nmax$ increases,
and similarly for $Q_{0,n}$ values, resulting also in closely matching
ratios $Q_{0,p}/Q_{0,n}$.  This is consistent with~--- though hardly a
conclusive indicator of~--- related underlying intrinsic structures
for these two bands, at least as far as quadrupole correlations are
concerned.

More generally, there is a clear separation of the bands into two
clusters in the $(Q_{0,p},Q_{0,n})$ space of Fig.~\ref{fig-Q0}.  The
natural parity yrast bands (plus the $\isotope[9]{Be}$ natural parity
excited band) all have ratios $Q_{0,p}/Q_{0,n}\gtrsim1$, \textit{i.e.}, lying
just above the diagonal dashed line in Fig.~\ref{fig-Q0}.  (These
bands all fall into either the first or second categories, according
to band termination.)  The remaining natural parity excited bands and
the unnatural parity bands (which fall into the second and third
categories) are instead clustered well below (or, equivalently, to the right of) the
diagonal dashed line.  For each of these bands, $Q_{0,n}$ is about twice
$Q_{0,p}$.  Incomplete convergence makes it unreliable to
compare $Q_{0,p}$ and $Q_{0,n}$ values directly across different bands
in Fig.~\ref{fig-Q0}, since the quadrupole observables for these
different bands may exhibit different convergence rates.  Nonetheless, comparison of
bands within the same nucleus, in calculations at the same $\Nmax$,
are suggestive.  They indicate that change in the ratio of $Q_{0,p}$ and $Q_{0,n}$ between
the two clusters of bands arises primarily from an approximate doubling in $Q_{0,n}$,
rather than from any significant change in $Q_{0,p}$.

It is interesting to consider the present results for $Q_{0,p}$ and
$Q_{0,n}$ in light of the results of antisymmetrized molecular
dynamics (AMD) calculations, in particular, for $\isotope[10]{Be}$,
from Ref.~\cite{kanadaenyo1999:10be-amd}.  In the AMD framework, a
cluster structure arises consisting of an $\alpha+\alpha$ dimer plus
two ``valence'' neutrons.  In the yrast band, the two valence neutrons
are predominantly in $\pi$ orbitals, extending perpendicular to the
symmetry axis of the $\alpha+\alpha$ dimer.  In the excited band, the
two valence neutrons are predominantly in $\sigma$ orbitals.  These
latter orbitals extend along the symmetry axis, giving rise to a more
pronouncedly prolate mass distribution.  Such a change in neutron
distribution between bands is at least qualitatively consistent with the increased
neutron intrinsic quadrupole moment for the excited band in the
present calculations.  It also provides an explanation for the lower
rotational constant (higher moment of inertia) of the excited band, as
arising from a more prolate mass distribution induced by the valence
neutrons.  The AMD calculations also indicate that the yrast
$\pi$-orbital states are largely $p$-shell in character, while the
$\sigma$-orbital states display enhanced clustering and draw heavily
on $2\hw$ and higher excitations in a shell-model picture.  This
proposed difference in structure is consistent with the differing
termination behaviors identified for the $\isotope[10]{Be}$ yrast and
excited bands in the present calculations.

The magnetic dipole observables likewise are capable of providing
insight into the underlying structure, through comparison of the
strengths of the different dipole terms.  (The dipole strengths taken
individually, on an absolute scale, could also yield insight through
comparisons with appropriate models of the intrinsic structure.)  In
particular, the dipole observables obtained with the different dipole
term operators $\Dlp$, $\Dln$, $\Dsp$, and $\Dsn$ allow the proton and
neutron, as well as orbital and spin, contributions to the angular
momentum structure of the band members to be probed independently.

Let us therefore examine the moments and transition matrix elements
arising from the individual dipole terms in greater detail, beginning
with the odd-mass isotopes.  Recall that, in a rotational
interpretation, the moments [Figs.~\reffigstransodd(c)] receive
contributions from the core rotor~--- yielding a term linear in
$J$~--- as well as intrinsic matrix elements (the direct and cross
terms), while only these latter contribute to the transition matrix
elements [Figs.~\reffigstransodd(d)] .  The numerical values of the
$a_0$, $a_1$, and, for $K=1/2$ bands, $a_2$ coefficients extracted
from the rotational fits for the odd-mass isotopes are summarized in
Table~\ref{tab-bands-m1}.
\begin{table*}
  \caption{Magnetic dipole best-fit coefficients for the calculated
  bands in the odd-mass $\isotope{Be}$ isotopes, as obtained with $\Nmax=10$ for
  natural parity or $\Nmax=11$ for unnatural parity, corresponding to
  the rotational curves in Figs.~\reffigstransodd{}.  Natural parity
  bands (upper table) and unnatural parity bands (lower table) are
  shown separately.  Coefficients are in units of $\mu_N$.}
\label{tab-bands-m1}
\begin{center}
\ifproofpre{}{\scriptsize}
\newcommand{\missing}{\multicolumn{1}{c}{---}}  
\begin{ruledtabular}
\begin{tabular}{rrddd@{~}ddd@{~}ddd@{~}ddd}
&&
\multicolumn{3}{c}{$\Dlp$} &
\multicolumn{3}{c}{$\Dln$} &
\multicolumn{3}{c}{$\Dsp$} &
\multicolumn{3}{c}{$\Dsn$} 
\\
\cline{3-5}
\cline{6-8}
\cline{9-11}
\cline{12-14}
 & $K$ & 
\multicolumn{1}{c}{$a_0$} & \multicolumn{1}{c}{$a_1$} & \multicolumn{1}{c}{$a_2$} &
\multicolumn{1}{c}{$a_0$} & \multicolumn{1}{c}{$a_1$} & \multicolumn{1}{c}{$a_2$} &
\multicolumn{1}{c}{$a_0$} & \multicolumn{1}{c}{$a_1$} & \multicolumn{1}{c}{$a_2$} &
\multicolumn{1}{c}{$a_0$} & \multicolumn{1}{c}{$a_1$} & \multicolumn{1}{c}{$a_2$} 
\\
\hline
$\isotope[7]{Be}$ & 1/2 &
+0.620 & -0.304 & +0.460 & +0.345 & -0.129 & +0.143 & +0.022 & +0.000 & +0.013 & +0.012 & +0.432 & -0.615 \\
$\isotope[9]{Be}$ & 3/2 &
+0.558 & -0.860 & \missing & +0.361 & +0.480 & \missing & +0.024 & -0.031 & \missing & +0.057 & +0.412 & \missing \\
 & 1/2 &
+0.541 & -0.281 & +0.105 & +0.442 & +0.656 & -0.108 & +0.030 & +0.003 & +0.022 & -0.012 & -0.377 & -0.019 \\
$\isotope[11]{Be}$ & 1/2 &
+0.702 & -0.319 & -0.160 & +0.162 & +0.849 & +0.221 & +0.079 & -0.067 & -0.016 & +0.057 & -0.463 & -0.044 \\
 & 3/2 &
+0.230 & -0.304 & \missing & +0.690 & -0.041 & \missing & +0.012 & -0.003 & \missing & +0.067 & +0.348 & \missing \\
\hline
$\isotope[9]{Be}$ & 1/2 &
+0.319 & -0.196 & -0.380 & +0.642 & -0.197 & -0.175 & +0.010 & -0.008 & -0.013 & +0.029 & +0.401 & +0.569 \\
$\isotope[11]{Be}$ & 1/2 &
+0.404 & -0.124 & -0.585 & +0.528 & -0.243 & +0.190 & +0.016 & -0.003 & -0.020 & +0.052 & +0.370 & +0.415 
\end{tabular}
\end{ruledtabular}
\raggedright
\end{center}
\end{table*}


The dipole terms yielding the largest moments are generally the proton
and neutron orbital operators $\Dlp$ and $\Dln$, as evident both from
the values of the individual moments in the plots and the magnitudes
of the overall fit coefficients in Table~\ref{tab-bands-m1}.  The
moments from the orbital operators are generally linearly increasing
in $J$, suggesting that the moments primarily arise from a core rotor
contribution [see Fig.~\ref{fig-rotor-m1}(a)].  Note that the core
rotor term rapidly outstrips the intrinsic terms, with increasing $J$,
if the values of $a_0$ and of the intrinsic coefficients $a_1$ or
$a_2$ are comparable in magnitude.  A substantial but subordinate
contribution from the intrinsic direct term appears as a modification
of the slope of the moment curve at low $J$, together with a
displacement of the intercept to a nonzero value when the curve is
extrapolated to $J=0$.\footnote{However, care must be taken in
attempting to identify such a displacement simply from inspection of
the figures.  A nonzero intercept is especially hard to identify for
the $K=3/2$ bands, since the $J$ axes of the plots start at $J\approx
K$, not at the origin.  The coefficients in Table~\ref{tab-bands-m1}
should be relied upon instead.\label{fn-intercept}} A contribution
from the intrinsic cross term appears as staggering in the moment
curve.

Let us take the $\isotope[9]{Be}$ natural parity bands
(Figs.~\ref{fig-9be0-band1} and~\ref{fig-9be0-band2}) for
illustration.  As a starting point for interpretation, it is helpful
to keep in mind the model of $\isotope[9]{Be}$ as an $\alpha+\alpha$
dimer, which may be taken as the core rotor, plus an additional
neutron, which may be taken as the extra-core particle.  Recall that
the coefficient $a_0$ describing the contribution of the core rotor
term represents the effective gyromagnetic ratio of the core, that is,
except for inclusion of the dimensionful factor $\mu_N$
(Sec.~\ref{sec-rot-m1}).  The core rotor contributions for protons are
nearly the same, at $a_0\approx0.54\,\mu_N$ and $0.56\,\mu_N$, for the
ground and excited bands, respectively (see Table~\ref{tab-bands-m1}),
differing by only $\sim3\%$.  The core rotor for $\isotope[9]{Be}$
need not be identical to that of $\isotope[8]{Be}$, due to possible
modifications of the $\alpha+\alpha$ dimer by the presence of the
additional neutron.  However, these $\isotope[9]{Be}$ results are
loosely consistent with the value $a_0\approx0.49\,\mu_N$ obtained for
the yrast band of $\isotope[8]{Be}$.  The core rotor gyromagnetic
ratios for the neutrons show greater variation between the two bands,
at $a_0\approx0.36\,\mu_N$ and $0.44\,\mu_N$, respectively, perhaps
indicating that, contrary to the simplest interpretation, the last
neutron does modify or contribute to the ``core'' neutron structure.
Note that these results lie below the $\isotope[8]{Be}$ value.  It
would be reasonable, in the picture where the core rotor is the
$\alpha+\alpha$ dimer, and under the basic assumption that only the
extra-core particles should contribute to the intrinsic matrix
elements~\cite{rowe2010:collective-motion}, to expect the proton
orbital observables to result entirely from the core gyromagnetic
contribution, with no intrinsic contribution, while the neutron
orbital observables would have an intrinsic contribution from the
extra-core neutron.  However, from the $a_1$ and $a_2$ coefficients in
Table~\ref{tab-bands-m1}, it is apparent that both the proton and
neutron orbital operators yield sizable intrinsic matrix elements,
\textit{e.g.}, for the ground-state band, the values are
$a_1\approx-0.86\,\mu_N$ for the protons and $+0.48\,\mu_N$ for the
neutrons, so the intrinsic matrix element for the proton operator is
actually larger in magnitude than that for the neutron operator,
though, incidentally, of opposite sign.

Considering the spin dipole terms, if the protons are all confined to
$\alpha$ particles, their spins should be pairwise coupled to zero,
and we would expect vanishing spin dipole matrix elements for the
protons.  The $\Dsp$ observables in these bands are indeed essentially
vanishing, compared to the other magnetic dipole observables, as
apparent from Figs.~\ref{fig-9be0-band1} and~\ref{fig-9be0-band2} (or
Table~\ref{tab-bands-m1}).  For the neutron spin operator, the
extra-core neutron can contribute.  The neutron spin dipole
observables are dominated by the intrinsic contribution (specifically,
the direct term, even for the excited band, where the cross term could
contribute), with negligible core contribution.

Indeed, across the $\isotope{Be}$ isotopes, the dipole moments and
transition matrix elements for the proton spin operator $\Dsp$ are
consistently suppressed, as evident both from the comparatively small
values of the individual moments in the plots and from the coefficients
in Table~\ref{tab-bands-m1}, all $\lesssim0.1\,\mu_N$.  The largest
values arise in the $\isotope[11]{Be}$ yrast band terminating state
[Fig.~\ref{fig-11be0-band1}(c)], for which sizable deviations from
rotational expectations for the calculated electromagnetic observables
have already been noted.  The intrinsic matrix element contributions,
for the various dipole term operators, may be more cleanly examined in
the calculations for transition matrix elements
[Figs.~\reffigstransodd(d)], where they are not obscured by the core
rotor contribution.  The neutron spin operator $\Dsn$ never generates
a sizable core rotor contribution ($a_0$ coefficient) but does give
rise to substantial intrinsic matrix elements, comparable to those
obtained for the orbital dipole operators.  The only dipole term with
consistently negligible intrinsic matrix elements is the proton spin
term, as already noted.  The intrinsic matrix elements for the
remaining dipole terms ~--- proton orbital, neutron orbital, and
neutron spin~--- variously take on greater or lesser strengths for the
different bands under consideration.

For the even-mass isotopes [Figs.~\reffigstranseven(c)], in a
rotational picture, the dipole moments arise exclusively from the core
rotor contribution.  The calculated moments do indeed generally follow
the expected linear behavior with respect to $J$.  The exception is
the $\isotope[10]{Be}$ natural parity yrast band
[Fig.~\ref{fig-10be0-band1}(c)], for which the moments of the $J=2$
and $4$ states do not follow a simple linear relation.  However, the
quadrupole observables also deviated markedly from rotational
relations for these states.  For the bands which continue past the
maximal valence angular momentum~--- namely, the $\isotope[8]{Be}$
natural parity yrast band [Fig.~\ref{fig-8be0-band1}(c)], the
$\isotope[10]{Be}$ natural parity excited band
[Fig.~\ref{fig-10be0-band2}(c)], and the $\isotope[12]{Be}$ natural
parity excited band [Fig.~\ref{fig-12be0-band2}(c)]~--- the dipole
moments closely follow the expected linear dependence on $J$ (aside
from fragmentation at $J=2$ in $\isotope[10]{Be}$).  Note that there
is no discontinuity at the maximal valence angular momentum.  The spin
dipole terms are negligible compared to the orbital
dipole terms, consistent with the small core rotor contributions found
for these terms in the bands of the odd-mass isotopes.

We may finally compare gyromagnetic ratios for the proton and neutron
orbital dipole operators in these $K=0$ bands of the even-mass
isotopes.  They are strictly identical in $\isotope[8]{Be}$
[Fig.~\ref{fig-8be0-band1}(c)] by isospin symmetry.  For the two
remaining natural parity yrast bands, meaningful comparison is
difficult: recall the deviation from rotational behavior in the yrast
band of $\isotope[10]{Be}$ [Fig.~\ref{fig-10be0-band1}(c)] and the
severely truncated nature of the yrast ``band'' in $\isotope[12]{Be}$,
for which we can only consider the dipole moments for the $J=2$ state
(only the moment for the proton orbital dipole term is substantial, at
$\sim1.5\,\mu_N$, while those for all other dipole terms are
$\lesssim0.3\,\mu_N$).  In the $\isotope[10]{Be}$
[Fig.~\ref{fig-10be0-band2}(c)] and $\isotope[12]{Be}$
[Fig.~\ref{fig-12be0-band2}(c)] excited bands, the moments for the
neutron orbital dipole term are more than twice those for the proton
orbital dipole term, ostensibly reflecting the neutron excess in these
nuclei.  Recall that the individual dipole terms ($\Dlp$, $\Dln$,
$\Dsp$, and $\Dsn$ ) are proportional to the different angular
momentum operators ($\lvec_p$, $\lvec_n$, $\svec_p$, and $\svec_n$)
which together add up to the total angular momentum operator.
Consequently, the dipole moments calculated for these different dipole
terms are proportional to the contribution of the corresponding
angular momentum operator to the total angular momentum.\footnote{In
general, if a total angular momentum operator is obtained as a sum of
components $\Jvec=\sum_i
\Jvec_i$, then the fractional contribution of any particular component
operator $\Jvec_i$ to the total angular momentum of a state
$\tket{\psi_J}$ is given by
$f_i\equiv\tbracket{\Jvec_i\cdot\Jvec}/\tbracket{\Jvec\cdot\Jvec}=[J(J+1)(2J+1)]^{-1/2}\trme{\psi_J}{\Jvec_i}{\psi_J}$.
This latter reduced matrix element $\trme{\psi_J}{\Jvec_i}{\psi_J}$ also defines the dipole moment,
calculated taking $\Jvec_i$ as the dipole operator, to within
geometrical factors involving $J$.  Note that the angular momentum
fractions $f_i$ necessarily sum to unity, by the definition of
$\Jvec$ as the sum of the $\Jvec_i$.}  Thus, \textit{e.g.}, for the $\isotope[12]{Be}$ excited
band, the values $a_0\approx+0.27\,\mu_N$, $+0.66\,\mu_N$,
$+0.012\,\mu_N$, and $+0.07\,\mu_N$, obtained for these dipole terms,
respectively, indicate that the angular momenta of the band members
all receive approximately proportionate contributions, of $\sim27\%$
from proton orbital motion, $\sim66\%$ from neutron orbital motion,
$\sim1\%$ from proton spin, and $\sim7\%$ from neutron spin.

\subsection{Rotational energy parameters}
\label{sec-disc-params}

Let us finally now focus on the global properties of the rotational
spectra, namely, the band energy parameters.  Our purpose is both to
explore what these parameters may indicate about the structure of the
calculated bands and to obtain points of comparison to the experimental data for
proposed rotational bands in the $\isotope{Be}$ isotopes.
\begin{figure*}
\centerline{\includegraphics[width=\ifproofpre{0.8}{1.0}\hsize]{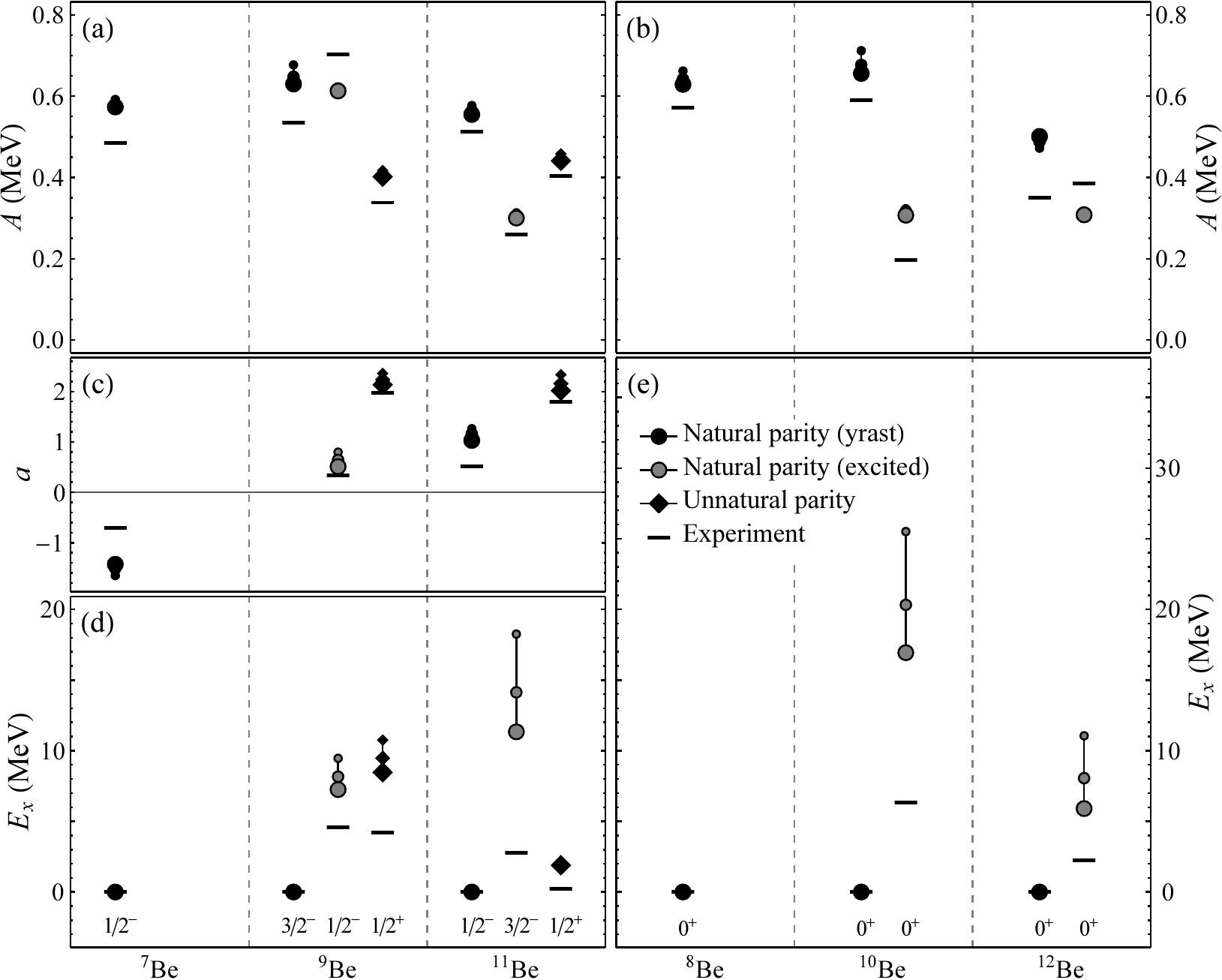}}
\caption{Band energy parameters for all bands considered in the
  present work, for the odd-mass $\isotope{Be}$ isotopes~(left) and
  even-mass $\isotope{Be}$ isotopes~(right).  Parameters considered
  are: the rotational energy constant $A$~(top), the Coriolis
  decoupling parameter $a$~(middle), and the band excitation energy
  $E_x$~(bottom), which is defined as the band energy $E_0$
  for the given band [see~(\ref{eqn-EJ}) and~(\ref{eqn-EJ-stagger})]
    relative to that of the natural parity yrast band.  Bands are distinguished
as natural parity yrast (circles, solid), natural parity excited
(circles, shaded), and unnatural parity yrast (diamonds).  
Values are shown for successive $\Nmax$ values ($6\leq \Nmax \leq 10$
for natural parity or $7\leq \Nmax \leq 11$ for unnatural parity) to
provide an indication of convergence (the larger symbols
indicate higher $\Nmax$ values).  The parameter values extracted from
experimental bands are indicated by horizontal lines.  The $K^P$
assignments for the bands are indicated at bottom.
}
\label{fig-params}      
\end{figure*}

Recall that the parameters appearing in the rotational energy
relations~(\ref{eqn-EJ}) and~(\ref{eqn-EJ-stagger}) are the band
energy $E_0$, the rotational constant $A$, and, for $K=1/2$ bands, the
Coriolis decoupling parameter $a$.  In a plot of energies
\textit{vs.}\ $J(J+1)$ (as in Figs.~\ref{fig-energy-odd}
and~\ref{fig-energy-even}), these parameters represent the ``height''
or energy intercept of the band (this is \textit{not} simply the band
head energy unless $K=0$), the ``slope'' of the band, and the
``staggering'' of the band, respectively.  The values of these
parameters, as extracted from the energies of the calculated band members (as described in
Sec.~\ref{sec-results-energy}), are shown in Fig.~\ref{fig-params},
with the odd-mass isotopes at left and even-mass isotopes at right.
The band energy [Fig.~\ref{fig-params}(d,e)] is presented as a band
\textit{excitation} energy $E_x$, obtained by taking the band energy
$E_0$ for the given band relative to that of the natural parity yrast band.  Results
obtained from calculations with successive values of $\Nmax$
($\Nmax=6$, $8$, and $10$ for natural parity bands, or $\Nmax=7$, $9$,
and $11$ for unnatural parity bands) are represented by symbols of
increasing size.  Note that the band parameters at the highest $\Nmax$
are determined by the same energy fits as shown in
Figs.~\ref{fig-energy-odd} and~\ref{fig-energy-even}.

Ideally, comparison of the calculated and experimental band
parameters provides a direct test of the degree to which the nuclear
many-body problem with the chosen internucleon interaction (here,
JISP16) reproduces the rotational dynamics actually occurring in the
physical $\isotope{Be}$ isotopes.  However, this comparison is subject
to limitations from both calculational and experimental considerations.

From the calculational side, 
we must consider the extent to which the values for band parameters obtained from the
present truncated calculations are converged.  The $\Nmax$ dependence
of the relative energies of calculated band members was already
explored in
$\isotope[9]{Be}$ (Sec.~\ref{sec-results-conv}), providing
some insight into what type of evolution with $\Nmax$ may be
expected.  The
dependence of the parameter values on $\Nmax$ shown in
Fig.~\ref{fig-params} more directly provides an indication of the robustness of these
results with respect to the $\Nmax$ truncation (as discussed in
further detail for each of the parameters below).

From the experimental side, clear-cut identification of rotational band members in
the $\isotope{Be}$ isotopes is challenging, since prospective band
members may include wide or poorly-resolved resonances, spin-parity
assignments are missing or uncertain for many of the known states, and
electromagnetic transition data are largely unavailable~\cite{npa2002:005-007,npa2004:008-010,npa1990:011-012}, although
reaction amplitudes or decay widths (\textit{e.g.},
Refs.~\cite{vonoertzen1996:be-molecular,vonoertzen1997:be-alpha-rotational,suzuki2013:6he-alpha-10be-cluster}) can provide 
structural indicators relevant to band identification.  
Experimental candidates for rotational bands in the
$\isotope{Be}$ isotopes ($8\leq A \leq 12$) are surveyed by Bohlen
\textit{et al.}~\cite{bohlen2008:be-band}.  Only two or three low-lying band
members need be identified to extract an estimate of the band
parameters.   However, the parameter values obtained for the
experimental bands are, naturally, sensitive to the choice of included
band members and the energies adopted for these.
Values obtained from fits to the
experimentally observed bands are shown in
Fig.~\ref{fig-params} (indicated by horizontal lines).  
In generating these fits, we have used the band members and energies
indicated in Table~\ref{tab-bands-expt}.
\begin{table*}
  \caption{Experimenal band members used for the rotational energy fits
    yielding the experimental rotational parameter values in Fig.~\ref{fig-params}.
    For each band member, the nominal angular momentum and parity (assignments are not all
    definite, as discussed in the indicated references) and excitation
    energy (in $\MeV$) are indicated.  }
\label{tab-bands-expt}
\begin{center}
\begin{ruledtabular}
\begin{tabular}{D{.}{~}{-1}ll}
\multicolumn{1}{c}{Band} &
Levels &
References
\\
\hline
\isotope[7]{Be} . 1/2^-_1 & 
$1/2^-$ 0.429, $3/2^-$ 0, $7/2^-$ 4.570 &
\cite{npa2002:005-007}
\\
\isotope[9]{Be} . 3/2^-_1 &
$3/2^-$ 0, $5/2^-$ 2.429, $7/2^-$ 6.380 &
\cite{npa2004:008-010,bohlen2008:be-band}
\\
. 1/2^-_1 &
$1/2^-$ 2.78, $3/2^-$ 5.59, $5/2^-$ 7.94 &
\cite{npa2004:008-010}
\\
. 1/2^+_1 &
$1/2^+$ 1.684, $3/2^+$ 4.704, $5/2^+$ 3.049 &
\cite{npa2004:008-010,bohlen2008:be-band}
\\
\isotope[11]{Be} . 1/2^-_1 & 
$1/2^-$ 0.320, $3/2^-$ 2.654, $5/2^-$ 3.889 & 
\cite{npa1990:011-012,bohlen2003:be-transfer-ispun02,bohlen2008:be-band}
\\
. 3/2^-_1 &
$3/2^-$ 3.955, $5/2^-$ 5.255 &
\cite{npa1990:011-012,bohlen2008:be-band,bohlen1999:multinucleon-transfer-exotic}
\\
. 1/2^+_1 &
$1/2^+$ 0, $3/2^+$ 3.400, $5/2^+$ 1.783 &
\cite{npa1990:011-012,bohlen2003:be-transfer-ispun02,bohlen2008:be-band}
\\
\hline
\isotope[8]{Be} . 0^+_1 &
$0^+$ 0, $2^+$ 3.03, $4^+$ 11.35 &
\cite{npa2004:008-010,bohlen2008:be-band}
\\
\isotope[10]{Be} . 0^+_1 &
$0^+$ 0, $2^+$ 3.368, $4^+$ 11.760 &
\cite{npa2004:008-010,bohlen2008:be-band,bohlen2007:10be-pickup}
\\
. 0^+_2 & 
$0^+$ 6.179, $2^+$ 7.542, $4^+$ 10.150 &
\cite{npa2004:008-010,bohlen2008:be-band,freer2006:10be-resonant-molecule,suzuki2013:6he-alpha-10be-cluster}
\\
\isotope[12]{Be} . 0^+_1 & 
$0^+$ 0, $2^+$ 2.102 &
\cite{npa1990:011-012,bohlen2008:be-band}
\\
. 0^+_2 &
$0^+$ 2.24, $2^+$ 4.560 &
\cite{npa1990:011-012,bohlen2008:be-band,fortune1994:12be-tp,bohlen2008:12be-transfer,shimoura2003:12be-fragmentation}
\end{tabular}
\end{ruledtabular}
\end{center}
\end{table*}


The values for the rotational parameter $A$ extracted from the
calculations [Fig.~\ref{fig-params}(a,b)] appear to be sufficiently
stable with respect to $\Nmax$ to warrant at least a qualitative
analysis for all the bands.  Recall the convergence pattern
observed for the band members of $\isotope[9]{Be}$ in
Fig.~\ref{fig-energy-9be-conv}.  Although the energies of the band
members shift by several $\MeV$ with increasing $\Nmax$, the relative
energies within the band, which determine the slope $A$, remain
comparatively unchanged.  The largest $\Nmax$ dependence of $A$ is
found for the $\isotope[8]{Be}$, $\isotope[9]{Be}$, and
$\isotope[10]{Be}$ natural parity yrast bands.  However, each
successive step in $\Nmax$ brings a change which is smaller by a
factor of at least $1.5$, suggesting a reasonably rapid approach to a
converged value.

The value of the rotational parameter varies by a factor of $\sim2$
across the different calculated bands.  Indeed, this variation is
apparent by inspection of Figs.~\ref{fig-energy-odd}
and~\ref{fig-energy-even}, from the range of slopes of the energy fit
lines.  We may observe that the variation in $A$ across the bands
follows a trend which correlates with the clustering of bands
according to
$Q_{0,p}/Q_{0,n}$ ratios (see Fig.~\ref{fig-Q0}) discussed in
Sec.~\ref{sec-disc-intrinsic}: into bands with approximately equal
proton and neutron quadrupole moments ($Q_{0,p}/Q_{0,n}\gtrsim 1$) and
those with an enhanced neutron quadrupole moment ($Q_{0,p}/Q_{0,n}\sim
0.5$).  From Fig.~\ref{fig-params}(a,b), the former bands are seen to
be the ``steep'' bands ($A\gtrsim 0.5\,\MeV$), while the latter bands
are seen to be the ``shallow'' bands ($A\lesssim 0.5\,\MeV$).  The
distinction is particularly clear if one compares the yrast and
excited bands \textit{within} a single isotope in
Fig.~\ref{fig-params}(a,b).  Although the connection between
quadrupole deformation (measured by $Q_0$) and rotational moment of
inertia (hence, $A$) is only uniquely defined if model assumptions
about the nature of the rotational motion are
imposed (\textit{e.g.}, Refs.~\cite{bohr1998:v2,eisenberg1987:v1}),
the observed correlation between $Q_{0,p}/Q_{0,n}$ and $A$ is reasonable from
simple arguments.  If the enhanced neutron quadrupole moment
represents a greater extension of the neutron distribution along the
symmetry axis (\textit{e.g.}, due to neutrons in $\sigma$ orbitals) as
interpreted in
Sec.~\ref{sec-disc-intrinsic}, this may be expected to lead to a
greater moment of inertia for rigid rotation, and thus a reduced rotational energy scale
$A$ (Sec.~\ref{sec-rot-energy}).

A wide range in values for the rotational parameter $A$ is also found
experimentally in the $\isotope{Be}$ isotopes, as surveyed by Bohlen
\textit{et al.}~\cite{bohlen2008:be-band}.  Comparing the parameter
values for the calculated and experimental bands, and excluding the
bands in $\isotope[12]{Be}$ from this discussion (to be considered
further below), we observe that the values for $A$ extracted from the
highest $\Nmax$ calculation lie within $\sim0.1\,\MeV$ of the
experimental values.  Qualitatively, the pattern of which bands fall
into the ``steep'' \textit{vs.}\ ``shallow'' clusters is consistent between the
calculations and experiment.  With increasing $\Nmax$, the $A$
parameters for the calculated bands (still excepting
$\isotope[12]{Be}$) are decreasing. Except in the case of the
$\isotope[9]{Be}$ natural parity excited band, this brings the
calculated values further toward the experimental values.

The Coriolis staggering for the calculated $K=1/2$ bands, measured by
the decoupling parameter $a$ [Fig.~\ref{fig-params}(c)], varies 
in both amplitude and sign.  The values obtained for $a$ are sufficiently
stable with respect to $\Nmax$ to permit an examination of the trends
in its value across the bands. The energy staggering in the calculated
$\isotope[7]{Be}$ natural parity yrast band
[Fig.~\ref{fig-energy-odd}(a)] is such that the $J=1/2$, $5/2$,
$\ldots$ levels are raised in energy, and the $J=3/2$, $7/2$, $\ldots$
levels are lowered. This corresponds to a negative value of the
decoupling parameter ($a\approx-1.4$ at $\Nmax=10$).  Note that the
staggering is sufficiently pronounced that the two lowest-$J$ band
members are inverted, as is experimentally observed for this
nucleus~\cite{npa2002:005-007}.  Positive values of the decoupling
parameter are obtained for all other candidate $K=1/2$ bands.  Of
these, the natural parity bands (namely, in $\isotope[9]{Be}$
[Fig.~\ref{fig-energy-odd}(b)] and $\isotope[11]{Be}$
[Fig.~\ref{fig-energy-odd}(d)]) exhibit lesser staggering ($a\lesssim
1$), while the unnatural parity bands (namely, in $\isotope[9]{Be}$
[Fig.~\ref{fig-energy-odd}(c)] and $\isotope[11]{Be}$
[Fig.~\ref{fig-energy-odd}(e)]) exhibit more marked staggering
($a\approx2$).  This categorization of the bands by sign and magnitude
of staggering is consistent with the experimentally
observed pattern.\footnote{\label{fn-nilsson}See also the discussion in Ref.~\cite{caprio2013:berotor}
for a comparison to the Coriolis staggering expected from the 
Nilsson model~\cite{nilsson1955:model,mottelson1959:nilsson-model}, \textit{i.e.}, based on a single unpaired particle in an
axially-symmetric deformed mean field. A more detailed and realistic mean-field analysis might be carried out
by estimating the
signature splitting from a cranked Nilsson
model~\cite{ragnarsson1981:nilsson-11b-11c}.}
 Furthermore, with increasing $\Nmax$, the calculated
values are all decreasing in magnitude, which serves to bring them closer to
the experimental values.

The excitation energies $E_x$ for the calculated bands
[Fig.~\ref{fig-params}(d,e)] vary considerably in their rates of
convergence.  For instance, the excitation energy of the unnatural
parity band in $\isotope[11]{Be}$ decreases by less than $0.1\,\MeV$,
as $\Nmax$ increases from $7$ to $11$, while that of the natural
parity excited band in $\isotope[10]{Be}$ decreases by $\sim9\,\MeV$,
as $\Nmax$ increases over a comparable range, from $6$ to $10$.  However, the rate of
convergence of $E_x$ follows a pattern consistent with the classification of the
bands by the nature of their termination in
Sec.~\ref{sec-disc-patterns}.  In particular, the bands which either
terminate or depart from the rotational energy formula at
the maximal valence angular momentum (\textit{i.e.}, bands in either
of the first two categories defined in Sec.~\ref{sec-disc-patterns})
have excitation energies which exhibit the least dependence on
$\Nmax$~--- specifically, these are the $\isotope[9]{Be}$ natural
parity excited band, $\isotope[9]{Be}$ unnatural parity band, and
$\isotope[11]{Be}$ unnatural parity band.  These may be thought
of as bands which, at
least to some extent, respect the $p$-shell closure.
Note that all excitation
energies are being taken with respect to the natural parity yrast band,
which likewise falls into one of the first two categories, so the
convergence of these
excitation energies serves to compare the convergence of two bands with
ostensibly \textit{similar} shell structures.
In contrast, bands which exhibit no discontinuity in energies at the maximal valence
angular momentum (\textit{i.e.}, bands in the third category) have
excitation energies with the largest $\Nmax$ dependence~---
specifically these are the remaining natural parity excited
bands, in $\isotope[10]{Be}$, $\isotope[11]{Be}$, and
$\isotope[12]{Be}$.

In comparing the excitation energies with experiment, let us first
consider the bands with the smaller $\Nmax$ dependence, for which a
more concrete comparison can be made.  The near degeneracy of the
natural and unnatural parity yrast bands in $\isotope[11]{Be}$ is
qualitatively reproduced.  Quantitatively, the experimental excitation
energy of the unnatural parity band is
$\sim0.22\,\MeV$,\footnote{Although the experimental $1/2^+$
  \textit{ground state} of $\isotope[11]{Be}$ has \textit{unnatural
    parity}~\cite{npa1990:011-012}, a situation described as ``parity
  inversion'' (Sec.~\ref{sec-results-calc}), the \textit{lowest energy
    band} (as defined by comparing $E_0$ parameters) is actually the
  \textit{natural parity} $K^P=1/2^-$ yrast band.  The perhaps
  counterintuitive distinction between ``ground state band'' and
  ``lowest energy band'' arises from the difference in staggering
  between the bands: the larger staggering of the unnatural parity
  band depresses the energy of the $1/2^+$ band head state below that
  of the $1/2^-$ band head state, despite the higher overall band energy.} while the excitation energy for
this band from the \textit{ab initio} calculations appears to be
robustly $\sim1.9\,\MeV$.  In $\isotope[9]{Be}$, the calculated
natural parity excited band and unnatural parity band lie within
$\sim1.2\,\MeV$ of each other, at the highest $\Nmax$ considered,
approximately reproducing the near degeneracy found in experiment.
These calculated energies are $\sim3$--$4\,\MeV$ above the
experimental values, but are descending toward the experimental values
with increasing $\Nmax$.

Moving on to the bands with larger $\Nmax$ dependence, the calculated
excitation energies still lie $\sim4$--$10\,\MeV$ above the
experimental excitation energies, at the highest $\Nmax$ considered, but are rapidly falling with $\Nmax$.  It is therefore
difficult to make a concrete comparison.  
For $\isotope[12]{Be}$, in
particular, the excitation energy of the calculated excited band is
nearly halved~---  from $\sim11\,\MeV$ to
$\sim6\,\MeV$~--- as $\Nmax$
increases from $6$ to $10$.  It is therefore reasonable to expect the
calculated excited band to fall to near degeneracy with the
calculated ground state band.  The experimental excited band in
$\isotope[12]{Be}$ is likewise low-lying, at
$\sim2.2\,\MeV$.\footnote{More highly excited $0^+$ bands in
  $\isotope[12]{Be}$, with
  $E_0\approx6.4\,\MeV$~\cite{bohlen1999:11be-12be-molecular-enpe99}
  and $E_0\approx10.8\,\MeV$~\cite{freer1999:12be-breakup-molecular},
  have also been reported.}  
Strong mixing may therefore be expected~--- for both the calculated
and experimental bands~--- 
which could significantly affect the extracted values for the band
parameters.  
  Recall that the calculated yrast band for
$\isotope[12]{Be}$ [Fig.~\ref{fig-energy-even}(c)] is radically truncated, terminating at the maximal $p$-shell
valence angular momentum $J=2$.  Although experimental
candidate states for $4^+$ and even $6^+$ members of both the yrast
band~\cite{fortune1994:12be-tp,npa-738-333-2004-bohlen-be-c-transfer}
and excited band~\cite{bohlen2008:12be-transfer} have been proposed
(see Ref.~\cite{bohlen2008:be-band}), these identifications are not
based on firm spin-parity assignments and therefore do not provide
definitive grounds for comparison.  In the shell model
framework, the ground state and excited $0^+$ states in
$\isotope[12]{Be}$ have been interpreted as being mixtures of $0\hw$
closed  $p$-shell configurations and $2\hw$ configurations involving
promotion of two neutrons to $sd$-shell orbitals (\textit{e.g.},
Refs.~\cite{barker1976:a12-shell,fortune1994:12be-tp}).  
At least schematically,
such shell model configurations may be associated with the
$p$-shell terminating and non-terminating bands in the present calculations.
Experimental results from, \textit{e.g.}, knockout~\cite{navin2000:12be-knockout,pain2006:12be-knockout}, transfer~\cite{kanungo2010:12be-transfer}, and charge
exchange~\cite{meharchand2012:12be-charge-exchange} reactions have
been interpreted as suggesting significant mixing of these $0\hw$ and
$2\hw$ configurations in both the ground and excited $0^+$ 
states.

\section{Conclusion}
\label{sec-concl}

The emergence of rotational patterns is observed in \textit{ab initio}
NCCI calculation with realistic interactions, despite the principal
challenge in identifying collective structure in NCCI calculations~---
namely, the weak convergence of many of the relevant observables.
Eigenvalues and other calculated observables are dependent upon both
the truncation $\Nmax$ and the oscillator length parameter (or $\hw$)
for the NCCI basis.  Although it may be possible to extrapolate the
values of calculated observables to their values in the full, infinite-dimensional
space~\cite{maris2009:ncfc,furnstahl2012:ho-extrapolation,coon2012:nscm-ho-regulator},
such methods are still in their formative stages and have not been
developed for the crucial electric quadrupole observables.  It is
therefore particularly notable that quantitatively well-developed and
robust signatures of rotation may be observed in the present results.
That this is possible reflects the distinction between convergence of
\textit{individual} observables, taken singly, and convergence of
\textit{relative} properties, such as ratios of excitation energies or
ratios of quadrupole matrix elements.  It is these latter relative
properties which are essential to identifying rotational dynamics and
which are found to be sufficiently converged to yield stable
rotational patterns at currently achievable $\Nmax$ truncations~---
along with the well-converged magnetic dipole observables.

We find that rotational structure is pervasive in the NCCI
calculations of the yrast and near-yrast regions of the $p$-shell
$\isotope{Be}$ isotopes.  (Comparable results for $\isotope[12]{C}$
are reported in Ref.~\cite{maris2012:mfdn-hites12}.)  With suitable
basis extrapolation methods, we may hope to determine the extent to
which \textit{ab initio} calculations can provide quantitatively
precise predictions of rotational band properties (\textit{e.g.}, the
rotational formula energy parameters, including Coriolis decoupling).
Even the present calculations, unconverged and unextrapolated, suggest
a notable degree of qualitative consistency with the experimentally observed
bands (Fig.~\ref{fig-params}).

While the emergence of rotation appears to be robust across different
\textit{ab initio} interactions~\cite{maris2013:ncci-chiral-ccp12}, it remains to be seen in what ways the
quantitative details of the rotation may be sensitive to the
interaction.  More broadly, different
\textit{ab initio} computational approaches can more or less readily
access different correlations within the nuclear wave functions.
Therefore, it is of particular interest to examine the emergence of
rotation in approaches other than the NCCI framework.  For instance,
taking $\isotope[7]{Be}$ (or its mirror nucleus $\isotope[7]{Li}$) as
an example, quantum Monte Carlo calculations readily reproduce the
$3/2$-$1/2$-$7/2$-$5/2$ yrast angular momentum
sequence~\cite{pieper2004:gfmc-a6-8}, reflective of a $K=1/2$ band
with strong negative Coriolis staggering.  Electromagnetic
observables of the type considered in the rotational analysis can
also readily be calculated by such methods for this
nucleus~\cite{pervin2007:qmc-matrix-elements-a6-7}.

Although the presence of rotational patterns suggests the separation
of the wave function into rotational and intrinsic factors, it leaves
open the question of the underlying stucture of the intrinsic state,
as noted in Sec.~\ref{sec-disc}.  Hints to this intrinsic structure may be
obtained from the intrinsic observables
(Sec.~\ref{sec-disc-intrinsic}), as well as the rotational parameters
of the bands (Sec.~\ref{sec-disc-params}).  For the $\isotope{Be}$ isotopes, we
may seek to determine the extent to which aspects of the rotational
structure may be understood within different physical frameworks,
including cluster structure (\textit{e.g.}, $\alpha+\alpha$,
$\alpha+n+\alpha$, \textit{etc.}), Nilsson-like single-particle motion
in a mean field, and valence shell structure (as suggested by the
termination effects).  Elliott $\grpsu{3}$
symmetry~\cite{elliott1958:su3-part1,harvey1968:su3-shell} provides
the classic theoretical explanation of the emergence of rotation
within the valence shell, while $\grpsptr$
symmetry~\cite{rowe1985:micro-collective-sp6r} provides a natural
context for the emergence of collective deformation and rotational
degrees of freedom in the full multi-shell configuration space.

One may observe that the present discussion represents a
phenomenological rotational analysis, in the traditional experimental
sense, but of a large set of observables taken from calculations of
the \textit{ab initio} nuclear many-body problem.  Having full access
to the calculated wave functions permits analysis of an extended set
of rotational observables, many of which are difficult or impossible
to access experimentally.  For example, we have presented results
independently probing proton and neutron degrees of freedom, and
orbital and spin degrees of freedom, through multipole operators
involving the corresponding contributions individually.

Furthermore, having direct access to the calculated wave functions, we
may also hope to extract information on the collective structure of
the rotational nuclear eigenstates from other measures of the wave
function correlations, such as density
distributions~\cite{cockrell2012:li-ncfc}, spin and orbital angular
momentum contributions~\cite{johnson-PREPRINT-spin-orbit}, and symmetry
decompositions~\cite{dytrych2008:sp-ncsm-deformation}.  In this
regard, it is important to note that collective $\grpsu{3}$
correlations, consistent with the nuclear symplectic model, have been
clearly demonstrated in calculations for $\isotope[6]{Li}$,
$\isotope[6]{He}$, and $\isotope[8]{Be}$, carried out directly in a
symmetry-adapted $\grpsu{3}$-based coupling
scheme~\cite{dytrych2013:su3ncsm}.


\section*{Acknowledgements} 

Discussions of experimental data with R.~Smith, M.~Freer, and
Tz.~Kokalova are gratefully acknowledged.  We thank A.~E.~McCoy and R.~Smith for
comments on the manuscript.  This material is based upon work
supported 
by the US~Department of Energy under Grants
No.~DE-FG02-95ER-40934, DESC0008485 (SciDAC/NUCLEI), and
DE-FG02-87ER40371, by the US~National Science Foundation under
Grant No.~0904782, and by the Research Corporation for Science
Advancement through the Cottrell Scholar program. This research used resources of the National Energy
Research Scientific Computing Center (NERSC) and the Argonne
Leadership Computing Facility (ALCF), which are US~Department of
Energy Office of Science user facilities, supported under Contracts
No.~DE-AC02-05CH11231 and DE-AC02-06CH11357, and computing resources
provided under the INCITE award ``Nuclear Structure and Nuclear
Reactions'' from the US~Department of Energy, Office of Advanced
Scientific Computing Research.


\appendix
\section{Single-particle estimate for intrinsic quadrupole moment}
\label{app-weisskopf}

In heavier mass regions, rotational collectivity is commonly measured
(\textit{e.g.}, Refs.~\cite{bohr1998:v2,casten2000:ns}) by comparison
of rotational transition strengths to the Weisskopf single-particle
estimate.  This is more properly a measure of the \textit{number} of
participating nucleons and of the quadrupole \textit{deformation} of
the intrinsic wave function, rather than of its \textit{rotational}
nature \textit{per se}, \textit{i.e.}, the separation of degrees of
freedom embodied in~(\ref{eqn-psi}).  Nonetheless, it is worthwhile to
keep the single-particle scale of transition strengths in mind, as
this also provides a natural scale for the deviations from rotational
strengths to be expected from noncollective admixtures in rotational
states.

Briefly, Weisskopf~\cite{weisskopf1951:estimate} estimates a typical
scale
\begin{math}
\scrA
\approx
e\tfrac35(4\pi)^{-1/2}R^2,
\end{math}
for the matrix element $\tme{\varphi_b}{Q_{2\mu}}{\varphi_a}$ of the
electric quadrupole operator for a single-particle transition between
two states $\varphi_a$ and $\varphi_b$, where we use $R=r_0A^{1/3}$
with $r_0=1.2\,\fm$.  This suggests the Weisskopf estimate $B_W\equiv
\scrA^2$ for $B(E2)$ strengths, commonly termed the \textit{Weisskopf unit} (W.u.).

However, quadrupole moments and $E2$ transition strengths vary greatly
within a rotational band simply due to the angular momentum factors
in~(\ref{eqn-e2-rme}) (see Fig.~\ref{fig-rotor-e2}).  Therefore, while
one could choose one particular quadrupole moment $Q(J)$ or transition
strength $B(E2;J\rightarrow J-\Delta J)$ within a band for comparison
to Weisskopf's estimate (\textit{e.g.}, the $2\rightarrow0$ transition
strength is commonly quoted for $K=0$
bands~\cite{raman2001:systematics}), this comparison cannot be made
consistently across bands of different $K$.

We therefore find it more meaningful to apply Weisskopf's estimate
directly to the \textit{intrinsic} matrix element of the quadrupole
operator, \textit{i.e.}, taking this to be of single-particle strength
and seeing what magnitude this would imply for the \textit{intrinsic}
quadrupole moment $Q_0$, thereby defining a single-particle scale
relative to which enhancement can be judged.\footnote{For instance,
for a well-deformed rotational nucleus with $A\approx150$, a typical
value of
$B(E2;2\rightarrow0)\approx100\,\mathrm{W.u.}$~\cite{raman2001:systematics}
thus corresponds to $Q_0/Q_{0,W}\approx 22$.  Here we have used
$B(E2;2\rightarrow0)=(eQ_0)^2/(16\pi)$ 
for a $K=0$ band,
by~(\ref{eqn-BE2}).}  Taking
$\tme{\phi_K}{Q_{20}}{\phi_K}\approx\scrA$ in~(\ref{eqn-e2-rme}) gives
a single-particle estimate
\begin{equation}
\label{eqn-Q0W}
Q_{0,W}=\tfrac35\left(\tfrac45\right)^{1/2}r_0^2A^{2/3}.
\end{equation}

For the $\isotope{Be}$ isotopes considered here, the value of
$Q_{0,W}$ ranges from $\sim2.8\,e\fm^2$ for $A=7$ to $\sim4.1\,e\fm^2$
for $A=12$.  Of course, the traditional nuclear radius formula
embodied in Weisskopf's estimate, and thus in~(\ref{eqn-Q0W}), is of
only limited validity in these light nuclei.


\providecommand{\APSLONG}{}
\providecommand{\ELSEVIER}{}




\end{document}